\documentclass[twocolumn]{aastex631}

\usepackage{natbib}
\usepackage{amsmath, mathtools, amsthm, amssymb}
\usepackage{hyperref}
\usepackage{graphicx}
\usepackage{threeparttable}

\shorttitle{Double-exponential disk galaxies from TNG100}
\shortauthors{Sarkar et al.}

\begin{document}
% Editage delete, add, comment

%%%%%%%%%%%%%%%%%%%%%%%%%
%%%%%%% 	Title, Author, Affil- Settings
%%%%%%%%%%%%%%%%%%%%%%%%%
\title{Massive double-exponential galaxies with extended LSB emission: An IllustrisTNG exploration II - are they unique?}  %a comparative study with the overall disk populations}

\email{suchira.sarkar@theory.tifr.res.in}
\email{kanak@iucaa.in}
\email{anshuman@iucaa.in}
%%	Author1
\author{Suchira Sarkar}
\affil{Tata Institute of Fundamental Research, Mumbai- 400005, Maharashtra, India}

%%	Author2
\author{Kanak Saha}
\affil{Inter-University Centre for Astronomy \& Astrophysics, Pune-411007, Maharashtra, India}

\author{Anshuman Borgohain}
\affil{Inter-University Centre for Astronomy \& Astrophysics, Pune-411007, Maharashtra, India}

\begin{abstract}
The massive disk galaxies (stellar mass $>=\mathrm{10^{11}M_{\odot}}$) in the local universe ($z<0.1$) encode challenging aspects of galaxy formation physics. We have studied one class of such galaxies - the giant low surface brightness galaxies (gLSBs) that show a hybrid morphology - a central Sersic, an inner high surface brightness, exponential stellar disk surrounded by an outer, extended low surface brightness disk, using large volume IllustrisTNG-100 simulation. This is motivated by the recent observation of several such systems and an earlier work by \citet{Sarkar2025} from IllustrisTNG-50. We find that despite using a $\sim$ 10 times larger volume simulation, only $\sim$11\% of massive disk galaxies show such morphology, consistent with findings from TNG50. This shows robustly that such galaxies are not common in the local universe. We compared them with the typical disk galaxies from observations, and single-exponential gLSBs from TNG100. We show that while the inner disk components lie well within the central surface brightness vs. radial scale length distribution of the typical disks, the outer LSB disks lie significantly at the outskirt, occupying a unique space. Their median bulge-to-total ratio is $\sim$ 0.2 making them disk dominated, similar to the typical disks and the single-exponential gLSBs. We find their median dark matter mass fraction to be high ($\sim$ 0.9), typical of gLSBs. The outer LSB disks reach fainter central surface brightness and larger scale length values than the single-exponential gLSBs. Our exploration will be useful in understanding their formation mechanism across cosmic time. 
\end{abstract}


\section{Introduction}
The formation and evolution of massive disk galaxies of stellar mass $>=$ $10^{11}\mathrm{M_{\odot}}$ in the nearby universe (z $<$ 0.1) is a challenging phenomenon to explore in the field of galaxy formation and evolution. At low redshift, the galaxy stellar mass function, which represents the number density of galaxies per logarithmic mass bin as a function of stellar mass, shows a sharp fall-off around the stellar mass $\sim$ $10^{10.6-10.8}$ - $10^{11}$ $\mathrm{M_{\odot}}$ \citep{Baldry2012,liwhite2009}. At such high mass end, the formation and evolution of galaxies are significantly determined by the interplay between strong AGN feedback from the central supermassive black holes, and galaxy mergers across cosmic time \citep{Bower2006, Croton2006, McIntosh2008, Robotham2014, Beckman2017, Adams2021, Jacksonetal2022}. Therefore, studying massive disk galaxies in the local universe can tell us about the rich galaxy formation theory. 

One class of disk galaxies of stellar mass of $>=10^{11}$ $\mathrm{M_{\odot}}$ are the giant low surface brightness galaxies (gLSBs). They harbor disk with central surface brightness fainter than $\mathrm{\sim 22.5 \ mag \ arcsec^{-2}}$ in optical B-band (fainter than the typical night sky background) \citep{McGaugh1996,Rosenbaumetal2009, Das2013, Hagenetal2016}, and radial, exponential scale length $>$ 10 kpc \citep{Sprayberry1995,Das2013, Saburovaetal2023}. There are several gLSBs that show a complex, layered morphology. The galaxy has a central bulge/bar, and the stellar disk is best described as a combination of an inner, high surface brightness exponential disk surrounded by an outer, extended low surface brightness envelope. This came into light for the first time with the serendipitous discovery of Malin 1 \citep{Bothunetal1987}. The stellar disk in Malin 1 extends almost upto 100 kpc at z =0.08, tracing it large spiral structure \citep{MooreParker2006, Galazetal2015, Boissieretal2016}. The central region of the galaxy,however, appears like a bright lenticular system with a bulge, bar and a high surface brightness inner disk component \citep{Barth2007, Sahaetal2021}. A detailed surface brightness modeling of Malin 1 were performed by \cite{Sahaetal2021} to bring out the complex hybrid morphology using Canada-France-Hawaii Telescope imaging data. A few other massive, double-exponential disk galaxies, which are less extreme in size than Malin 1, have been observed with similar hybrid morphology (e.g, UGC 1382 by \cite{Hagenetal2016, Saburova2026}; NGC 2841 by \cite{Zhangetal2018}, UGC 1378 by \cite{Saburovaetal2019}). For a detailed discussion see \citet{Sarkar2025}. In a very recent literature, \citet{Kolganov2026} has found a small fraction of gLSBs observed by the Subaru telescope to have HSB plus extended LSB disks. These discoveries invoke several interesting questions:

1.) How frequent should they be among the massive disk galaxy populations in the local universe ?\\

2.) How unique are their properties compared to the typical disk galaxy populations in the local volume as well as the typical single-exponential gLSBs?\\  

3.) What gives rise to the formation of such layered, hybrid morphology and maintain the stability of the extended faint envelope till the nearby universe?\\ 

4.) Very importantly, what do they tell us about the galaxy formation physics at the very high mass end? \\

The lack of a statistically large sample indicates that these systems could be genuinely challenging to form or a large fraction of them could remain hidden due to the lack of detection of their faint outskirt. Thus our understanding of galaxy formation in the high mass regime remains potentially incomplete. This motivates the use of a large volume high resolution simulations to explore the above questions. Because simulations are devoid of observational biases as well as it can help us to understand the galaxy formation scenario by looking at the snapshots across cosmic time.

In an earlier work by \citet{Sarkar2025}, we used IllustrisTNG50, the state-of-the art, cosmological gravo-magneto-hydrodynamical simulation suit \citep{Nelsonetal2019,Pillepichetal2019} of volume 51.7 Mpc$^{3}$ to estimate the statistics of such galaxies, for the first time. We found that at $z$=0, 12\% of massive disk galaxies can show such hybrid morphology. We studied their star formation rate, color profiles and Baryonic Tully-Fisher relation. The main focus of the above paper was to find such galaxies from simulations for the first time. There was no attempt to understand how unique these galaxies are compared to the typical disk galaxies as well as the routinely observed single-exponential gLSBs in terms of the structural parameters space, bulge-to-total ratio, dark matter fraction etc. In this context, it is worth looking at a larger volume simulation with sufficiently high resolution, and understand the robustness of the theoretical estimate, eventually create a combined, robust sample to compare their properties with typical disk galaxies. This is essential to explore the formation and evolution scenario of such galaxies across simulation snapshots. In this paper, we use IllustrisTNG-100 (simulation volume 110.7 Mpc$^{3}$) \citep{Nelsonetal2018} to detect the massive, double-exponential galaxies and combine the sample with that found using TNG50. A large volume is expected to be helpful in probing massive galaxies in the local redshift. 

We also detect the massive, single-exponential gLSBs from TNG100 for comparison. We note that, \citet{Zhuetal2023} had identified gLSBs from IllustrisTNG100 with single exponential disk. Since the primary selection criteria was very different we do not expect to see a strong overlap between their gLSB sample and the gLSBs detected in this paper. We discuss this later in the paper as well. 

%The primary selection was based on the HI gas mass and HI disk size. The galaxies were then modeled for surface photometry, and, interestingly, they all turned out to be best modeled by a single, extended low surface brightness disk. We note, that importantly, the double-exponential HSB plus LSB galaxies were not studied. Also, their sample is not limited to the most massive galaxies, instead, it extends upto around $10^{10}\mathrm{M_{\odot}}$ which is much larger range for searching for gLSBs compared to what we select here. Therefore, we do not expect to see a large overlap between our single disk LSB sample and the gLSB sample shown in \citep{Zhuetal2023}. We mention any overlap that we find later in this paper. We note that, it is worth looking at the massive, single exponential gLSBs detected based on the stellar disk central surface brightness and scale length criteria, as done in this paper, so that we can compare the double-exponential gLSBs with them. 

The paper is organised as follows. We describe the data in Section 2, the method to find the double-exponential disk galaxies from simulation in Section 3, present the results in Section 4, comparison of physical properties in Section 5, and finally conclude in Section 6.

\section{Data}
IllustrisTNG simulation\citep{Marinaccietal2018, Naimanetal2018, Nelsonetal2018, Pillepichetal2018a, Pillepichetal2018b,Springeletal2018} is a suit of large-volume, cosmological gravo-magneto-hydrodynamical simulations run using the moving-mesh code AREPO \citep{Springel2010}, succeeded by the original Illustris simulation \citep{Geneletal2014, Vogelsbergeretal2014a, Vogelsbergeretal2014b, Sijackietal2015}. The underlying galaxy formation model is described in \citet{Weinbergeretal2017} and \citet{Pillepichetal2018a}. IllustrisTNG was calibrated to reproduce the galaxy stellar mass function at $z=0$, galaxy stellar-to-halo mass relation, galaxy stellar mass-size relation etc. The cosmologically motivated initial conditions, assuming an updated cosmology consistent with the Planck results \citep{Planck2016}, where $\Omega_{\Lambda,0}$=0.6911, $\Omega_{m,0}$=0.3089, $\Omega_{b,0}$=0.0486, h=0.6774, $\sigma_{8}$=0.8159, $n_{s}$=0.9667, were used for running the simulations. Three simulations were run with physical boxes with cubic volumes of roughly 50, 100, and 300 Mpc side lengths, referred to as TNG50, TNG100, and TNG300, respectively. TNG100-1 has a volume of 110.7$^{3}$ Mpc$^{3}$, mean baryonic mass and dark matter mass resolutions of $1.4\times 10^{6}\mathrm{M_{\odot}}$ \& $7.5\times 10^{6}\mathrm{M_{\odot}}$, respectively \citep{Nelsonetal2018,Pillepichetal2018b}. It is the highest resolution run among the three runs of TNG100. The Plummer equivalent gravitational softening of stars and dark matter (collisionless components), is 740 parsecs at $z$ = 0. The adaptive softening for gas component is set to a minimum of 185 comoving parsec. Haloes are identified here using a FoF algorithm, and galaxies, i.e, subhaloes within a halo are identified using the SUBFIND algorithm \citep{Springeletal2001}.

\section{Method}
\subsection{Selection of massive disk galaxies from IllustrisTNG100-1}
There are 694 galaxies in TNG100-1 at $z=0$ of subhalo stellar mass $>=10^{11}M_{\odot}$. We use their synthetic, idealized SDSS (Sloan Digital Sky Survey \citep{Yorketal2000})-g,r band equivalent images, made using radiative transfer code SKIRT, and publicly available as one of the TNG supplementary data catalogs \citep{Gomezetal2019}. Each image, containing one galaxy at a random orientation, was created assuming that the galaxy is observed at $z$ = 0.0485 by a local observer, and the pixel scale is set to 0.396 arcsec. We do not add any observational realism (e.g., sky background, noise, psf) to these images. For a detailed discussion, see \citet{Gomezetal2019} and \citet{Sarkar2025}. We determined disk galaxies among the above sample of 694 following the process described below. 

The importance of visual morphology has recently been discussed by \citet{Pillepich2024}, specifically for TNG simulation. %Here in this paper, we make visual analysis as the primary identification criteria. 
Here in this paper, the visual inspection of all the galaxy images has been done by two of the authors independently, and the morphology is finalised only when both authors agree. In the first step, we identify the galaxy images with the signature of strong tidal disturbances/interactions, jets, etc, and discard them (121 cases). Because, the above features can make their 2D light modeling, and robust determination of the faint disk in the galaxy's outskirts complicated. For simplicity, we discard 79 edge-on cases as well. %since it could be complicated to model the faint disk. 
To find out the disk galaxies among the rest of the sample, we check the following - if there is clear visual signature of spirals, bars, clumps, or rings then the galaxies can be considered as disk galaxies \citep{Cheng2011}. If such features are not present, if the light distribution of the galaxy appears mostly smooth and overall roundish then the galaxies are considered to be ellipticals \citep{Bailard2011}. %We note that a large fraction of ellipticals have isophotes that are either boxy isophotes or disk isophotes. (Bender \& Kormendy, 1996). 
Each galaxy was further checked in face-on and edge-on view using the visualisation tool available on the TNG website. A disk galaxy shows clear, thin disk-like structure in edge-on view. %whereas an elliptical galaxy shows ellipsoidal structure in edge-on view. 
Based on this, we find 139 galaxies to be ellipticals, 73 galaxies to be lenticulars with bulge/bar but lacking spiral arms in disk, and 134 galaxies to be spirals, and thus a sample of 207 visually identified disk galaxies. We further applied kinematic and Sersic index criteria on them. For the rest of the 148 galaxies the visual morphology do not agree among the authors, and those are not used in further analysis.

A galaxy can be kinematically classified as a disk galaxy if it is dominated by rotational motion compared to random motion of the stars. We used the publicly available catalog by \citet{Du2019} where galaxies of $M_{star}>10^{10}M_{\odot}$ from IllustrisTNG100-1 were identified as disk galaxies if their $k_{rot}$ ($< v_{\phi} ^{2}> / <v^{2}>$) $>=$ 0.5 measured using stellar particles within 30 kpc radius from the center of the galaxy, a routinely used criteria in literature. %Since the $K_{rot}$ values are not available, we cannot comment on how strongly rotation dominated they are.  the disk galaxy IDs are provided. We matched our visually selected disk galaxy sample with the above catalog and noted the matched cases. 
On the other hand, a single Sersic modeling of disk galaxies usually gives Sersic index $n<=2.5$ \citep{Lange2015,Sachdeva2020}. Hence, we run 2D Sersic modeling on the images of the 207 galaxies using GALFIT \citep{Pengetal2002}.

We find that 42 galaxies satisfy the above kinematic and Sersic index criteria, leading to sample 1 - the strongest sample of disk galaxies. 39 galaxies are found to belong to the kinematic catalog but have Sersic index $n>$ 2.5 (sample 2). We note that, a very bright bulge in a disk galaxy can give rise to a higher n value. %The range of the best-fit Sersic index for sample 2 is (). 
Another 39 galaxies (sample 3) do not belong to the kinematic catalog but have $n<=$2.5. %If the galaxies show see clear visual signature of disk morphology then it is justified to consider them as disks. 
We note that if there is a significant stellar halo in a disk galaxy then it can cause smaller $k_{rot}$ values. 
The remaining 87 disk galaxies do not satisfy either of the criteria. For simplicity, we only choose the spirals (30 cases) and make sample 4. The aim is not to include any elliptical galaxy in the sample mistakenly. Thus we construct a total of 150 massive disk galaxies as the parent disk galaxy sample for finding the double-exponential cases. 
This is around three times larger than the parent sample used from TNG50 in \citet{Sarkar2025}.

\subsection{Finding the double-exponential gLSBs, and the single-exponential gLSBs}
We follow an overall similar logic for the detection of the double-disk cases as described in \citet{Sarkar2025} with some differences in the technical aspects. Therefore, we have briefly described the method here. 

A Sersic plus single exponential disk modeling is performed on each galaxy image using GALFIT. The Sersic is used here to model the bulge and bar together \citep{Simard2011, PahwaSaha2018}. We extracted the radial surface brightness profile of the modeled Sersic and exponential components using their ellipticity, position angles, and images obtained from GALFIT. The sum of these profiles, i.e., the model surface brightness profile is overplotted on the radial surface brightness profile of the galaxy, extracted by fitting concentric elliptical isophotes along the semi-major axis on the galaxy image using IRAF \citep{Jedrzejewski1987}.

We note the galaxies for which the model profile leaves significant, systematic positive residue in the outskirt compared to the galaxy surface brightness profile. Such residuals may arise due to an outer disk component, spiral arms, contamination from the stellar halo light etc. Therefore, we add a second exponential disk in the model, and check if the residue gets overall randomised and reduced. At the same time, we visually inspect if the GALFIT residual maps show positive, negative residual values distributed in an overall randomised fashion in the galaxy outskirt, instead of a systematic positive bias. Finally, we verified that galaxy images show that there is disk-like light distribution till the outskirt. We discard the cases where an additional exponential component overfits the galaxy light profile, or, it could be fitted only at the expense of an unphysically shrinked inner disk.
Based on the above analysis, we find 16 galaxies ($\sim$ 11\% of the parent disk sample) out of the 150 galaxies to be best described by central Sersic plus inner HSB plus outer, extended LSB disk. We present their detailed GALFIT models, surface brightness profiles as well as their models with single disk component, and the measure of goodness of fit in the following section. 

The above statistics brings out an interesting result. Despite exploring a larger volume simulation than TNG50 ($\sim 10$ times), and $\sim$ 3 times larger sample of massive disk galaxies as the parent sample, we found a similar fraction of massive, double-exponential galaxies. This helps us to show robustly that such galaxies are not common in the local universe.

Finally, We identify 32 galaxies ($\sim$ 21\%, i.e, twice the sample of double-disk cases) which are best modeled as single exponential gLSBs. We note that \citet{Zhuetal2023} had used the HI mass and disk size as the primary selection criteria to find gLSBs using TNG100. The galaxies were then considered to be represented by Sersic plus single exponential disk. Also, the stellar mass in their sample extends upto around $10^{10}\mathrm{M_{\odot}}$, a much larger range compared to our sample of massive galaxies. We note that, it is worth looking at the massive, single exponential gLSBs detected solely based on the stellar disk central surface brightness and scale length criteria with detailed 2D light modeling, as done in this paper, so that we can compare the double-exponential gLSBs with them.

\section{Results for the double-exponential and the single-exponential gLSBs}

\begin{figure*}
\centering
\includegraphics[width = 0.31\textwidth, height=0.24\textwidth]{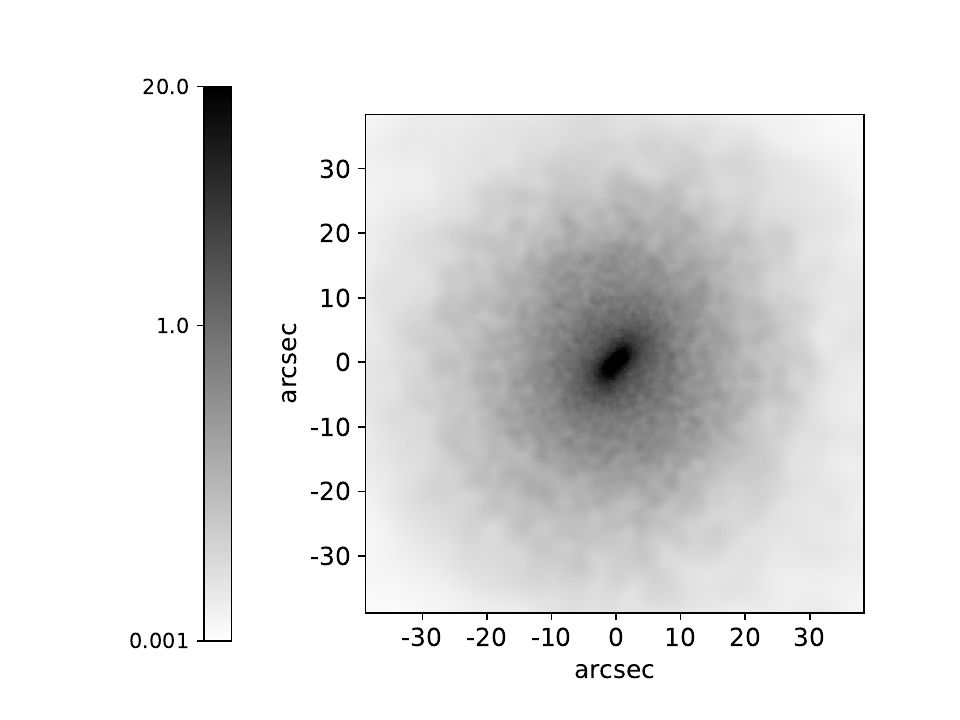}
\includegraphics[width = 0.31\textwidth, height=0.22\textwidth]{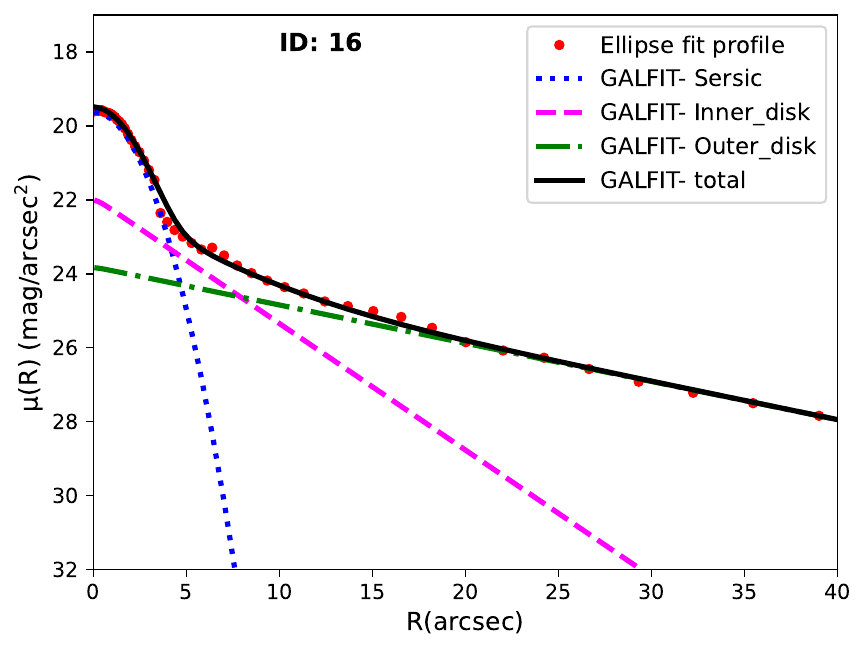}
\includegraphics[width = 0.31\textwidth, height=0.24\textwidth]{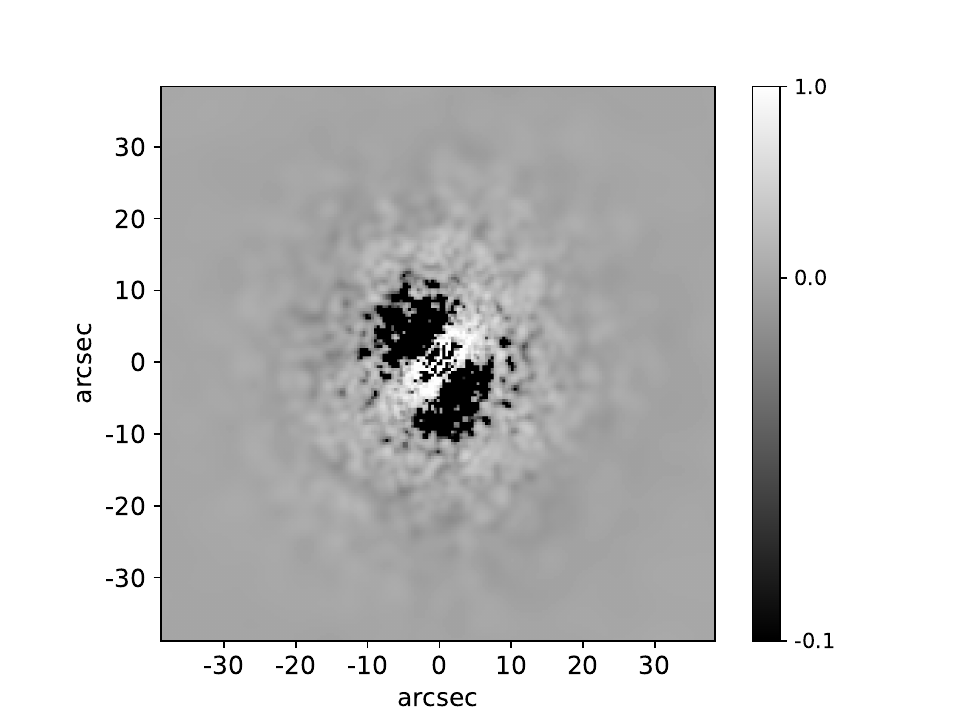}
\medskip
\includegraphics[width = 0.33\textwidth, height=0.24\textwidth]{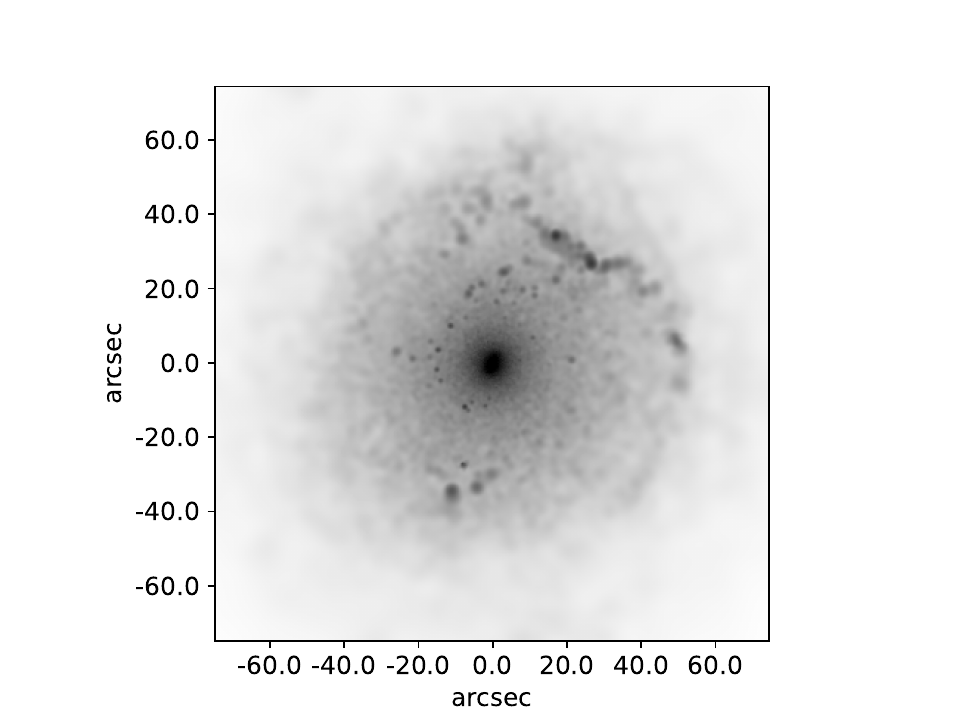}
\includegraphics[width = 0.31\textwidth, height=0.22\textwidth]{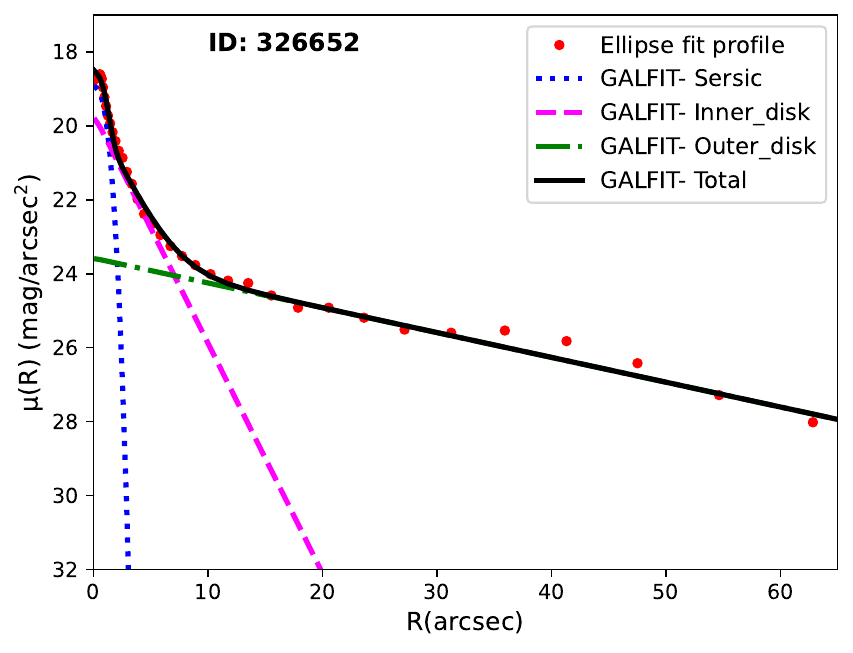}
\includegraphics[width = 0.33\textwidth, height=0.24\textwidth]{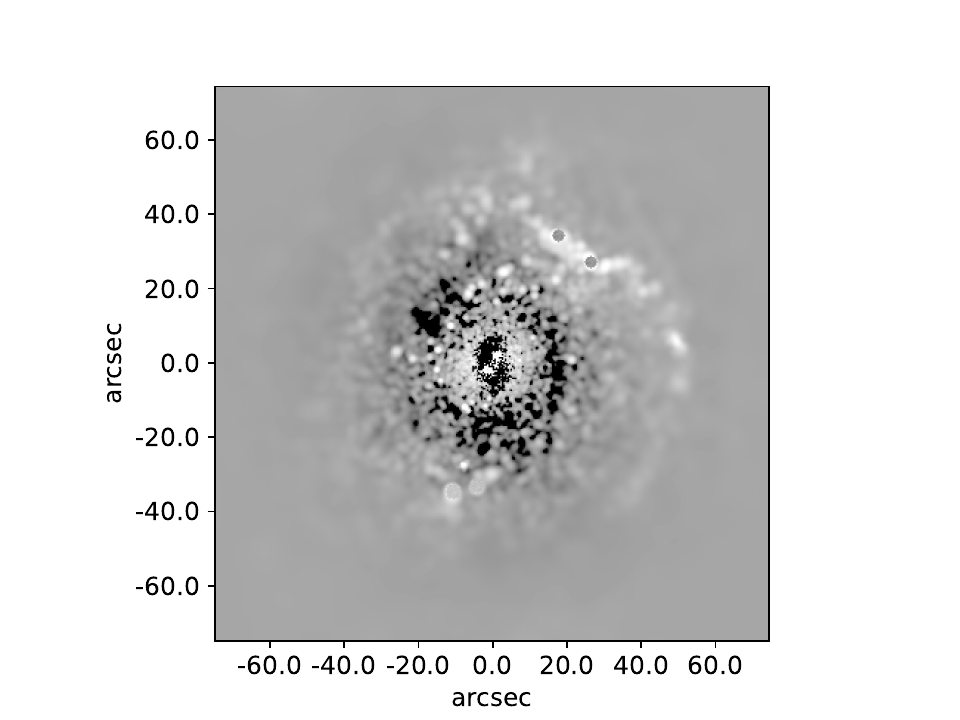}
\medskip
\includegraphics[width = 0.33\textwidth, height=0.24\textwidth]{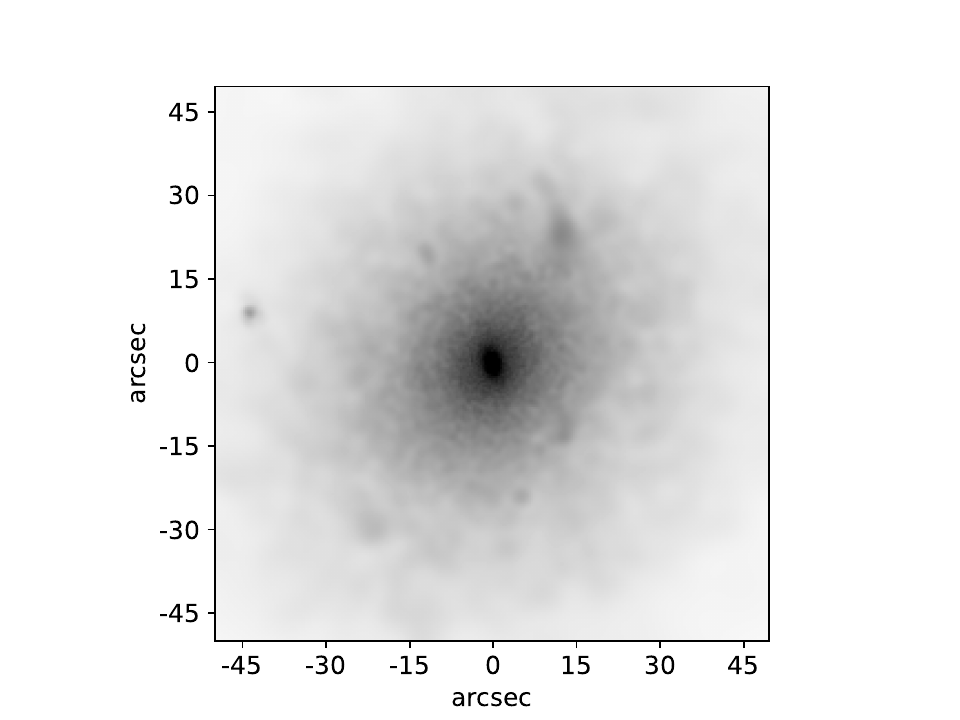}
\includegraphics[width = 0.31\textwidth, height=0.22\textwidth]{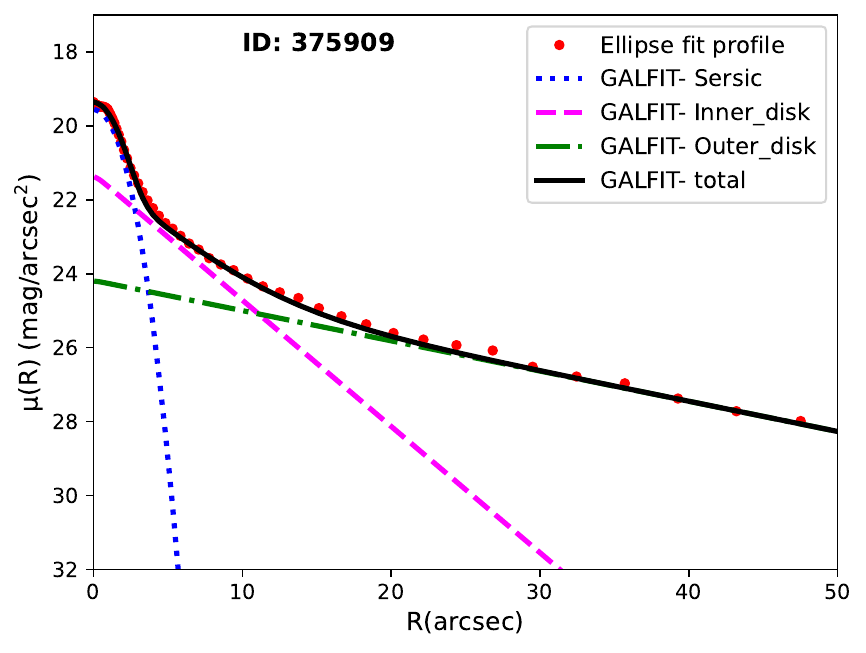}
\includegraphics[width = 0.33\textwidth, height=0.24\textwidth]{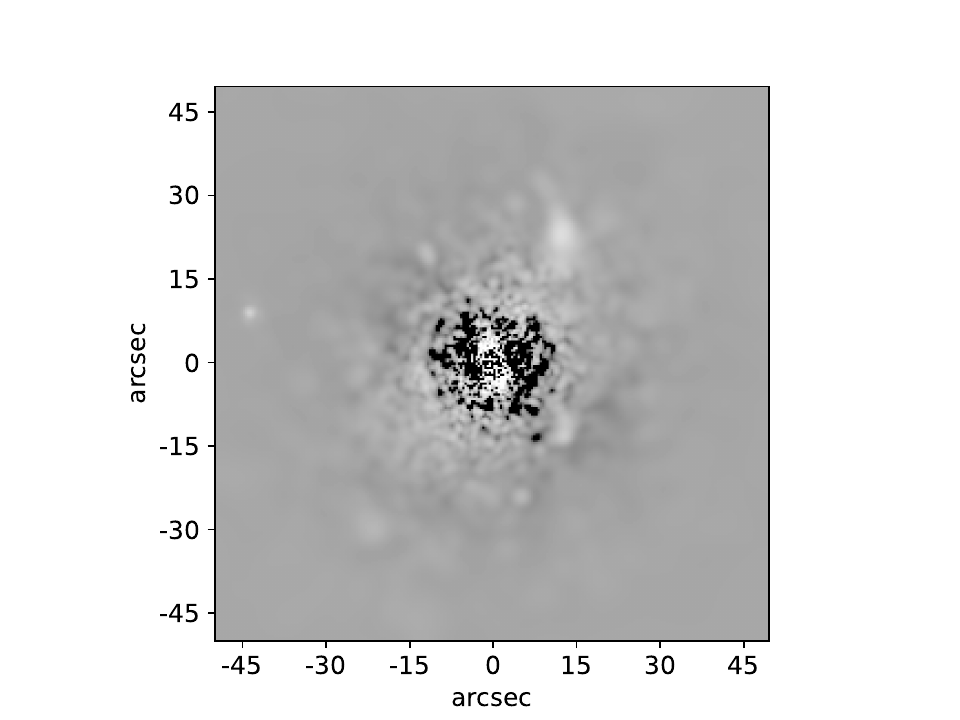}
\medskip
\includegraphics[width = 0.33\textwidth, height=0.23\textwidth]{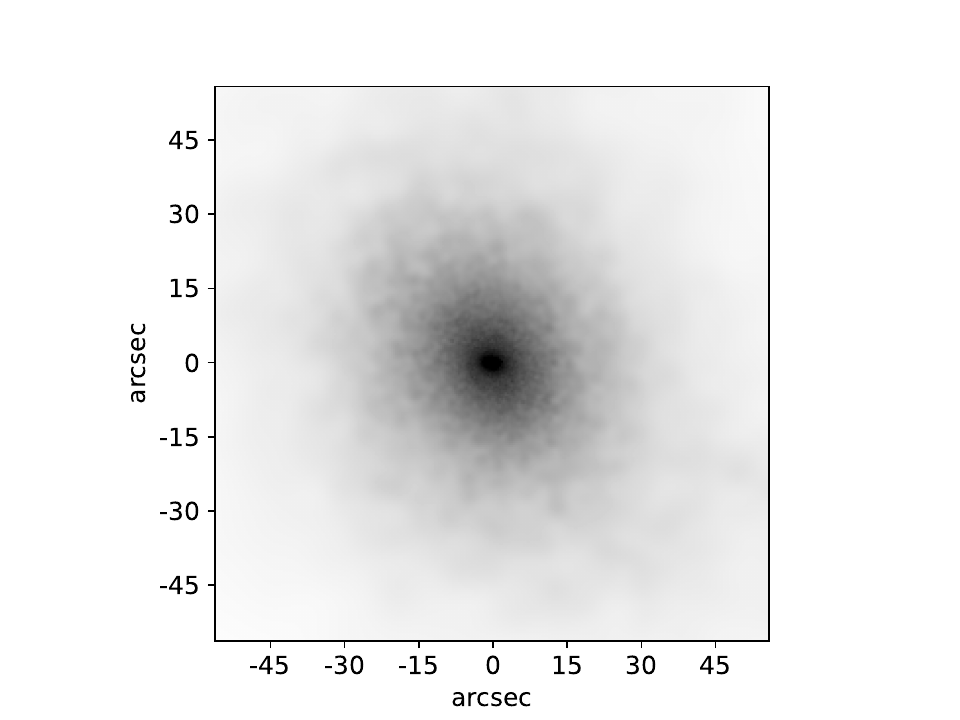}
\includegraphics[width = 0.31\textwidth, height=0.21\textwidth]{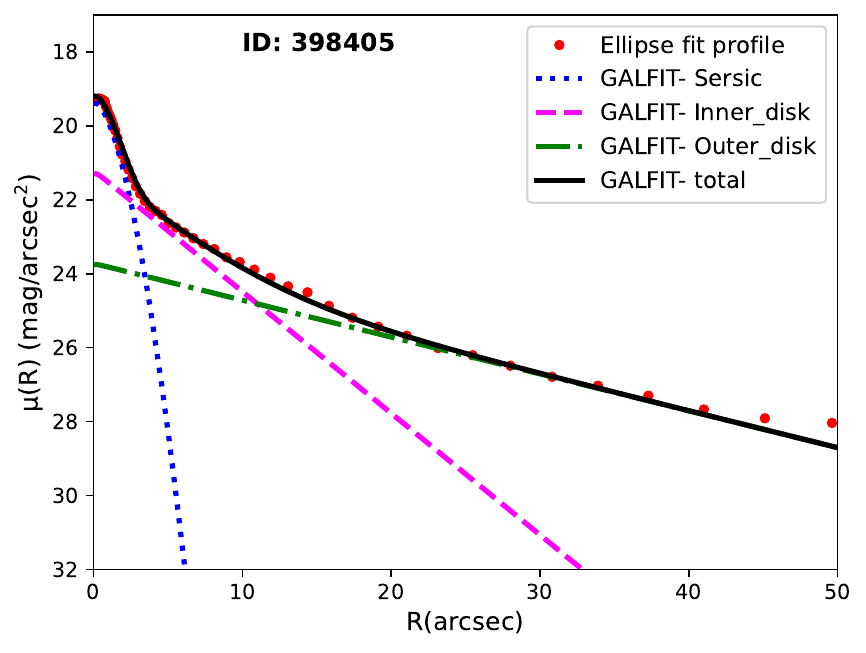}
\includegraphics[width = 0.33\textwidth, height=0.23\textwidth]{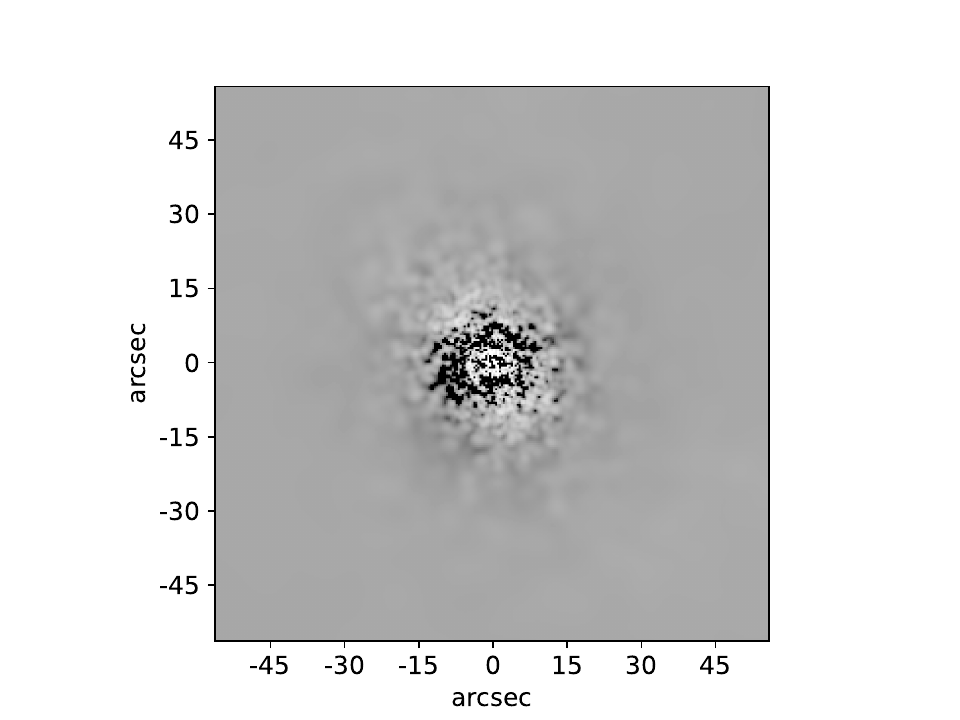}
\medskip
\includegraphics[width = 0.33\textwidth, height=0.23\textwidth]{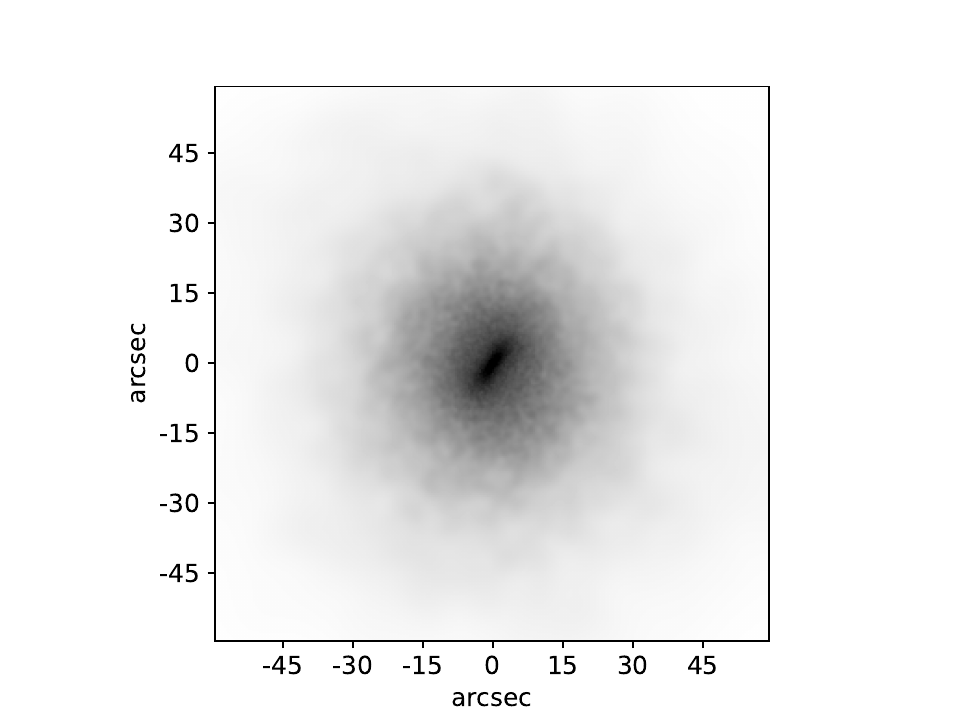}
\includegraphics[width = 0.31\textwidth, height=0.21\textwidth]{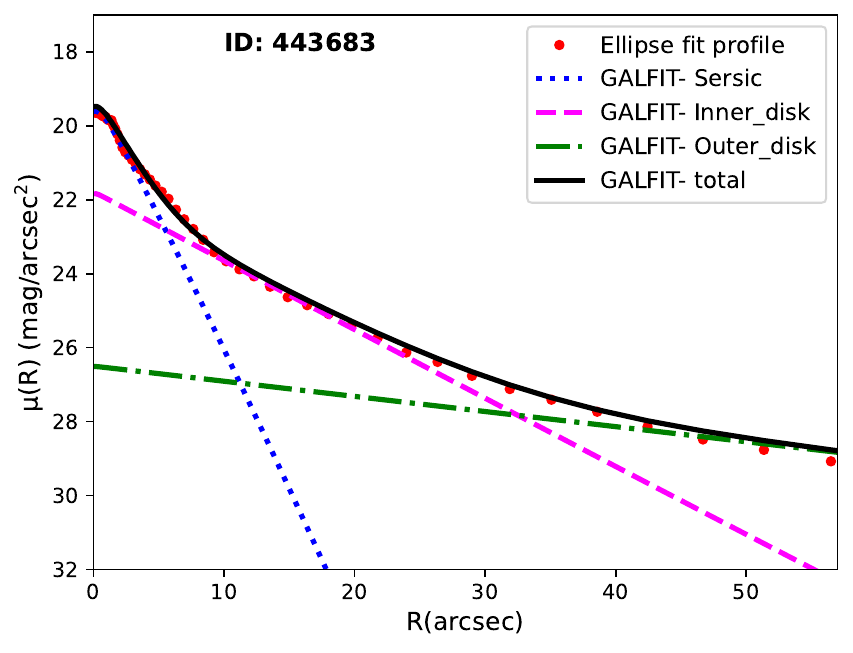}
\includegraphics[width = 0.33\textwidth, height=0.23\textwidth]{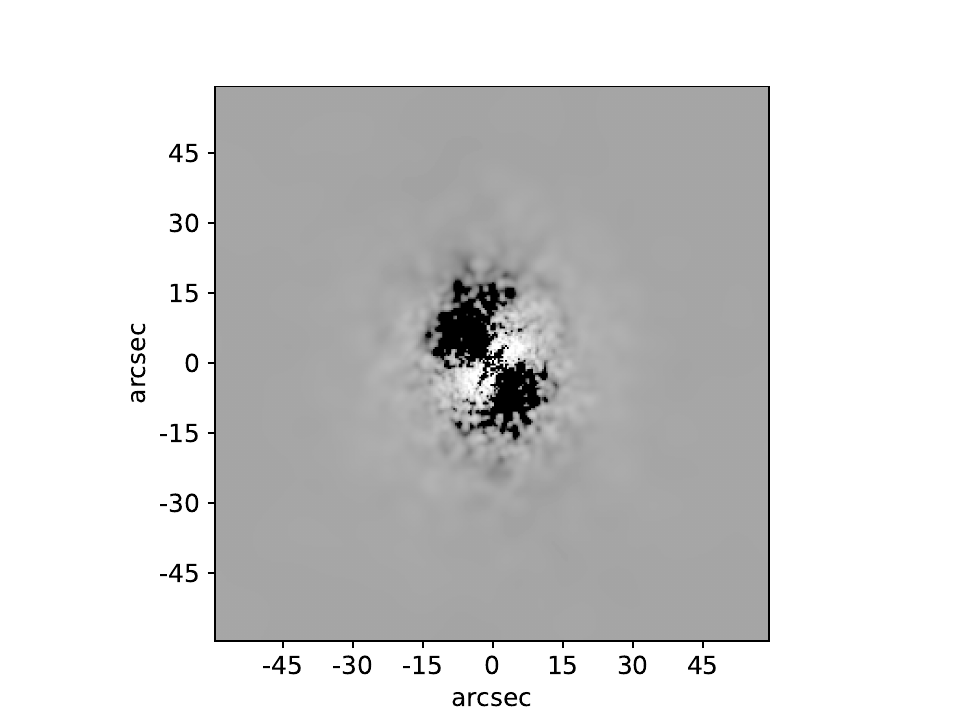}
%\captionsetup{belowskip=0pt}
\caption{Simulated galaxies (subhaloes) with double-exponential i.e., HSB plus LSB disk structures obtained from disk sample 1. The subhalo IDs are mentioned in each case. All the images are displayed in the image unit. \textbf{Left column}: Synthetic SDSS g-band idealized images of the galaxies, shown using the same color bar (in log stretch). \textbf{Middle column}: Plot of the radial surface brightness profiles of the galaxies (red data points) obtained from elliptical isophote fitting of the images. The profiles of the best-fit central S\'{e}rsic (blue dotted curve), inner (magenta dashed curve) exponential HSB, and outer exponential LSB disk (green dash-dotted curve), obtained from GALFIT modeling, are over-plotted on the galaxy surface brightness profiles. The total model, i.e., the sum of the components is denoted by the black solid curve. \textbf{Right column}: Residual images obtained from GALFIT modeling are shown using the same color bar (in log stretch).}
\label{fig:fig1}
\end{figure*}

We present the double-exponential disk galaxy images, surface brightness profiles, and GALFIT residual images in Figure \ref{fig:fig1},\ref{fig:fig3}, \ref{fig:fig4}, \ref{fig:fig5} for disk sample 1, 2, 3 \& 4, respectively. The best-fit central Sersic, inner \& outer exponential disk profiles are over-plotted on the radial surface brightness profile of the galaxies. We note that the model profiles overall reproduce the galaxy light profiles well. Since we do not model the spiral structure they are present in the residual images. However, this does not affect our LSB determination.

\begin{figure*}
\centering
\includegraphics[width = 0.33\textwidth, height=0.24\textwidth]{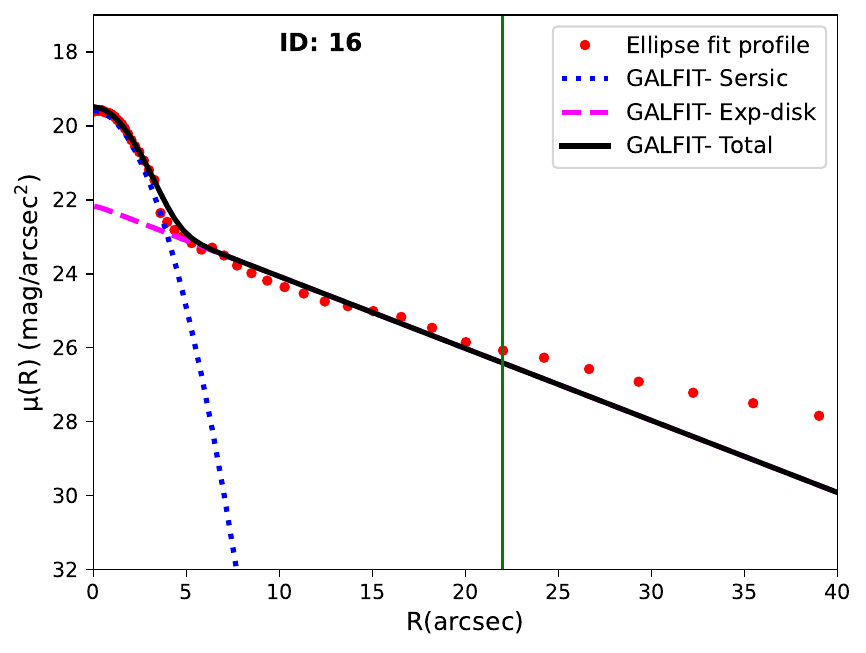}
\includegraphics[width = 0.33\textwidth, height=0.24\textwidth]{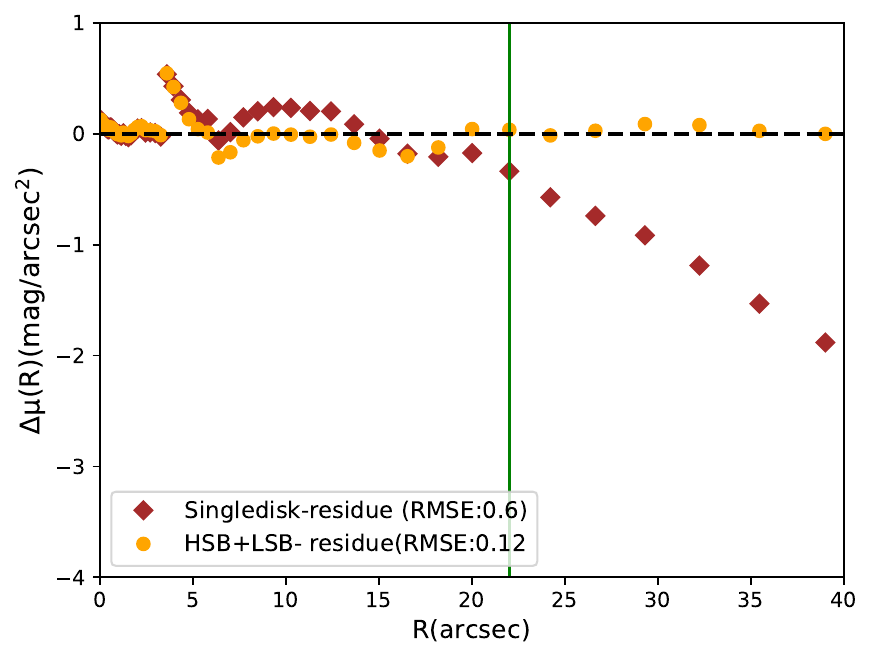}
\medskip
\includegraphics[width = 0.33\textwidth, height=0.24\textwidth]{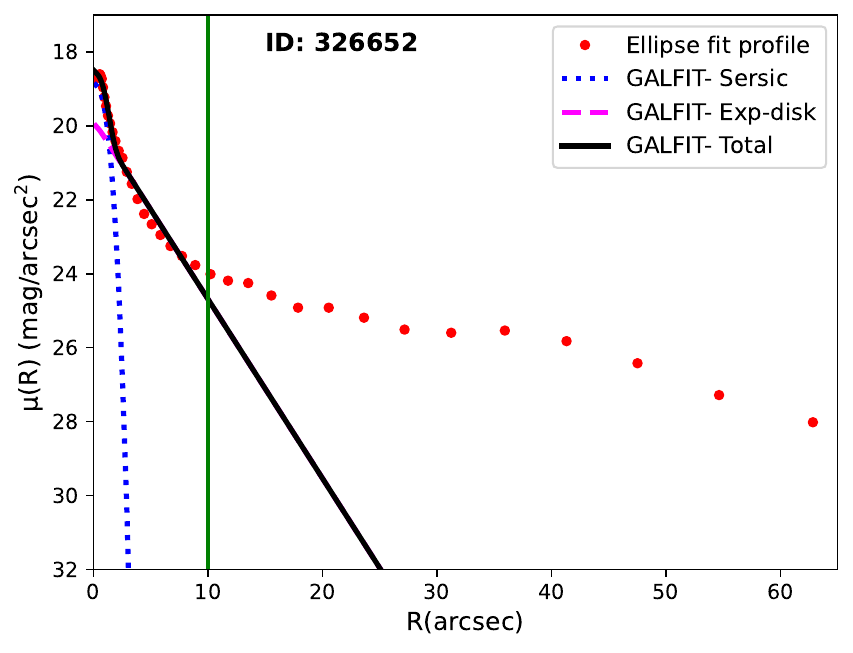}
\includegraphics[width = 0.33\textwidth, height=0.24\textwidth]{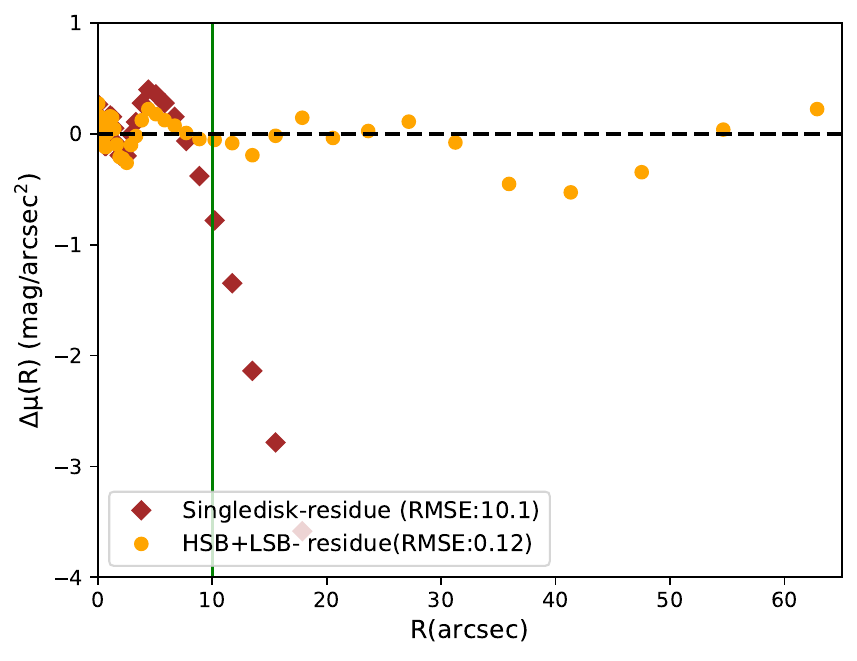}
\medskip
\includegraphics[width = 0.33\textwidth, height=0.24\textwidth]{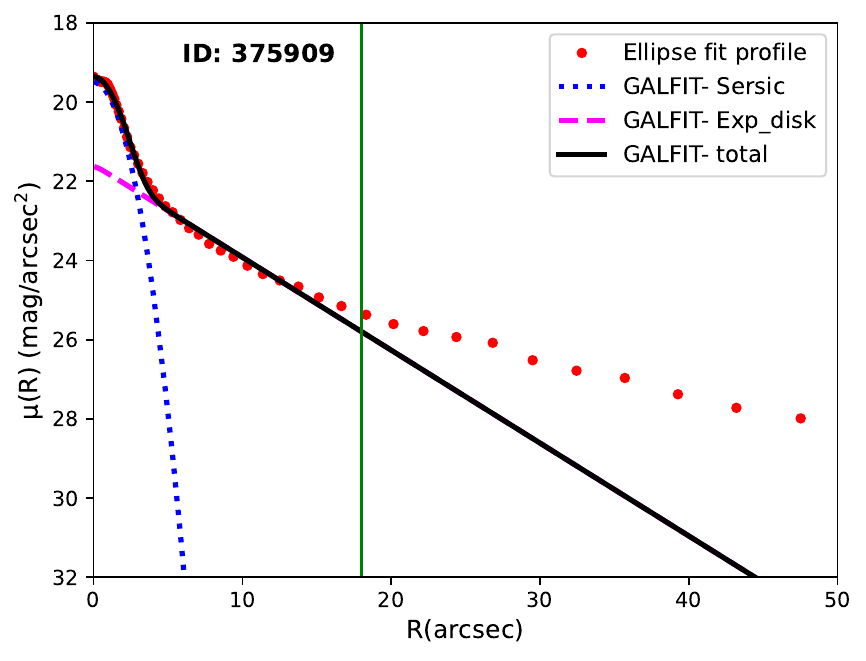}
\includegraphics[width = 0.33\textwidth, height=0.24\textwidth]{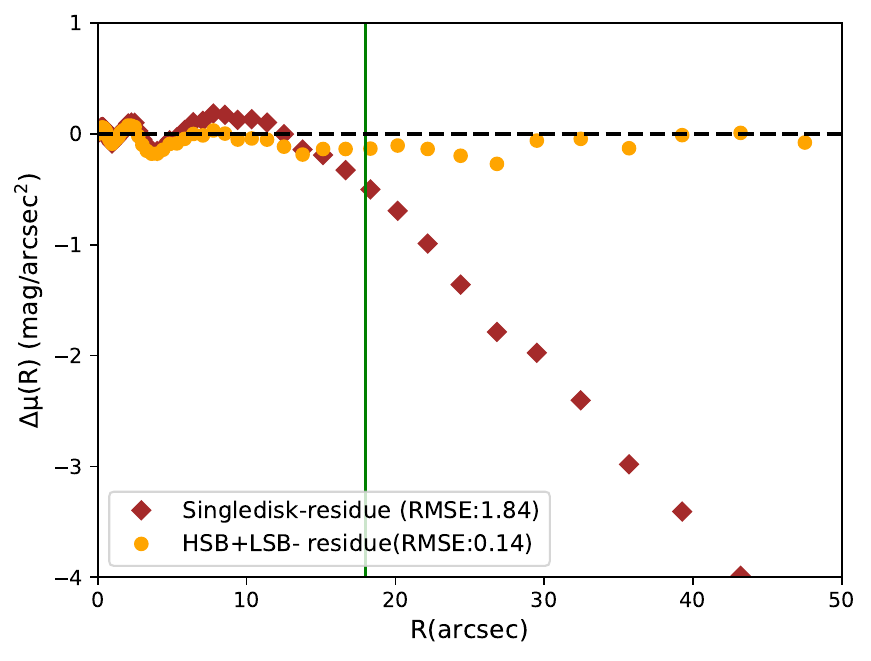}
\medskip
\includegraphics[width = 0.33\textwidth, height=0.24\textwidth]{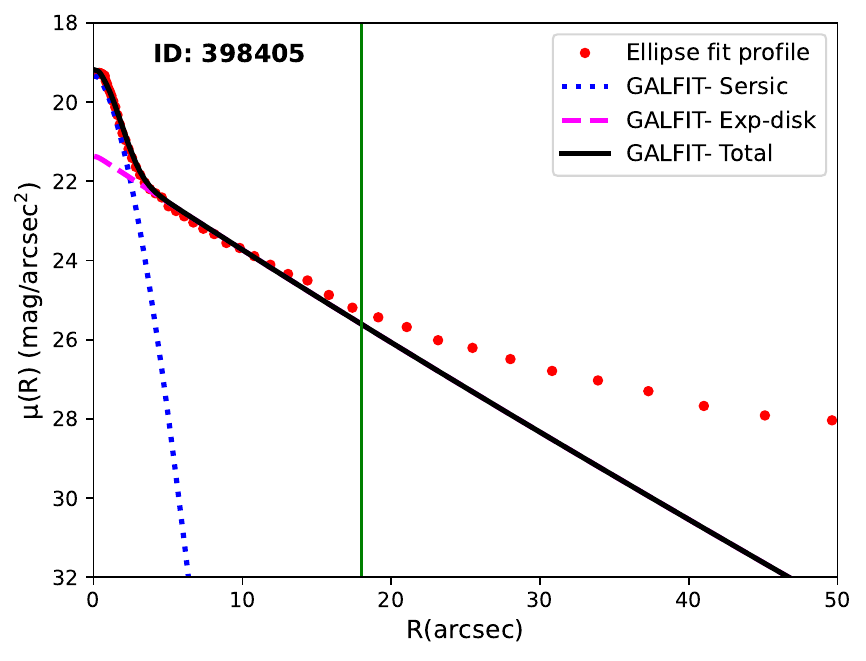}
\includegraphics[width = 0.33\textwidth, height=0.24\textwidth]{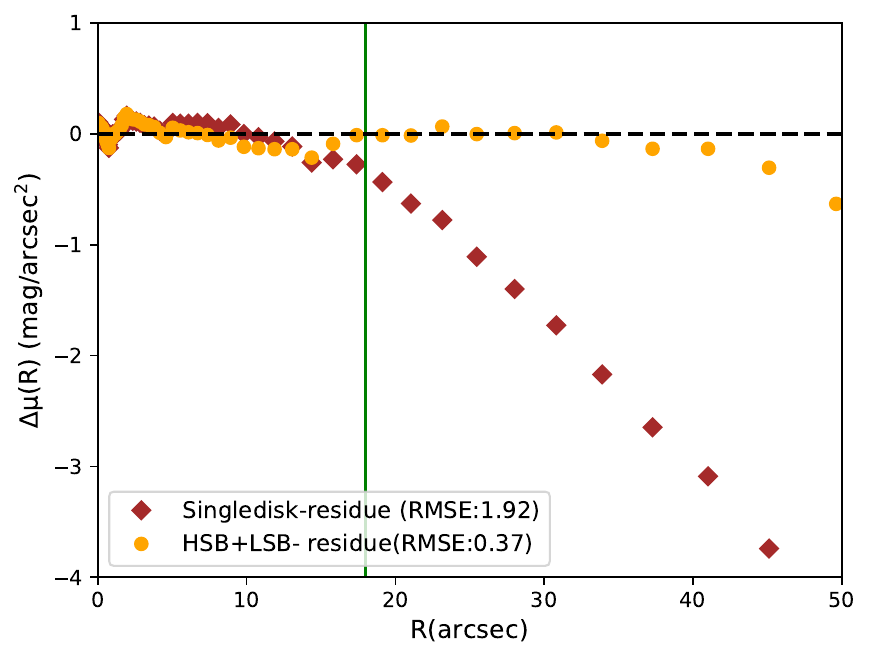}
\medskip
\includegraphics[width = 0.33\textwidth, height=0.24\textwidth]{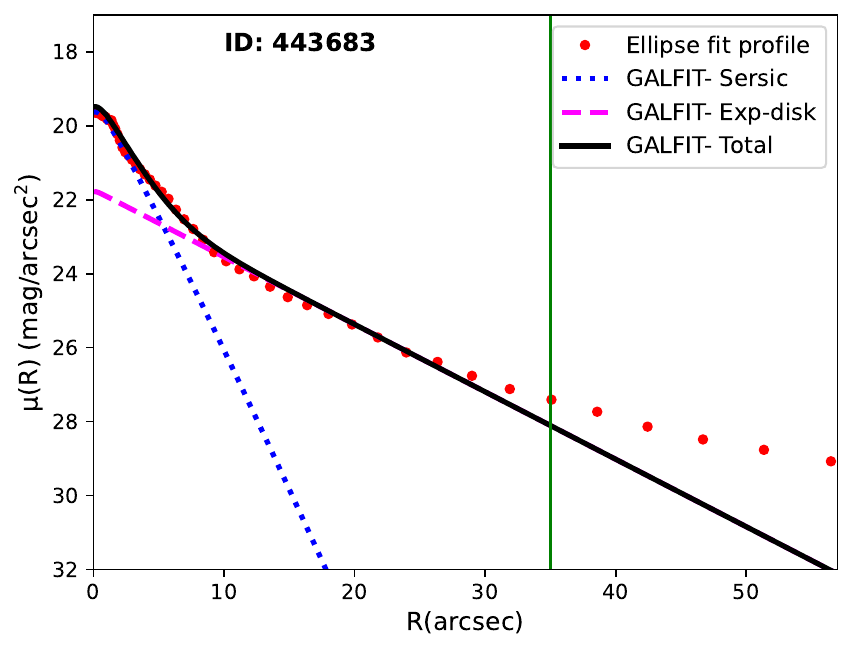}
\includegraphics[width = 0.33\textwidth, height=0.24\textwidth]{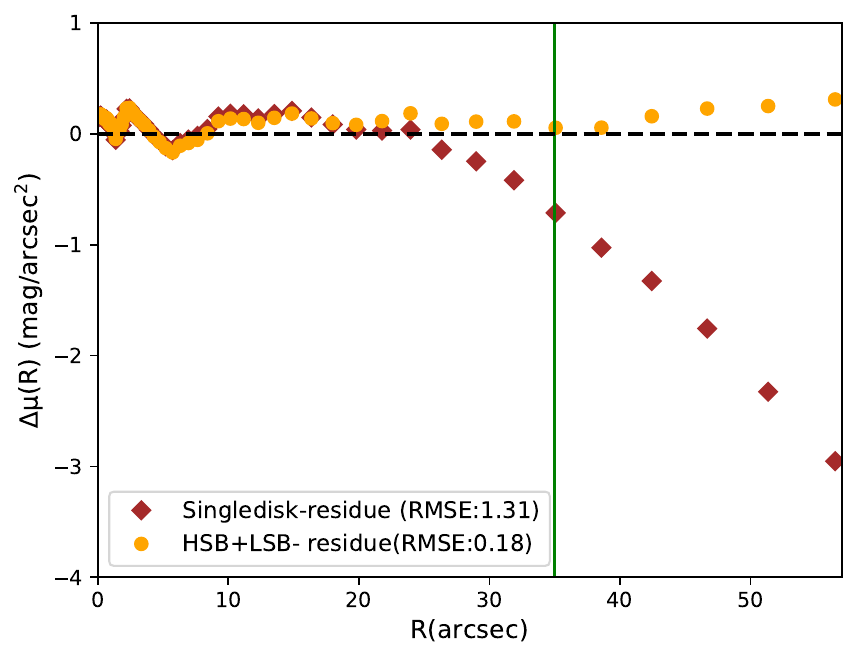}
\caption{\textbf{Left column}: The S\'{e}rsic plus single exponential disk modeling of each of the double-disk galaxies of disk sample 1. The black curve is the total model, i.e, sum of the components. The vertical green line denotes the semi-major axis distance from where the model profile starts deviating from the galaxy surface brightness profile (red data points). \textbf{Right column}: The difference between the total galaxy surface brightness profile and the modeled surface brightness profile representing the radial variation of the residue. The brown diamonds represent the residue corresponding to the single-disk model, and the orange circles represent the residue corresponding to the double-disk model. The root mean square errors (RMSE) calculated using the above residues are mentioned.}
\label{fig:fig2}
\end{figure*}
\noindent 

To show that an additional exponential component was indeed required we present the profile of the double-disk cases corresponding to the Sersic plus single exponential disk model in Figure \ref{fig:fig2} for sample 1, and in the Appendix for the rest of the samples. We denote the semi-major axis distance (by a vertical green line) beyond which the above model deviates clearly from the galaxy surface brightness profile. The residual flux, calculated as the difference between the galaxy and the model profile, corresponding to both single and double exponential models are presented. We also estimated RMS errors using them, which represent a global measure of the goodness of the models. Both of these together clearly show significant improvement in the residue, and hence the necessity of the second component. 

We note that, a few of our double-exponential galaxies are classified as single disk  gLSBs in \citet{Zhuetal2023} (IDs: 330955, 350486, 364807, 365493, 371259, 391260). They had assumed a single disk fitting for all galaxies and did not do a detailed modeling of each galaxy separately as shown here to find out the possibility of a Sersic plus HSB plus LSB component. 

\begin{figure*}
\centering
\includegraphics[width = 0.32\textwidth, height=0.26\textwidth]{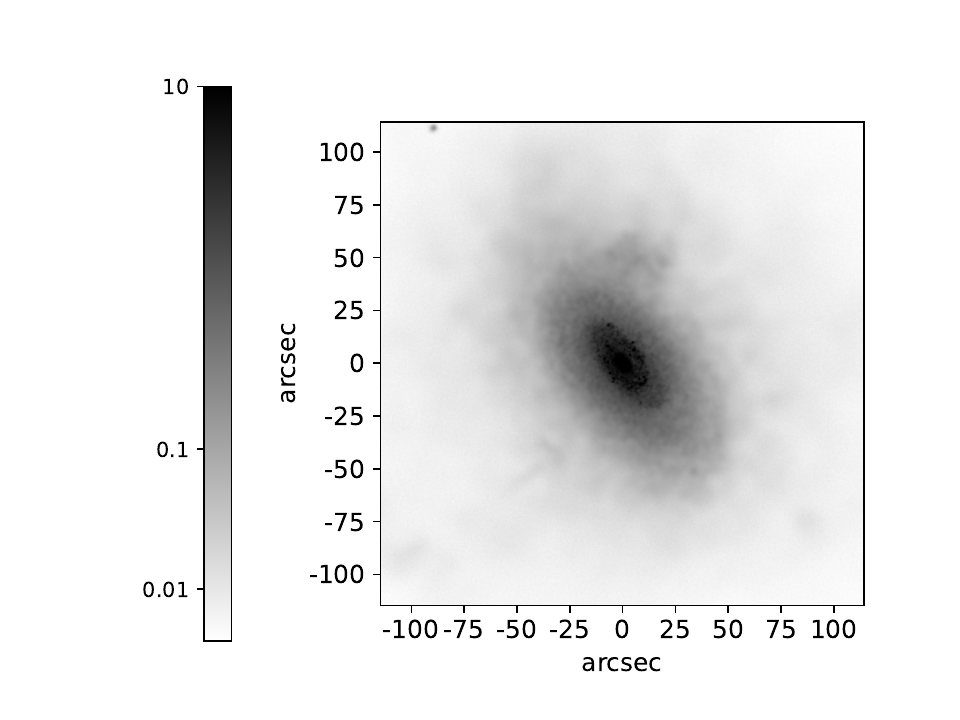}
\includegraphics[width = 0.31\textwidth, height=0.23\textwidth]{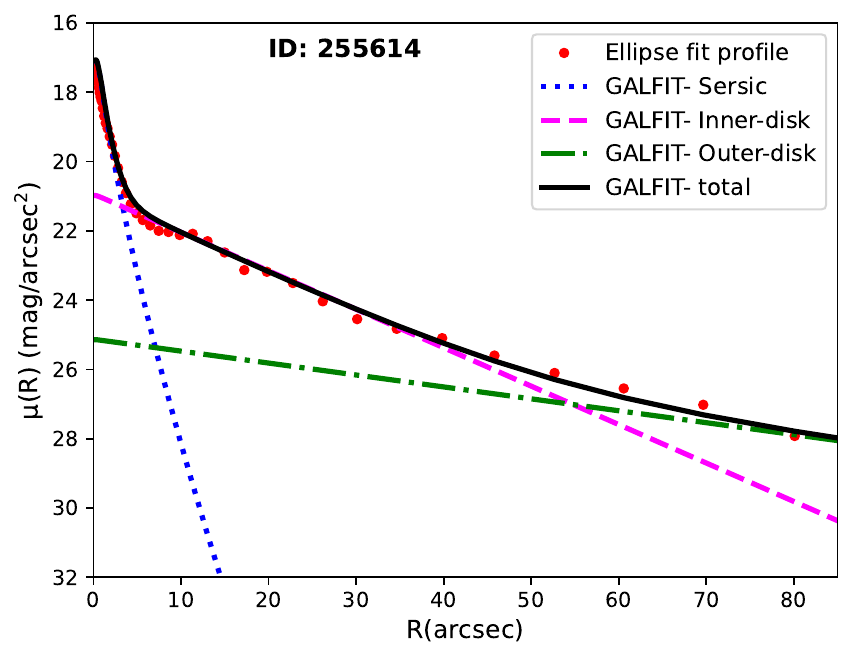}
\includegraphics[width = 0.32\textwidth, height=0.26\textwidth]{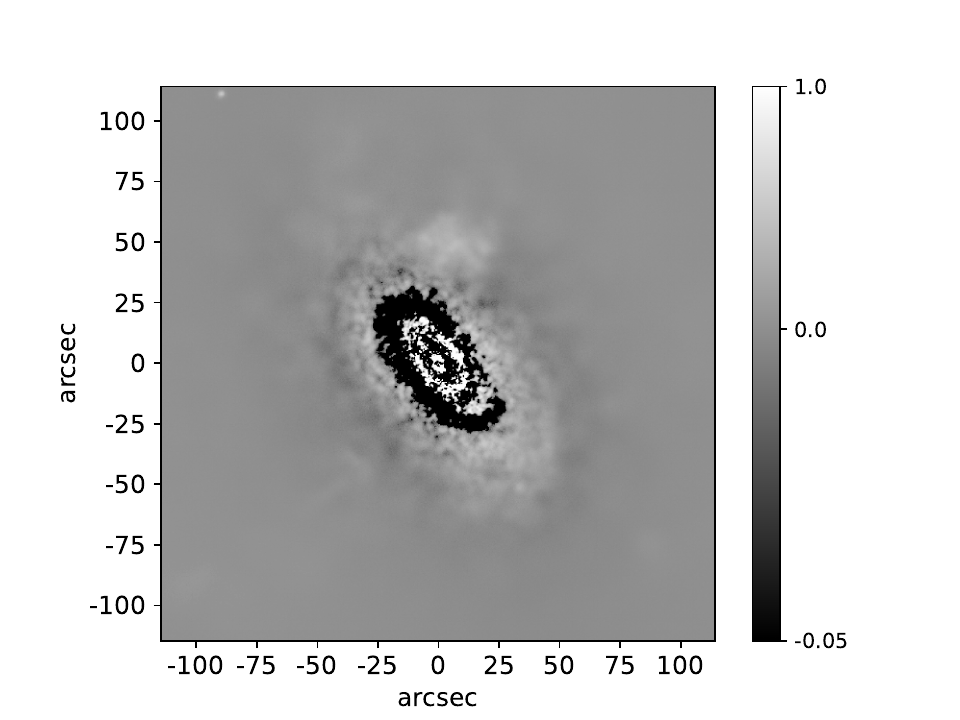}
\medskip
\caption{Double-exponential galaxy from Sample 2.}
\label{fig:fig3}
\end{figure*}

%\subsection{The structural parameters}
The structural parameters of the galaxies are calculated from their modeled magnitudes and radial scale lengths following the standard relations described in \citet{PahwaSaha2018}, \citet{Sarkar2025}. We present the parameters, namely, the central surface brightness of the Sersic component ($\mu_{0,sersic}$), effective radius ($R_{e}$), Sersic index($n$), and the central surface brightness of the disk ($\mu_{0,disk}$) (corrected for inclination and cosmological surface brightness dimming effect), radial scale lengths ($\mathrm{R_{D}}$) for the double-disk cases in Table 1. No correction was done to the Sersic parameters for simplicity, and because we are mainly interested in studying the disk in detail. The values in Table \ref{tab:table1} show that the inner disks are indeed HSB, while the outer disks are extended, LSBs.

\begin{table*}
%\rotate
\centering
\caption{Structural parameters of the massive, double-exponential disk galaxies, studied in this work, obtained from central S\'{e}rsic plus double-exponential disk modeling done by GALFIT.}
\begin{tabular}{cccccccc}
\hline
\hline
Subhalo ID & \multicolumn{3}{c}{{Central component(Bulge/Bar)}} & \multicolumn{2}{c}{Inner disk (ID)} & \multicolumn{2}{c}{Outer disk (OD)} \\
\hline
 & \multicolumn{3}{c}{S\'{e}rsic} & \multicolumn{2}{c}{Exponential} & \multicolumn{2}{c}{Exponential} \\
\hline
& $\mathrm{\mu_{0,sersic}}$ & $R_{e}$ & $n$  & $\mathrm{\mu_{0,c,ID}}$ & $R_{D,ID}$ & $\mathrm{\mu_{0,c,OD}}$ & $R_{D,OD}$  \\
&  (mag arcsec$^{-2}$) & (arcsec) &  & (mag arcsec$^{-2}$) & (arcsec)  & (mag arcsec$^{-2}$) & (arcsec) \\
\hline
(1)  & (2)   & (3)  & (4) & (5) & (6)  & (7) &(8) \\
\hline
Disk sample 1 &  &  &  &  SDSS g-band (Optical B band)  &  &   & \\
16 & 19.6 & 1.84 & 0.5 & 21.7(22.2) & 3.2 & 23.6(24.1) & 10.5\\
326652 & 18.9 & 0.83 & 0.5 & 19.4 (20.0) & 1.74 & 23.3(23.8) & 16.2 \\
375909 & 19.5 & 1.44 & 0.47 & 21.1(21.7) & 3.2 & 23.9(24.4) & 13.3 \\
398405 & 19.4 & 1.3 & 0.6 & 21.0(21.6) & 3.3 & 23.5(24.0) & 10.9 \\
443683 & 19.6 & 2.7 & 0.9 & 21.2(21.6) & 5.8 & 26.2(26.6) & 26.5\\
\hline
\hline
Disk sample 2 & & & &  &   &    &  \\
255614 & 17.1 & 1.2 & 1.4 &  20.7(21.2) & 9.1 & 24.9(25.1) & 31.5 \\
\hline
\hline
Disk sample 3 & & & &  &   &    &  \\
14 & 19.0 & 1.7 & 0.73 & 20.7(21.3) & 3.76  & 23.4(24.0)  & 14.0 \\
52623 & 19.9 & 1.8 & 0.5 & 21.1(21.3) &  3.9 & 23.1(23.3) & 13.9 \\
330955 & 19.5 & 2.95 & 0.82 & 21.2(21.7) & 4.4 & 22.9(23.3) & 19.1 \\
350486 & 19.0 & 3.12 & 0.92 & 21.5(22.0) & 4.97 & 24.8(25.3) & 37.4 \\
364807 & 19.5 & 2.61 & 0.91 & 22.1(22.3) & 5.4 & 24.4(24.5) & 25.6 \\
365493 & 19.3 & 2.04 & 0.6 & 21.4(21.6)  & 3.1  & 24.5(24.6) & 33.6 \\
371259 & 19.4 & 1.3 & 0.5 & 20.2(20.7)   & 2.3   & 23.8(24.2)  & 25.9 \\
391260 & 19.3 & 2.0 & 0.5 & 20.7 (21.3) & 2.6  &  23.9(24.4)  & 13.2 \\
399969 & 19.0 & 2.1 & 0.6  & 20.9 (21.4) & 2.83 & 23.3(23.9)  & 12.8 \\
\hline
\hline
Disk sample 4 & & & &  &   &    &  \\
217486 & 18.4 & 2.7 & 1.4 & 21.1(21.6) &  10.9 & 26.2(26.3)  & 57.4 \\
\hline
\hline
\end{tabular}
%\begin{tablenotes}
Note: Column (1) denotes the galaxy/Subhalo ID of the double-exponential disk galaxies. Cols.(2),(3) \& (4) show the central surface brightness value, the effective radius, and the S\'{e}rsic index, respectively, of the S\'{e}rsic component. Cols.(5) \& (6) show the central surface brightness value ($\mu_{0,c}$ ) (corrected for inclination and cosmological dimming effect) and the radial scale length, respectively, of the inner exponential disk. Cols.(7) \& (8) represent the same for the outer exponential disk. These values are given corresponding to the synthetic SDSS g band images here. The $\mu_{0}$ for the disks calculated for the optical B-band are given as well. No correction has been done to the Sersic parameters. 
%\end{tablenotes}
\label{tab:table1}
\end{table*}

\begin{figure*}
\centering
\includegraphics[width = 0.33\textwidth, height=0.24\textwidth]{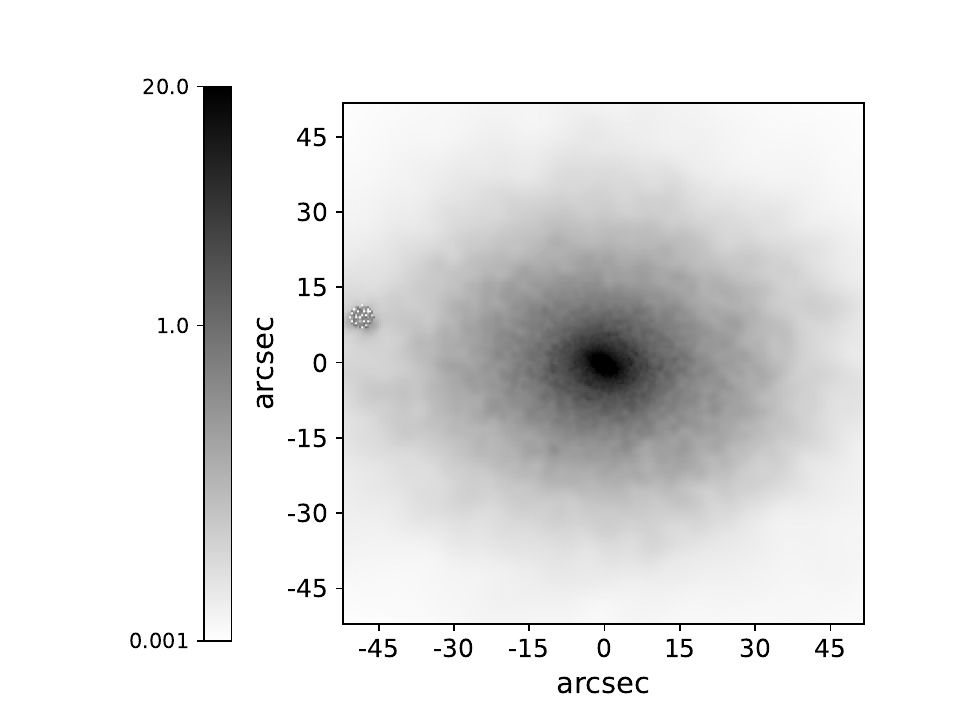}
\includegraphics[width = 0.31\textwidth, height=0.22\textwidth]{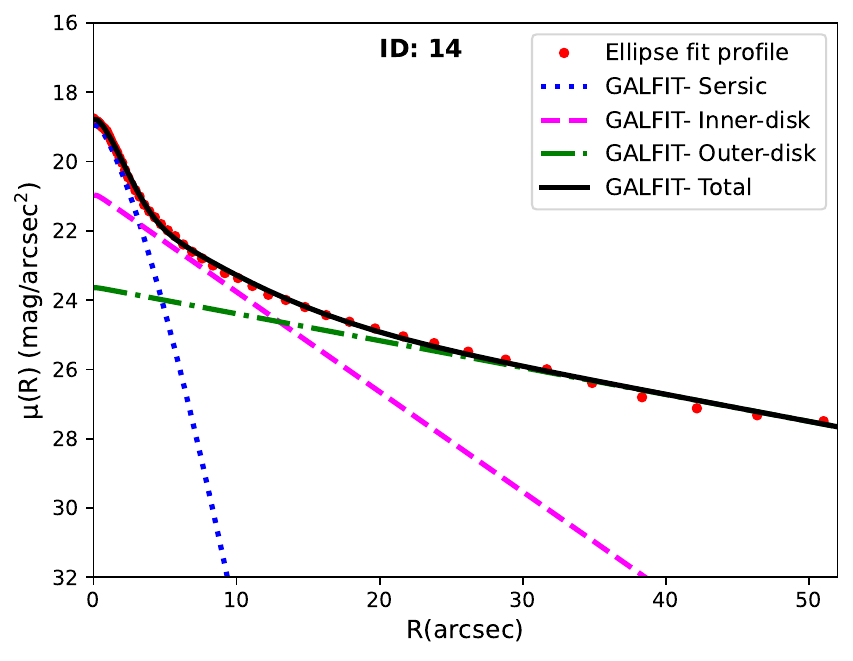}
\includegraphics[width = 0.33\textwidth, height=0.24\textwidth]{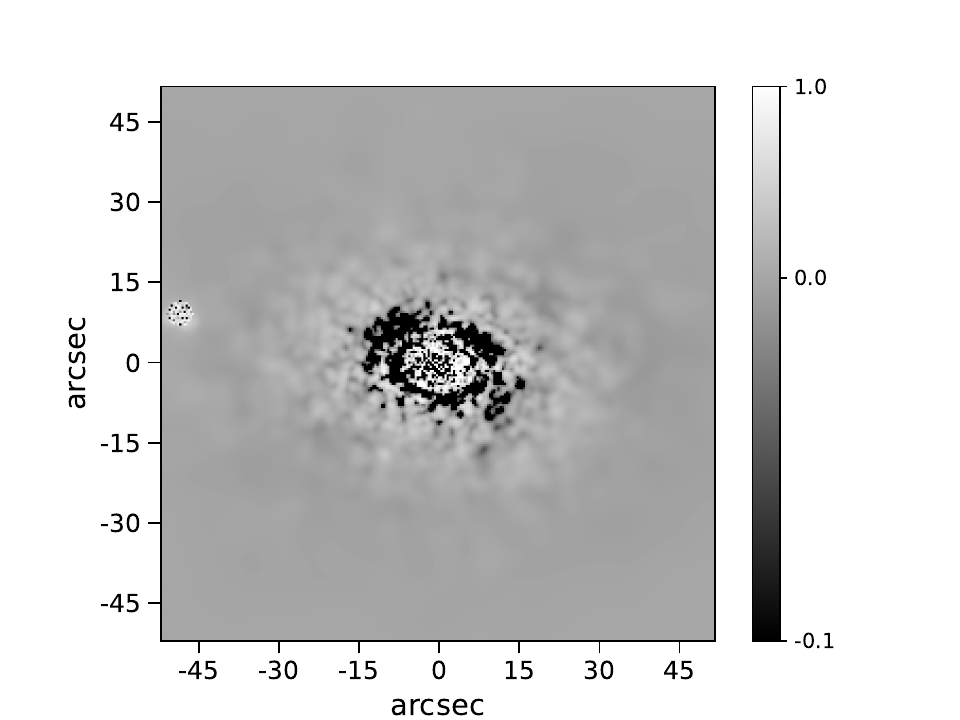}
\medskip
\includegraphics[width = 0.33\textwidth, height=0.24\textwidth]{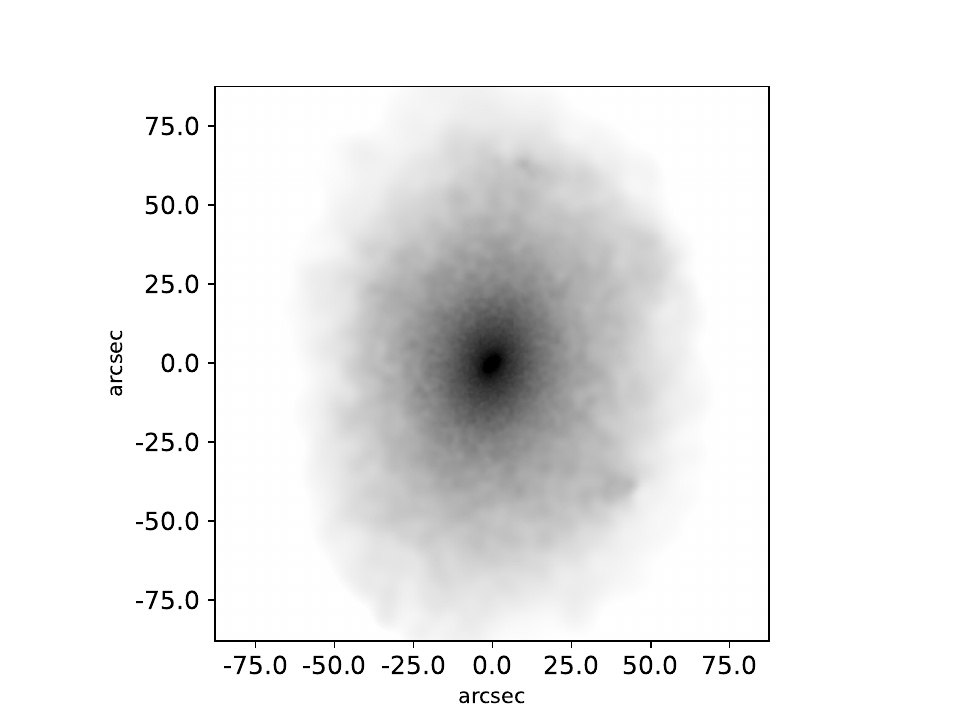}
\includegraphics[width = 0.31\textwidth, height=0.22\textwidth]{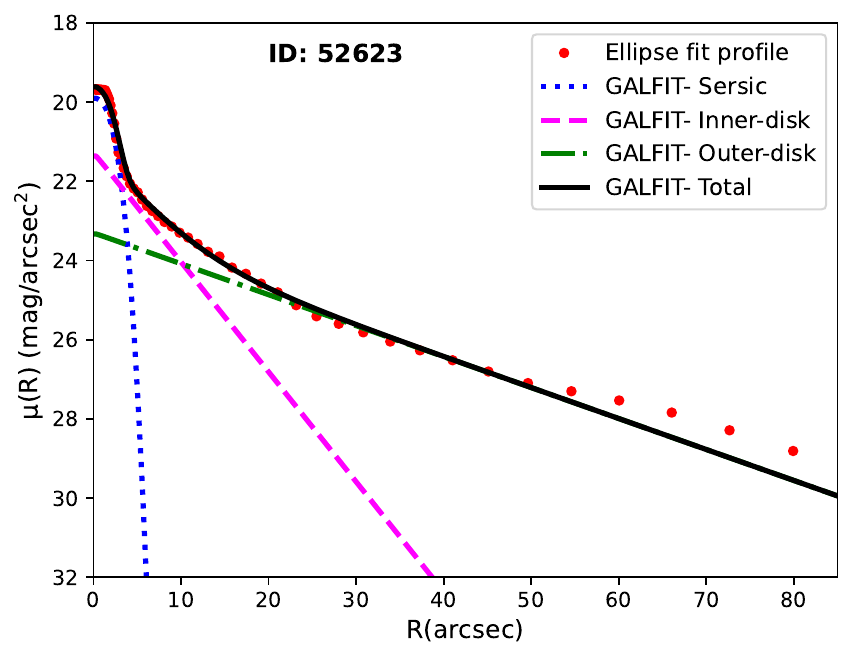}
\includegraphics[width = 0.33\textwidth, height=0.24\textwidth]{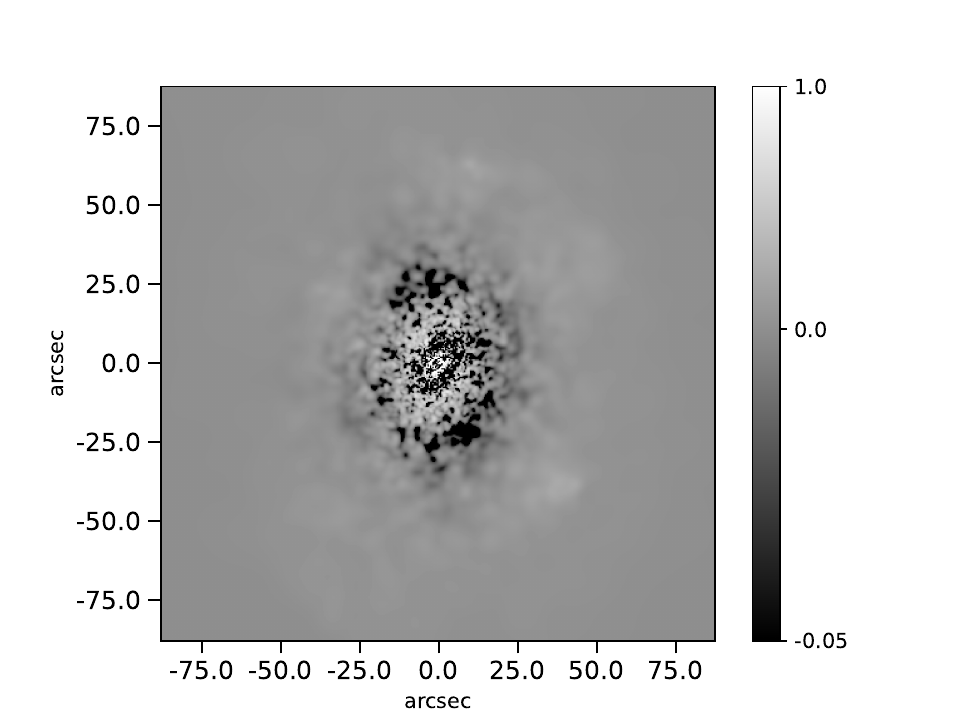}
\medskip
\includegraphics[width = 0.33\textwidth, height=0.24\textwidth]{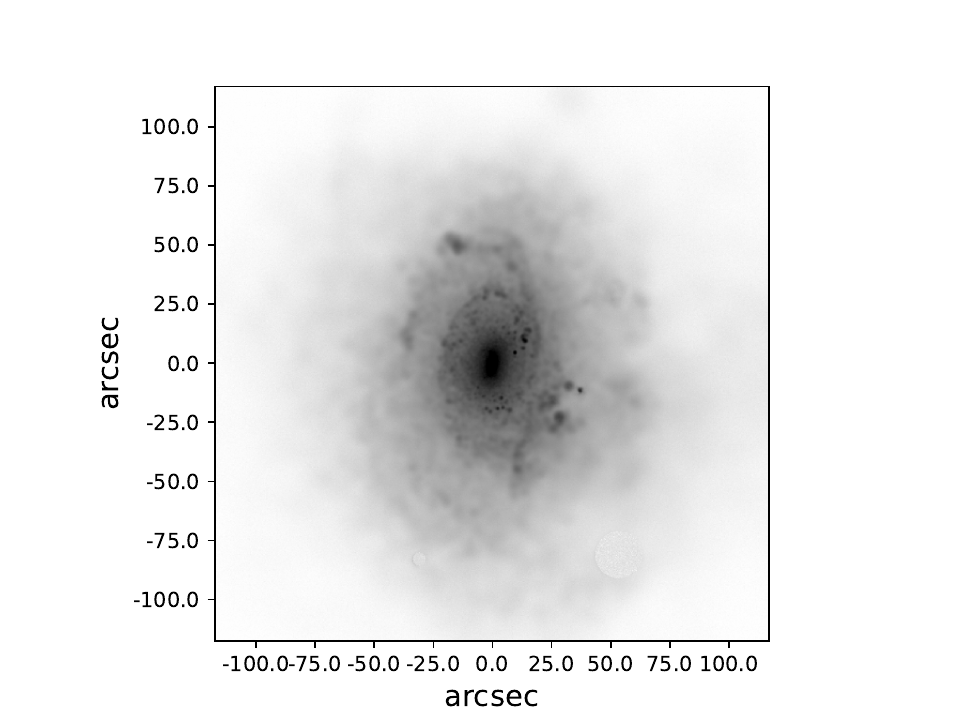}
\includegraphics[width = 0.31\textwidth, height=0.22\textwidth]{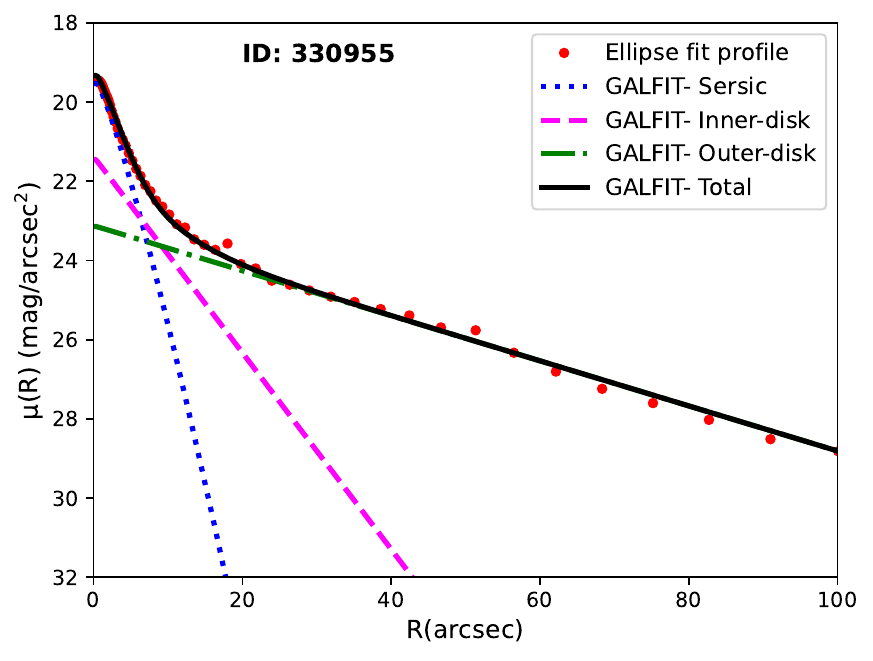}
\includegraphics[width = 0.33\textwidth, height=0.24\textwidth]{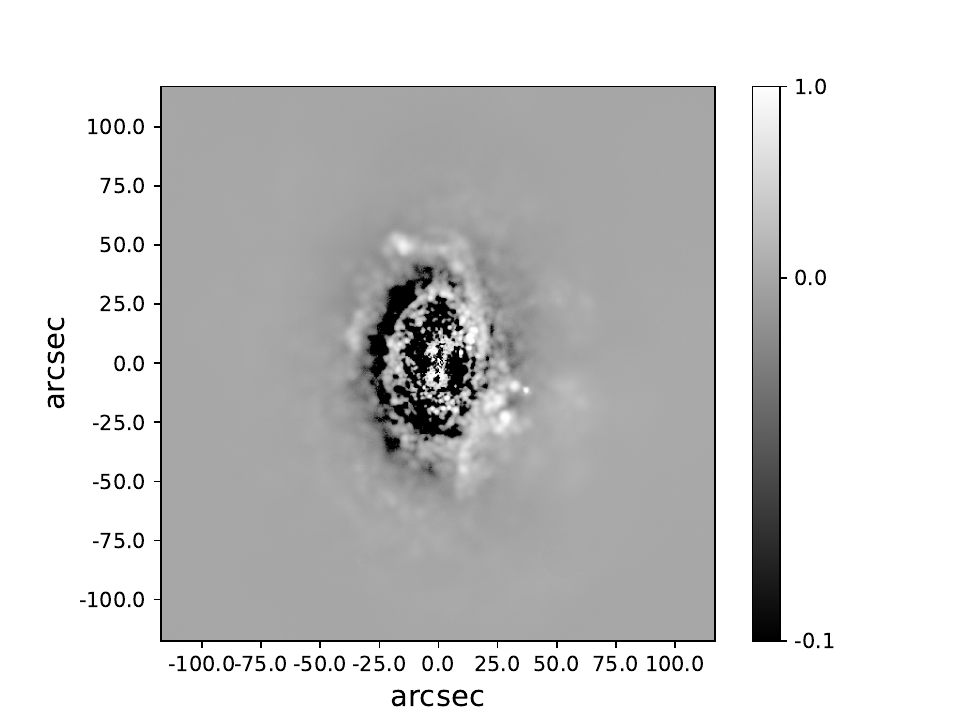}
\medskip
\includegraphics[width = 0.33\textwidth, height=0.24\textwidth]{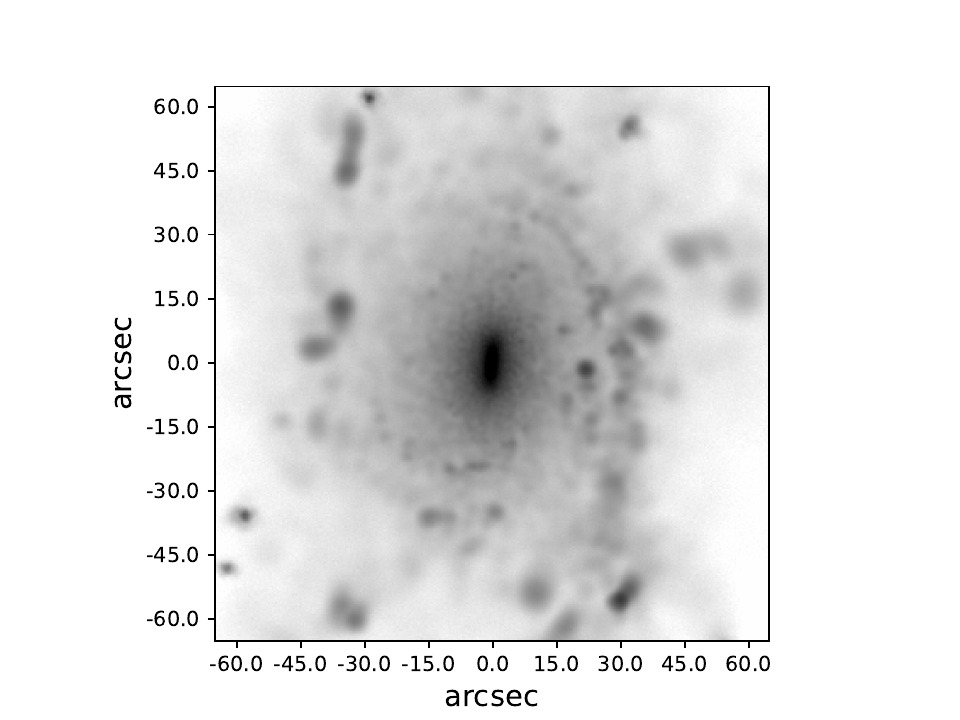}
\includegraphics[width = 0.31\textwidth, height=0.22\textwidth]{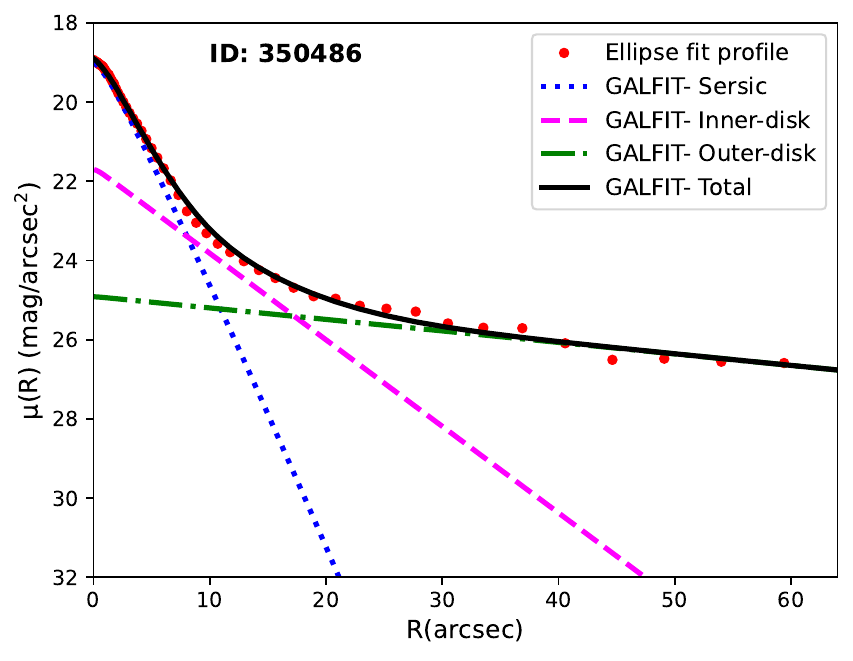}
\includegraphics[width = 0.33\textwidth, height=0.24\textwidth]{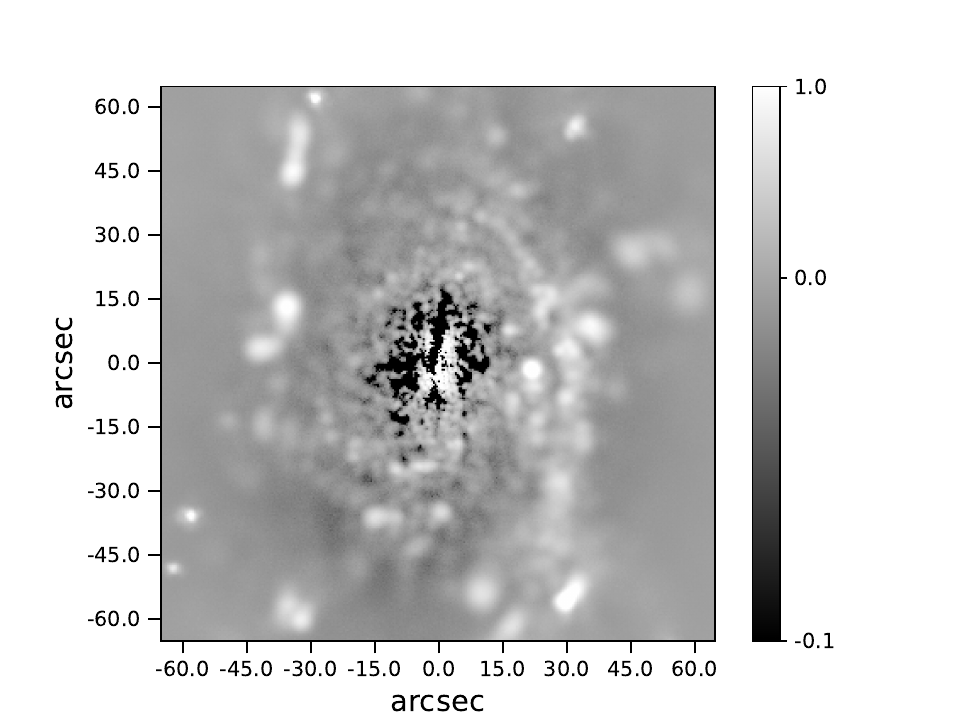}
\medskip
\includegraphics[width = 0.33\textwidth, height=0.24\textwidth]{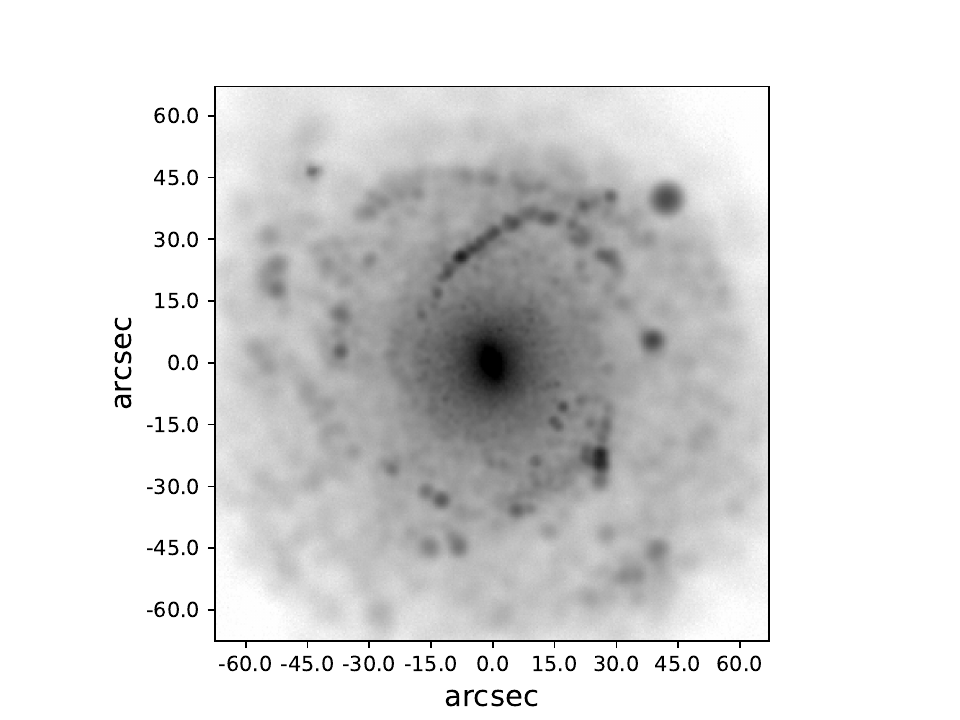}
\includegraphics[width = 0.31\textwidth, height=0.22\textwidth]{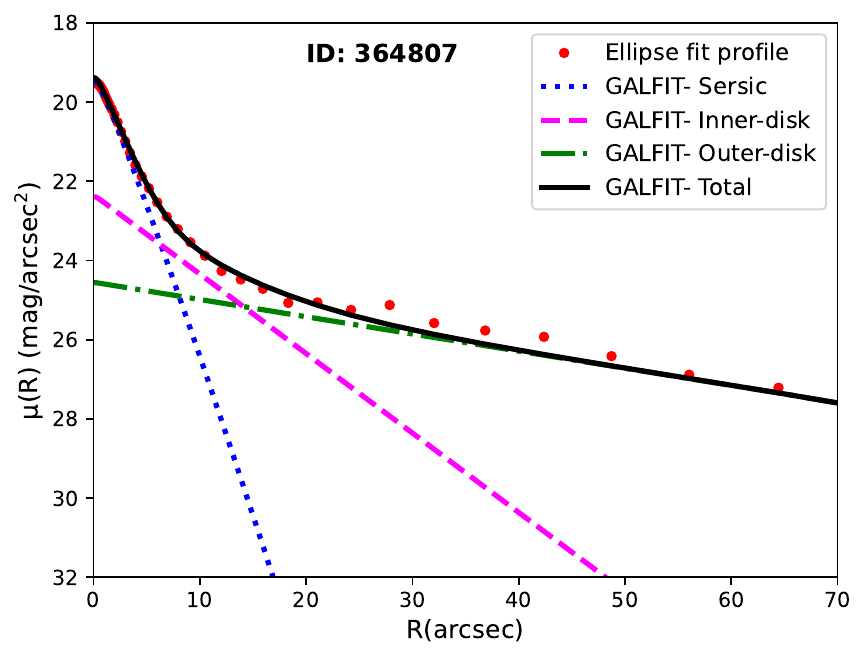}
\includegraphics[width = 0.33\textwidth, height=0.24\textwidth]{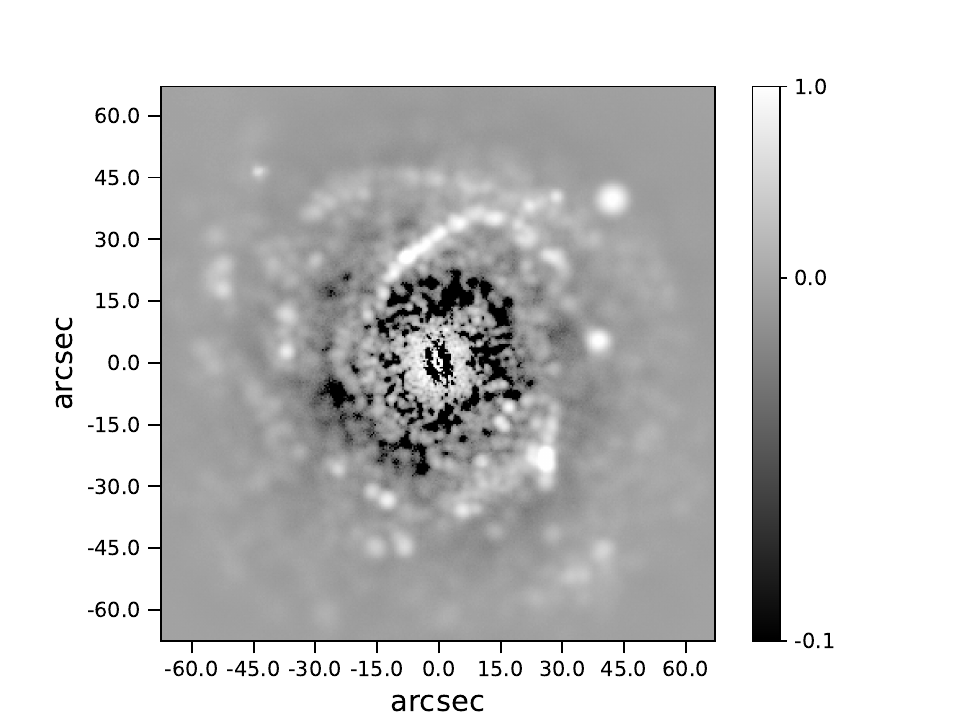}
\caption{Double-exponential galaxies : Sample 3. The colorbars corresponding to each residual map is provided as the colorbars are not the same for each case.}
\label{fig:fig4}
\end{figure*}
\noindent

\begin{figure*}
\centering
\includegraphics[width = 0.33\textwidth, height=0.24\textwidth]{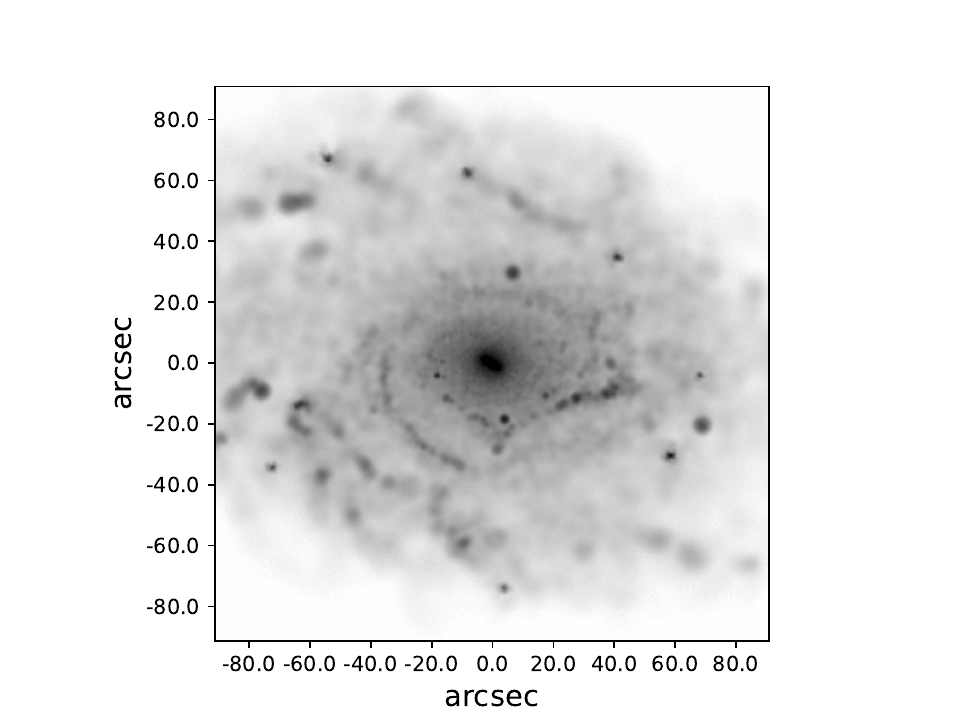}
\includegraphics[width = 0.31\textwidth, height=0.22\textwidth]{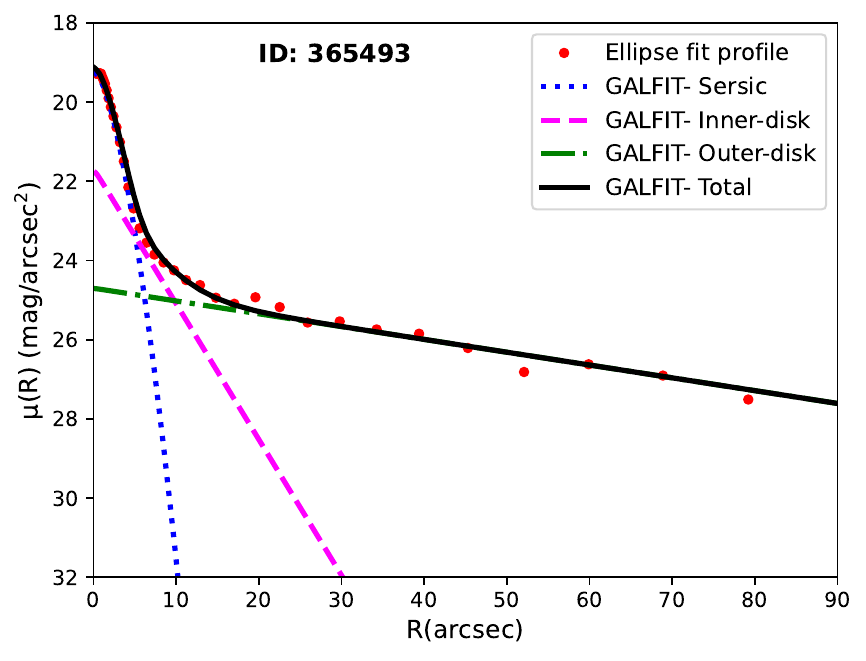}
\includegraphics[width = 0.33\textwidth, height=0.24\textwidth]{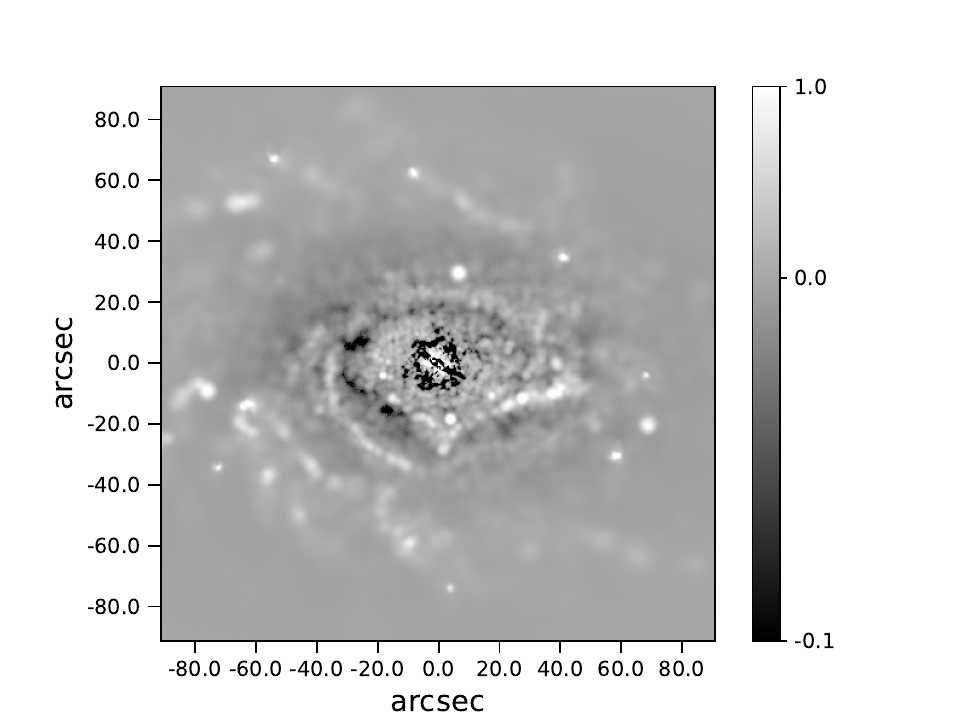}
\medskip
\includegraphics[width = 0.33\textwidth, height=0.24\textwidth]{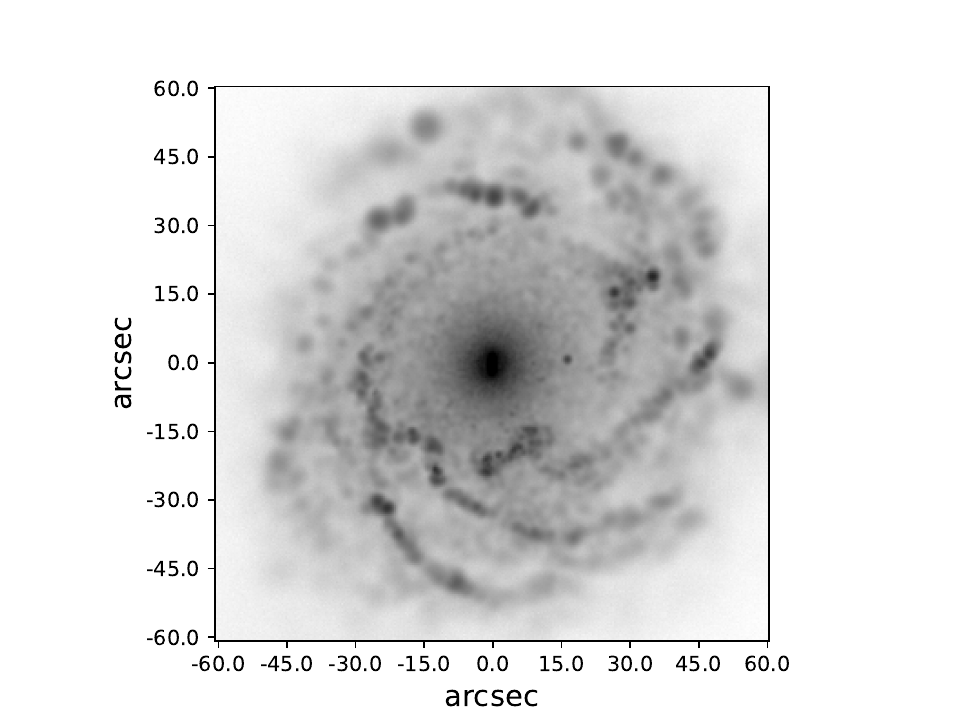}
\includegraphics[width = 0.31\textwidth, height=0.22\textwidth]{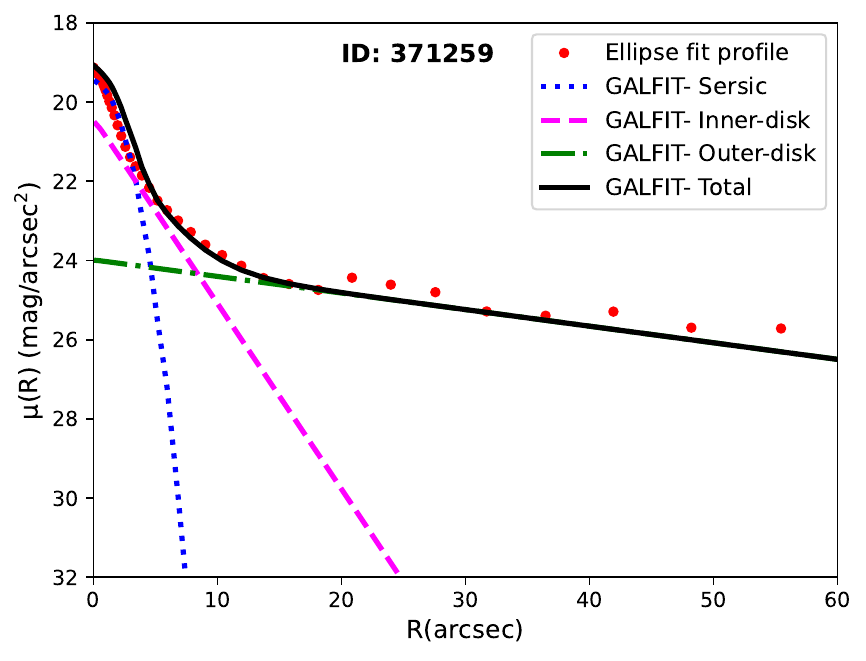}
\includegraphics[width = 0.33\textwidth, height=0.24\textwidth]{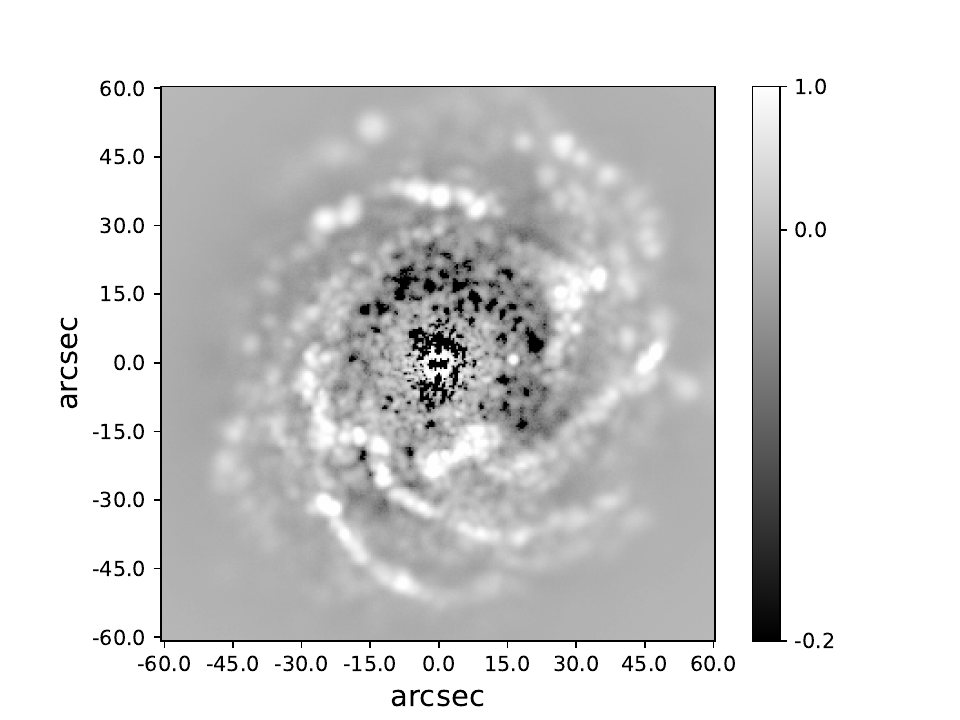}
\medskip
\includegraphics[width = 0.33\textwidth, height=0.24\textwidth]{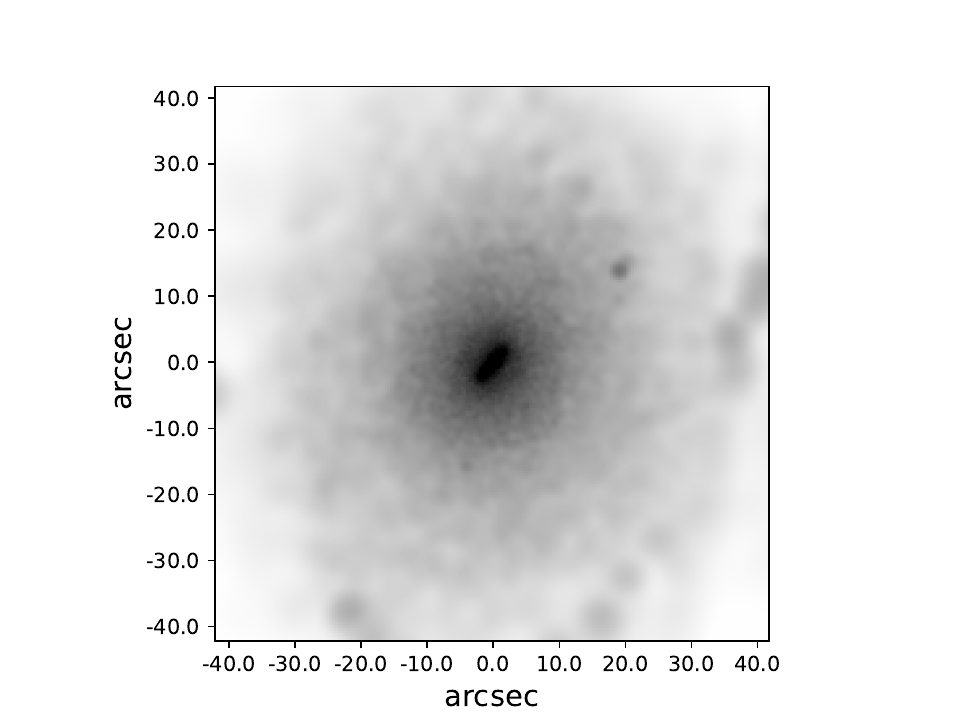}
\includegraphics[width = 0.31\textwidth, height=0.22\textwidth]{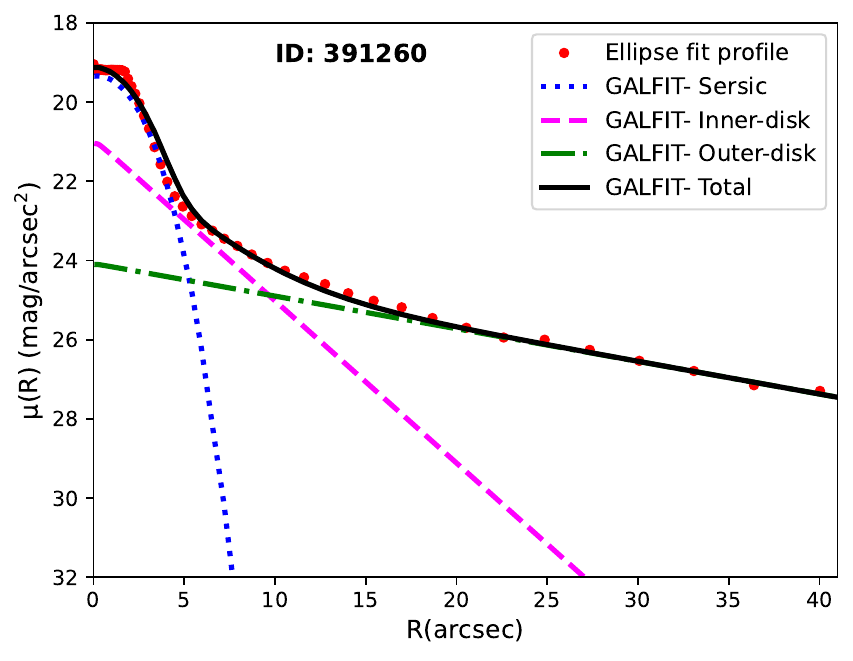}
\includegraphics[width = 0.33\textwidth, height=0.24\textwidth]{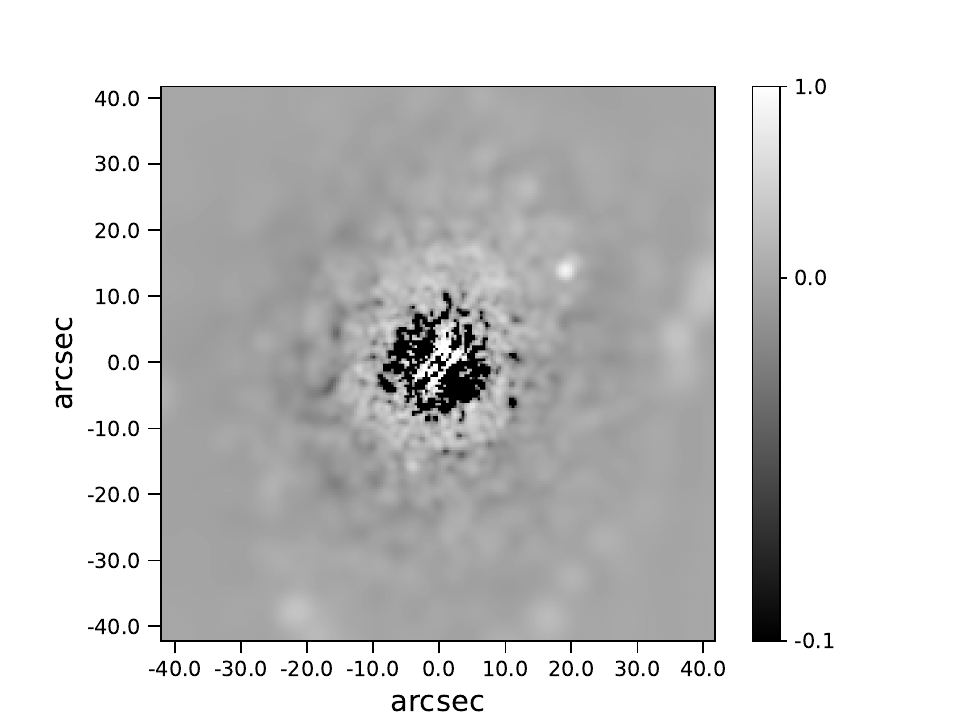}
\medskip
\includegraphics[width = 0.33\textwidth, height=0.24\textwidth]{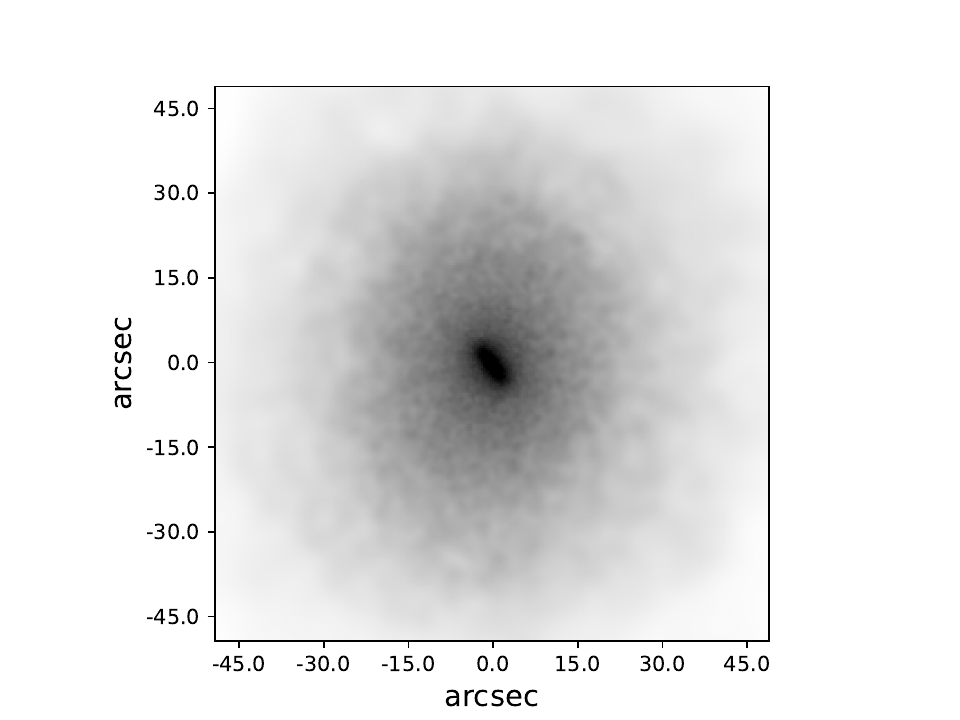}
\includegraphics[width = 0.31\textwidth, height=0.22\textwidth]{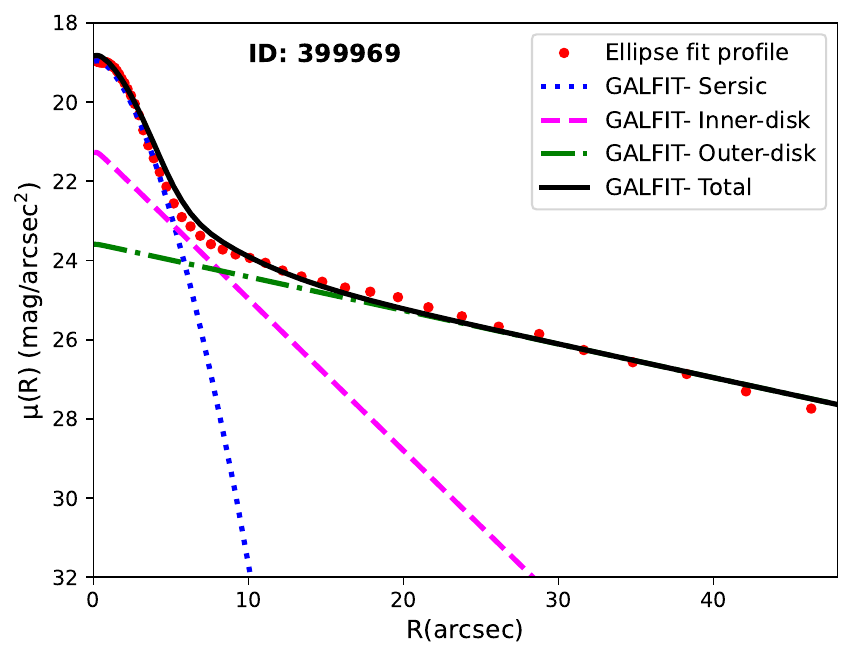}
\includegraphics[width = 0.33\textwidth, height=0.24\textwidth]{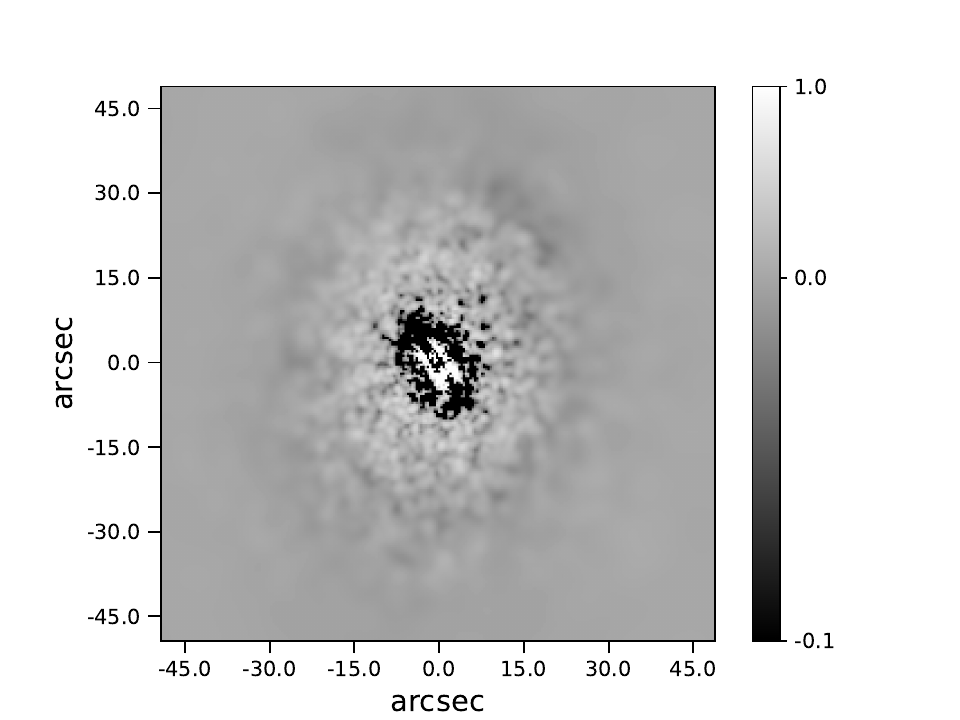}
\medskip
{\textbf{Figure 4.} Continued from the above figure}
\end{figure*}

\begin{figure*}
\centering
\includegraphics[width = 0.33\textwidth, height=0.26\textwidth]{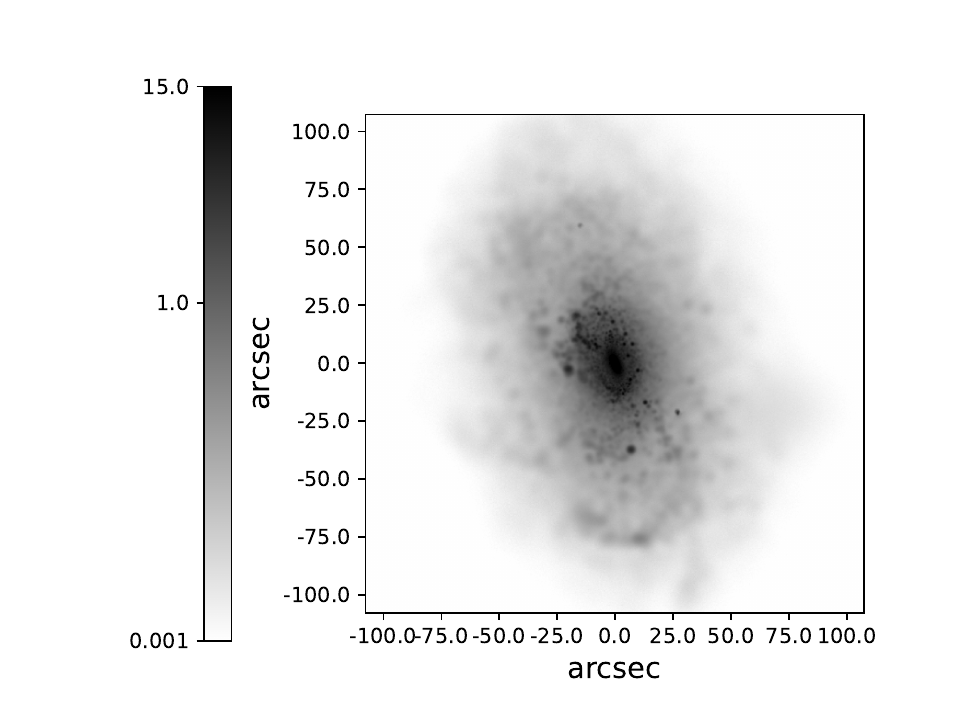}
\includegraphics[width = 0.31\textwidth, height=0.23\textwidth]{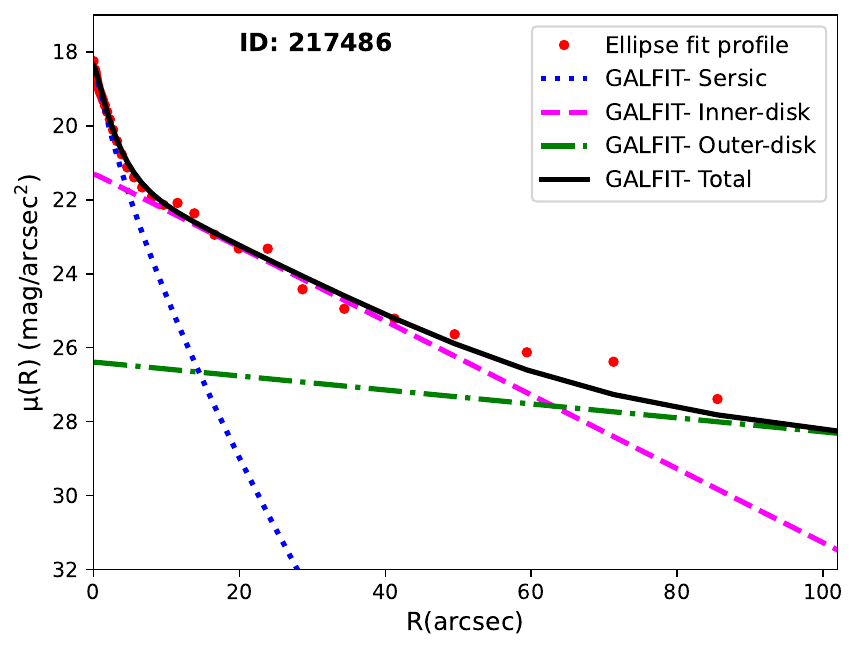}
\includegraphics[width = 0.33\textwidth, height=0.26\textwidth]{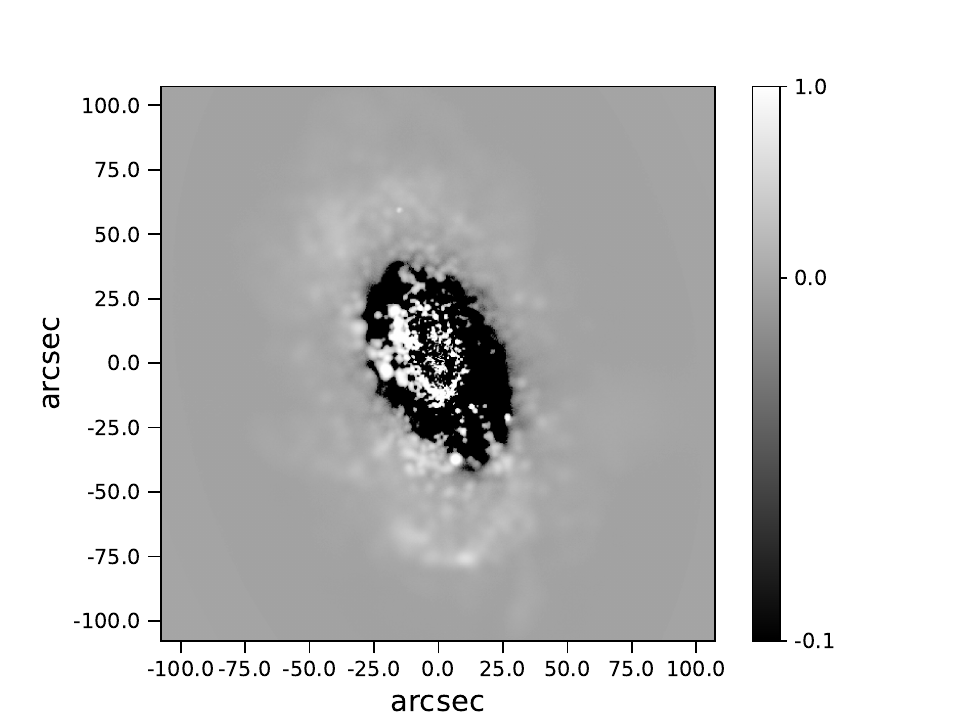}
%\medskip
%\includegraphics[width = 0.31\textwidth, height=0.25\textwidth]{217486_singledisk_profile_final.pdf}
%\includegraphics[width = 0.31\textwidth, height=0.25\textwidth]{217486_residue_final.pdf}
\caption{Double-exponential galaxy : Sample 4.}
\label{fig:fig5}
\end{figure*}

\begin{figure*}
\centering
\includegraphics[width = 0.47\textwidth, height=0.35\textwidth]{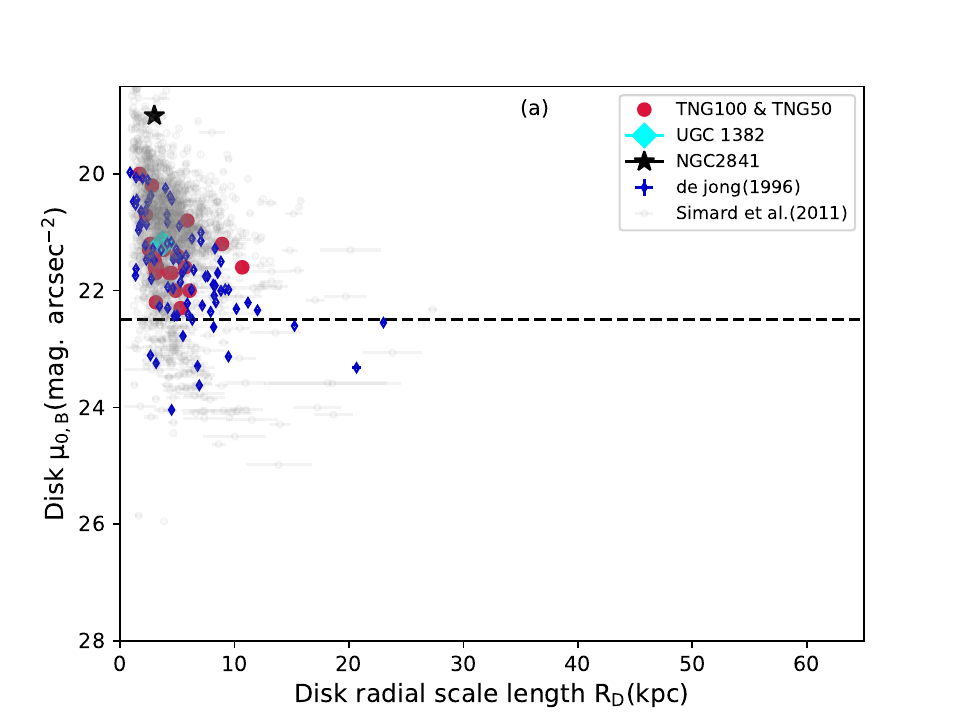}
\includegraphics[width = 0.47\textwidth, height=0.35\textwidth]{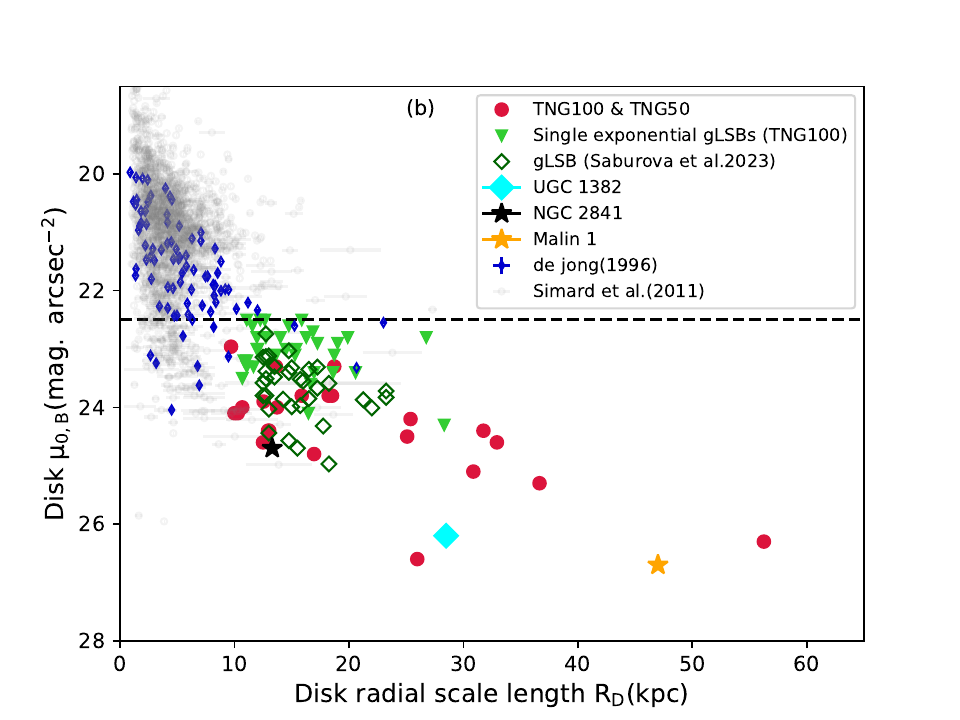}
\caption{Central surface brightness ($\mathrm{\mu_{0,B}}$) vs. radial scale length of the inner HSB (\textbf{Panel a}), and the outer, extended LSB disk (\textbf{Panel b}) of the double-exponential gLSBs from this work \& TNG50 \citep{Sarkar2025}. This is compared with the typical disk galaxies \citep{dejong_1996, Simard2011} (B band), single-exponential gLSBs from this work, observed gLSBs from \citet{Saburovaetal2023} (g-band), and the observed double-exponential gLSBs (UGC1382 \citep{Saburova2026, Hagenetal2016} in g \& B bands for the two components, respectively; both disks of NGC2841\citep{Zhangetal2018} in g-band; LSB disk of Malin 1\citep{Sahaetal2021} in g-band). The horizontal line divides the high and the low surface brightness regime.}
\label{fig:fig6}
\end{figure*}

\section{Physical properties of the double-exponential gLSBs and comparison with overall disk galaxy populations}

The structural parameters of disk galaxies that we observe today are the consequence of their mass assembly history. For example, the correlation between the disk central surface brightness and the radial scale length is linked to the total mass and angular momentum build-up of the galaxy \citep{Dalcanton1997}. At the same time, the light fraction contained in bulge and disk components, studied in terms of bulge-to-disk(B/D), bulge-to-total(B/T) flux ratio %, is affected by the formation and evolution of the galaxy. It also 
indicate the position of the galaxy in the Hubble sequence \citep{GrahamWorley2008,Gadotti2009, Simard2011}. Therefore, we studied the above for the double-exponential disk galaxies obtained in this paper and the sample obtained from TNG50 in \citet{Sarkar2025} together, and compare them with the overall disk galaxy populations from observations. We note that this important aspect has been explored little in the literature so far. The comparison is essential to eventually understand the formation and evolution of the double-exponential galaxies using simulations. For completeness, we also looked at the dark matter fraction of these galaxies. 

%we also make comparison with the simulated single-disk gLSBs. 

\subsection{Central surface brightness vs. radial scale length} 
We use the available B-band central surface brightness (corrected for inclination, galactic extinction), radial scale lengths of 84 local, spiral galaxies from \citet{dejong_1996}, corresponding to 2D modeling of exponential bulge plus exponential disk (the best representative model in their paper). Their redshifts, taken from NED, are used to add cosmological dimming effect in $\mu_{0}$ as well as convert the scale lengths in kpc.  

We use another local galaxy sample from \citet{Simard2011} catalog where bulge plus disk decomposition were carried out on 1.12 million SDSS galaxies in g,r bands. We selected disk galaxies among them using the provided values of the probability of the galaxies being a genuine Sersic plus exponential system over a purely Sersic system. We ensured that the available bulge Sersic index is within $0.5<n<4.0$, and the absolute magnitudes (corrected for extinction, k-correction) are finite, physical. %restricted the absolute model magnitudes of bulge, disk components to $-24<Mg,r<-14 \\ mag based on what Pahwa et al shows as maximum range$. 
This leads to a sample of $\sim$ 2086 disk galaxies. We obtain the apparent magnitudes from the above absolute magnitudes, and calculate  $\mu_{0,B}$ of the disks. 
  
A sample of gLSBs observed by the Subaru telescope \citep{Saburovaetal2023} are taken for comparison as well. The face-on {$\mathrm{\mu_{0,g}}$} (extinction, dimming corrected) of the LSB disks are all fainter than 22.7 which corresponds to 23 mag $arcsec^{-2}$ in B band assuming g-r=0.03, as mentioned in their paper. 
We note that k-corrections to $\mu_{0}$ were not mentioned in \citet{dejong_1996} and \citet{Saburovaetal2023}. However, such corrections are very small since all the galaxies are well within z=0.1.

Finally, we compare the HSB, LSB disks of UGC1382, NGC2841, and the LSB disk of Malin 1 in g/B bands. The HSB disk of Malin 1 has $\mu_{0}$ = 17.64 mag arcsec$^{-2}$, $R_{D}$ = 4.8 kpc (HST/F814W band) \citep{Sahaetal2021}. We note that although UGC1378 was considered to be a double-exponential gLSB by \citet{Saburovaetal2019}, the $\mu_{0,B}$ value of the outer disk (22.1 mag $arcsec^{-2}$) is brighter than the LSB disk criteria used in this paper. 

Figure \ref{fig:fig6} shows that the inner HSB disks of the simulated double-exponential cases are comparable to the overall disk galaxy populations. Interestingly, the outer LSB disks lie significantly at the outskirt of the distribution and therefore occupy a unique space. This indicates that such galaxies might have gone through distinct channel of formation and evolution compared to the typical disks. Interestingly, \citet{Kolganov2026} has shown a similar trend for double-exponential gLSBs from the Subaru observations.

We note that although the simulated outer LSB disks are overall similar to the observed single-disk gLSBs from \citet{Saburovaetal2023}, several of them are more radially extended. As usual, Malin 1 being the extreme example of observed double-exponential gLSBs, lies at the outermost part of the distribution. We find only one double-exponential galaxy from TNG100 with a similar scale length. We also find that the outer LSB disk sample can reach shallower central surface brightness and larger scale lengths than the single-disk gLSBs from TNG100.

\subsection{Light fraction from bulge, disk, and estimation of dark matter fraction}
We calculate the B/D, B/T ratios of the simulated galaxies using the g-band modeled magnitudes of their Sersic and two exponential components, obtained from GALFIT. 
A few of them appear to have prominent bar-like structure. For simplicity of modeling we had modeled the central part using a single Sersic function to account for both bulge and bar light together as mentioned earlier, similar to the decomposition followed by \citet{Simard2011}.% We note that \citet{Simard2011} had also used a similar argument to do simple bulge plus disk decomposition and calculated B/T fraction for the millions of SDSS galaxies from those models. 

%The median values for the double-exponential gLSBs turn out to be 0.24, 0.20 respectively. The corresponding values for the single-exponential gLSBs are 0.27, 0.21 respectively. This shows that both types of galaxies are disk dominated and fall in the late type class. 
 %Using those we calculated the corresponding B/D fractions assuming the sum of B/D and D/T to the 1.0. %The median values of B/D, B/T corresponding to the free bulge plus exponential disk model, i.e, the one used here for comparison, are 0.315, 0.24 respectively. 

The simulated double-exponential galaxies are disk dominated, with median $B/T = 0.20$ and $B/D=0.24$, consistent with late-type disk galaxies and systems hosting pseudo-bulges \citep{Fisher2008, Gadotti2009, kormendy2004, Weinzirl2009}. For the observed typical disk population of \citet{Simard2011}, the median $g$-band $B/T$ is 0.23, indicating no significant difference from the simulated galaxies in their relative bulge and disk contributions. The single-exponential gLSBs identified in this work show similar values, $B/T=0.21$ and $B/D=0.27$. 
%We note that the B/D ratios of the spirals in \citet{dejong_1996} were reported to be too small (ranges from 0.01 to 0.1 in B band), as discussed by them. 
For UGC 1382, using the GALFIT component magnitudes from \citet{Hagenetal2016} gives a slightly higher $B/T=0.31$ likely reflecting its prominent central bulge and lenticular disk component. We do not derive $B/T$ for Malin 1 and NGC 2841 because the component magnitudes are not available. 

It is well known that LSB galaxies are usually dynamically dominated by dark matter haloes from their innermost radii unlike the high surface brightness counterparts \citep{deblok1997,McGaugh1998, perezsodi2019}. We estimated the dark matter fractions of the simulated double and single-exponential gLSBs using the subhalo stellar, gas and dark matter masses available from TNG subhalo data catalogs. The median values for the two samples are found to be similar, 0.90, \& 0.92, respectively. We note that \citet{Hagenetal2016} had found a similar fraction (0.95) for UGC 1382.
 
%\subsection{Dark matter fraction}
%It is well known that LSB galaxies are usually dynamically dominated by dark matter haloes from their innermost radii unlike the high surface brightness counterparts \citep{deblok1997,McGaugh1998, perezsodi2019}. Here we tried to see if there is any difference in the dark matter mass fraction between the double-exponential and the single-exponential gLSBs. We used the subhalo stellar, gas and dark matter masses available from TNG subhalo data catalogs. The mean dark matter fractions turn out to be similar, 0.90, \& 0.92, respectively. we note that \citet{Hagenetal2016} had found a similar fraction (0.95) for UGC 1382.

%UGC 1382, the HI gas mass, the dynamical mass and thus dark matter fractions were all provided within 110 kpc while the stellar mass measurement was done upto 70 kpc \citep{Hagenetal2016}.

%For Malin 1 we take the HI gas mass as 6 $\times$ $10 ^{10} M_{\odot}$ and stellar mass to be 8.9 $\times$ $10 ^{11} M_{\odot}$(ref) (Pickering et al.1997, Saha et al., 2021). We calculate the dark matter halo mass using the best-fit pesudo-isothermal model from (ref)Lelli etal., 2010 who modeled the rotation curve almost upto 100 kpc. The upper limit of the mass turns out to be 1.14 $\times$ $10^{12} M_{\odot}$.

%We take the HI mass, and the best-fit modeled disk mass and the dark matter halo mass of UGC 1378 from Saburova et al. 2019 to calculate its baryon fraction and the dark matter fractions. 

\section{Conclusion} 
We have studied massive disk galaxies from the large volume IllustrisTNG100 simulation to find galaxies with inner high surface brightness plus outer, extended low surface brightness disk morphology. We also added the double-exponential cases found from IllustrisTNG50 in our previous paper to have a comprehensive understanding of such systems. Below we summarize the main results from this work.

\begin{itemize}

\item
We find that $\sim$ 11\% of the massive disk galaxies from IllustrisTNG100 are best described by S\'{e}rsic plus HSB plus extended LSB disks, obtained based on detailed 2D light modeling of the galaxy images. Interestingly, despite using around ten times larger volume simulation and around three times larger parent sample of massive disk galaxies, we found a similar fraction compared to that from IllustrisTNG50 (12\%). This shows that such galaxies are not common within the local universe.

\item
We compared the double-exponential galaxies with the typical, local disk galaxies from observations. In the central surface brightness vs. radial scale length parameter space the outer LSB disks lie significantly at the outskirts of the distribution, occupying a unique space, whereas the inner HSB disks fall well within the typical disk galaxy regime. A similar trend is seen in the very recent observations from the Subaru telescope. This indicates the possibility of a distinct formation and evolution scenario. Interestingly, the B/T ratio turns out to be similar to that of the typical disk galaxies. 

\item 
A comparison with the single exponential gLSBs from TNG100 show that the double exponential cases span upto larger scale lengths and shallower central disk surface brightness. Interestingly, they have similar B/T ratios.

\item 
The outer LSB disk parameters of our double-disk cases show an overall similar range compared to the prototypical examples from observations, such as UGC1382, Malin 1, NGC2841. 

\item 
For completeness, we also checked that the median dark matter fraction of the double-exponential sample is high (0.90), typical of gLSBs.

\end{itemize}

We intend to study this robust, combined sample of massive, double-exponential galaxies across simulation snapshots to explore their formation and evolution, e.,g, detect the emergence of the outer LSB disk. Our prediction of the low number statistics could be verified from observations with the upcoming LSST (Legacy Survey of Space \& Time) by the Vera C. Rubin observatory, which is supposed to probe the low surface brightness outskirts of galaxies in an unprecedented way.

\section*{Acknowledgments}
The research in this article has been carried out using the IllustrisTNG simulation data publicly available on the IllustrisTNG website. Software: GALFIT, IRAF, Astropy, DS9.

\appendix

\section{Single disk modeling of the double-exponential galaxies}
We show the single disk modeling of the double-exponential disks from this work and the improvement in the residual profile after addition of the second component for Sample 2, 3 \& 4 together.

\begin{figure*}
\centering
\includegraphics[width = 0.33\textwidth, height=0.24\textwidth]{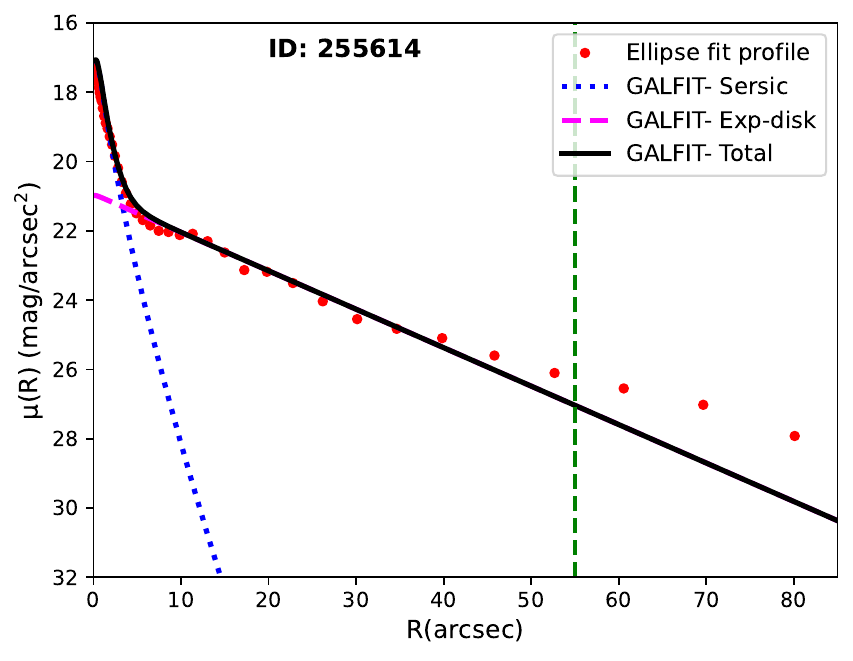}
\includegraphics[width = 0.33\textwidth, height=0.24\textwidth]{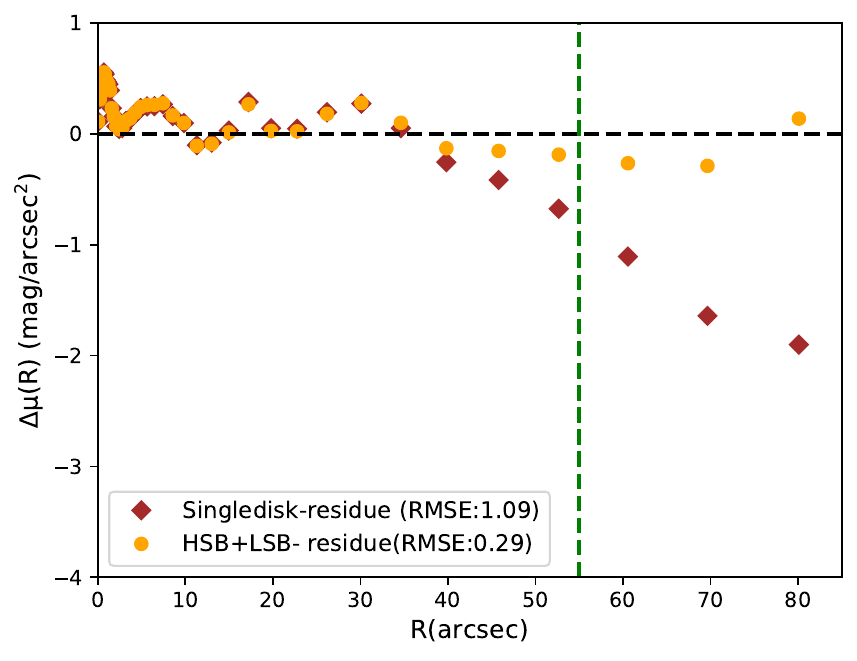}
\medskip
\includegraphics[width = 0.33\textwidth, height=0.24\textwidth]{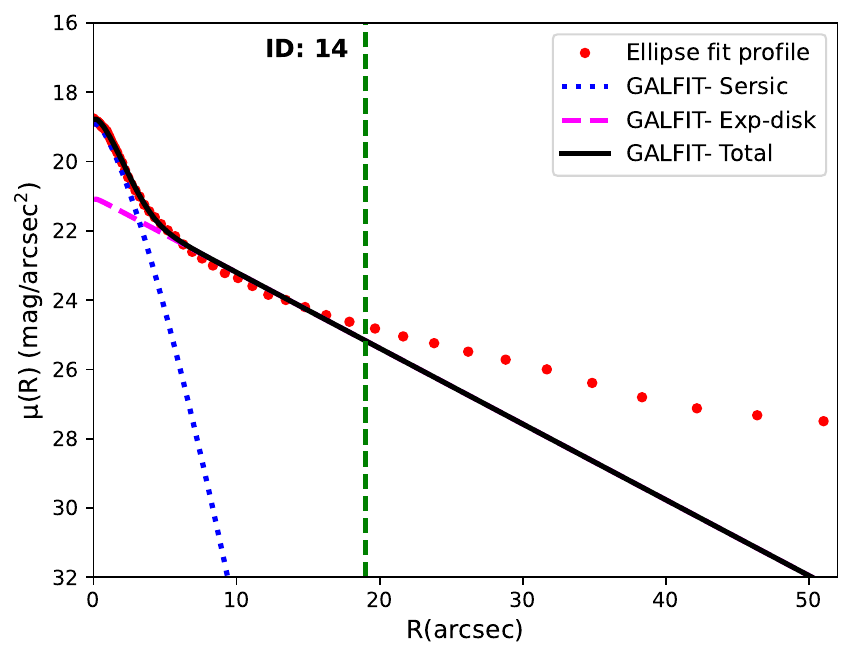}
\includegraphics[width = 0.33\textwidth, height=0.24\textwidth]{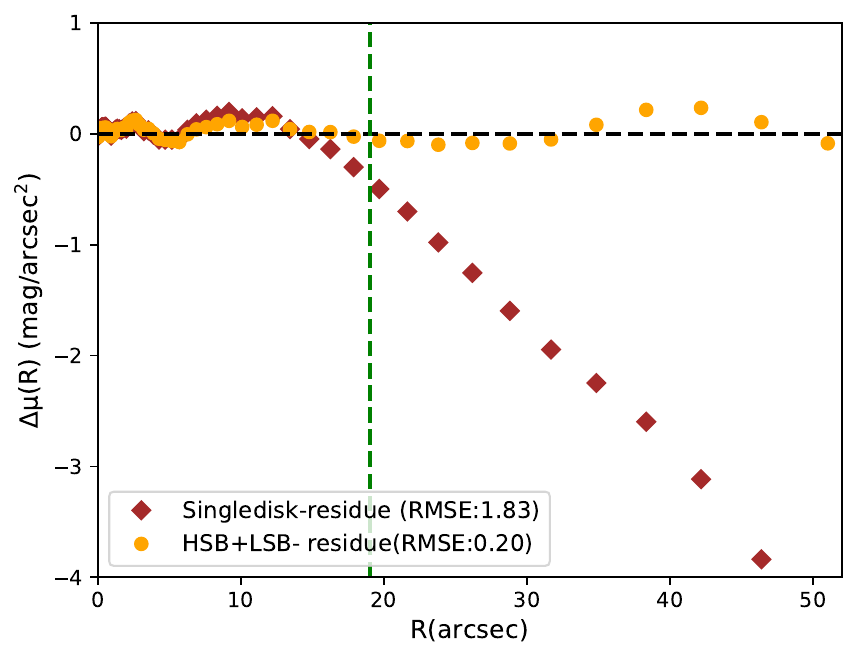}
\medskip
\includegraphics[width = 0.33\textwidth, height=0.24\textwidth]{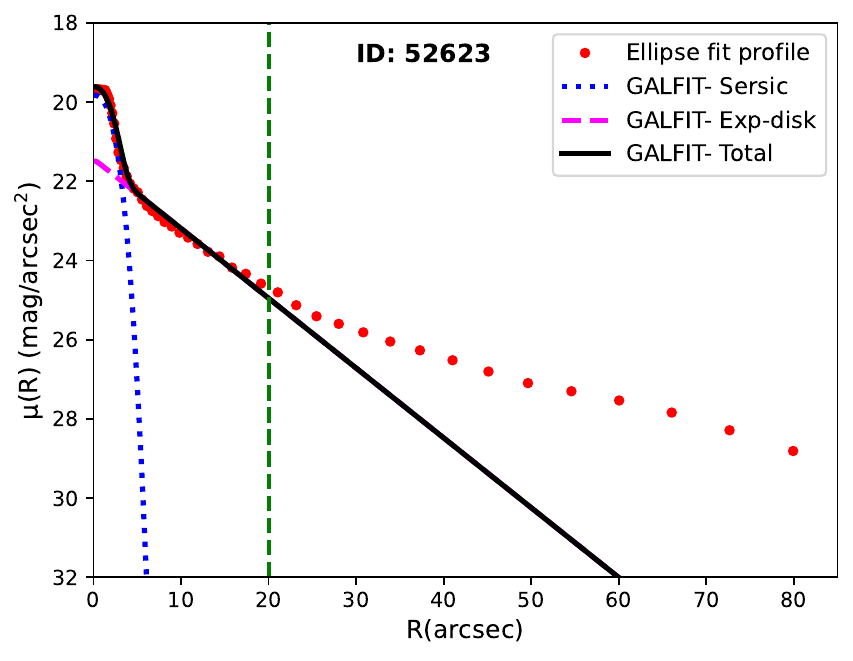}
\includegraphics[width = 0.33\textwidth, height=0.24\textwidth]{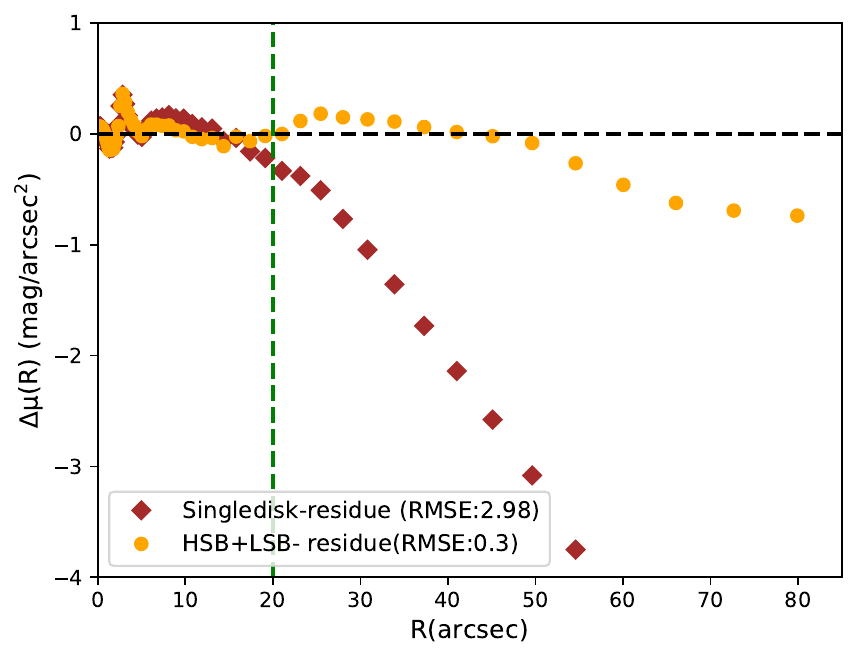}
\medskip
\includegraphics[width = 0.33\textwidth, height=0.24\textwidth]{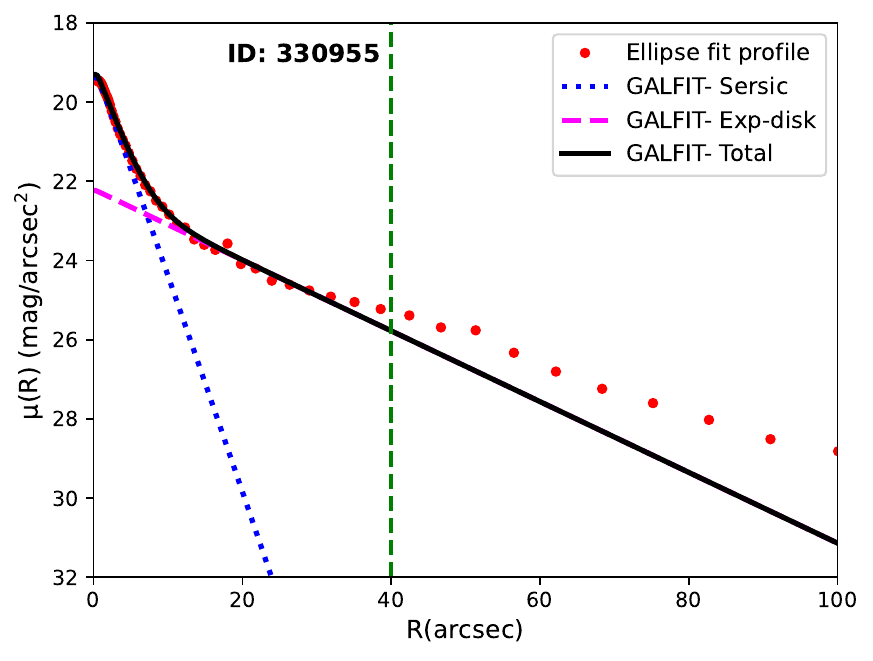}
\includegraphics[width = 0.33\textwidth, height=0.24\textwidth]{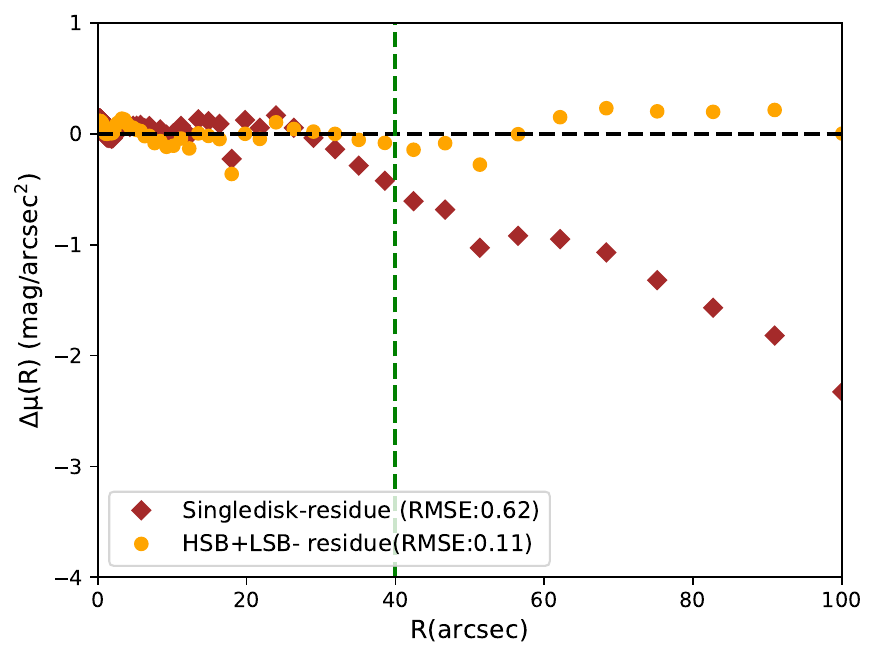}\par
\textbf{Figure A:} Single disk modeling of the double-disk gLSBs.
\end{figure*}

\begin{figure*}
\centering
\includegraphics[width = 0.33\textwidth, height=0.24\textwidth]{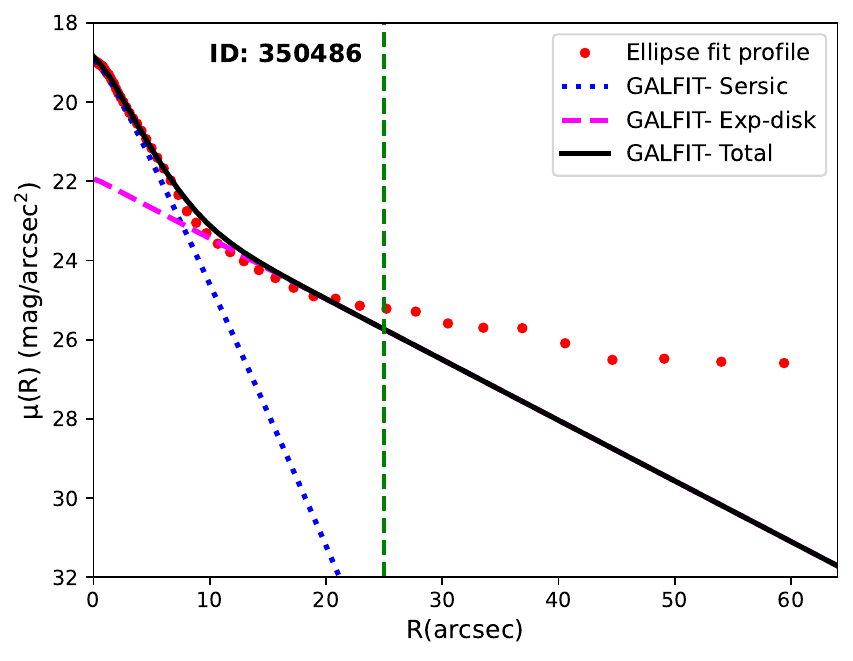}
\includegraphics[width = 0.33\textwidth, height=0.24\textwidth]{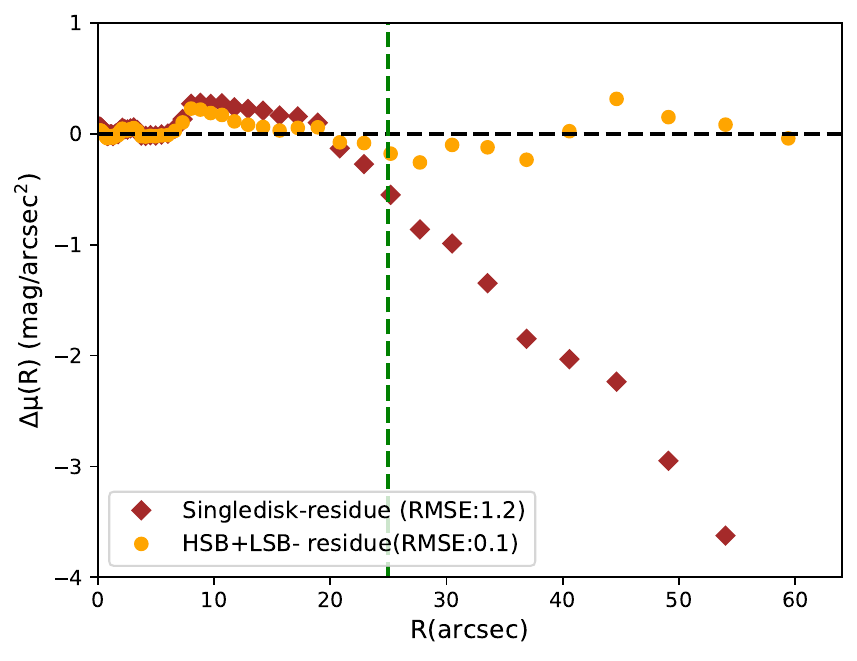}
\medskip
\includegraphics[width = 0.33\textwidth, height=0.24\textwidth]{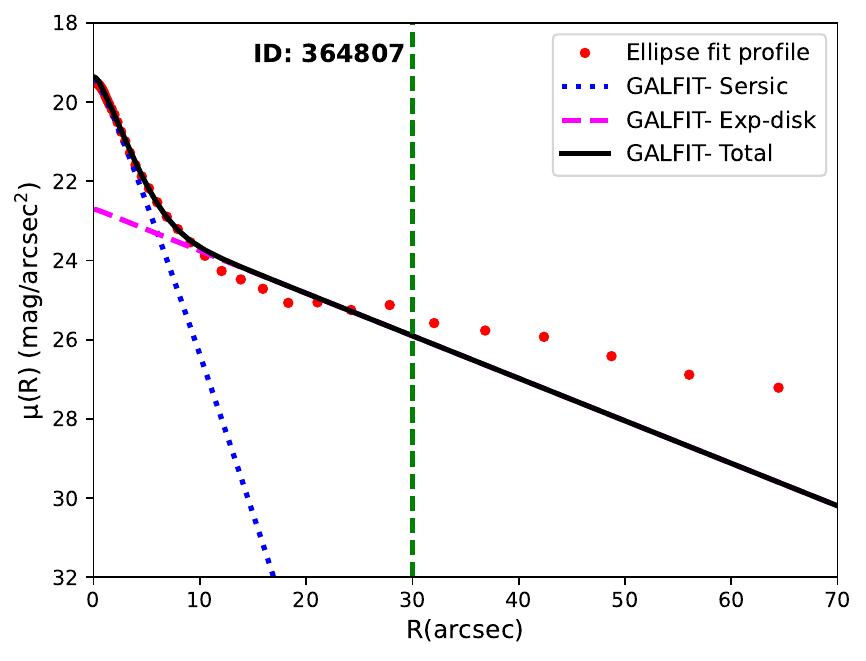}
\includegraphics[width = 0.33\textwidth, height=0.24\textwidth]{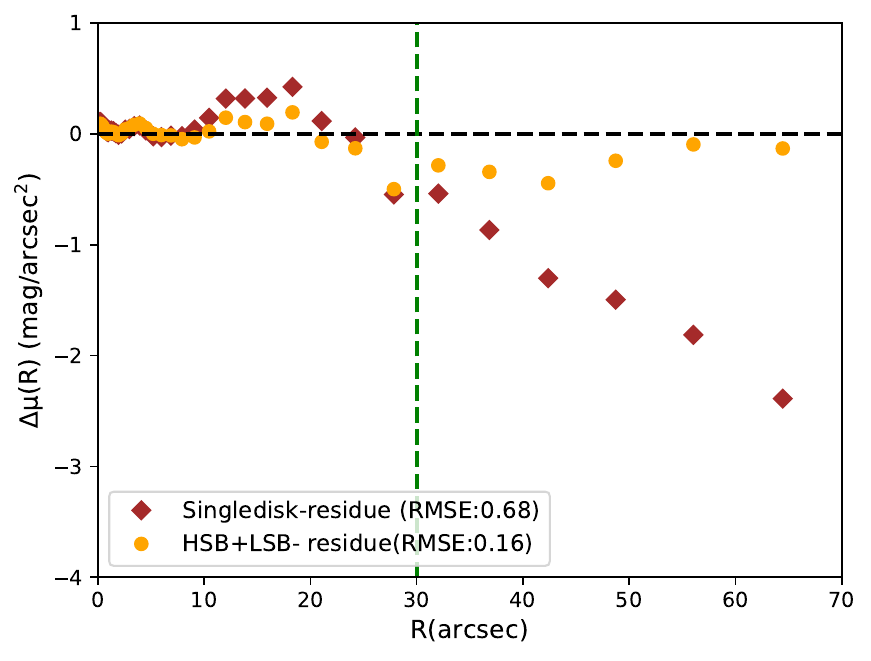}
\medskip
\includegraphics[width = 0.33\textwidth, height=0.24\textwidth]{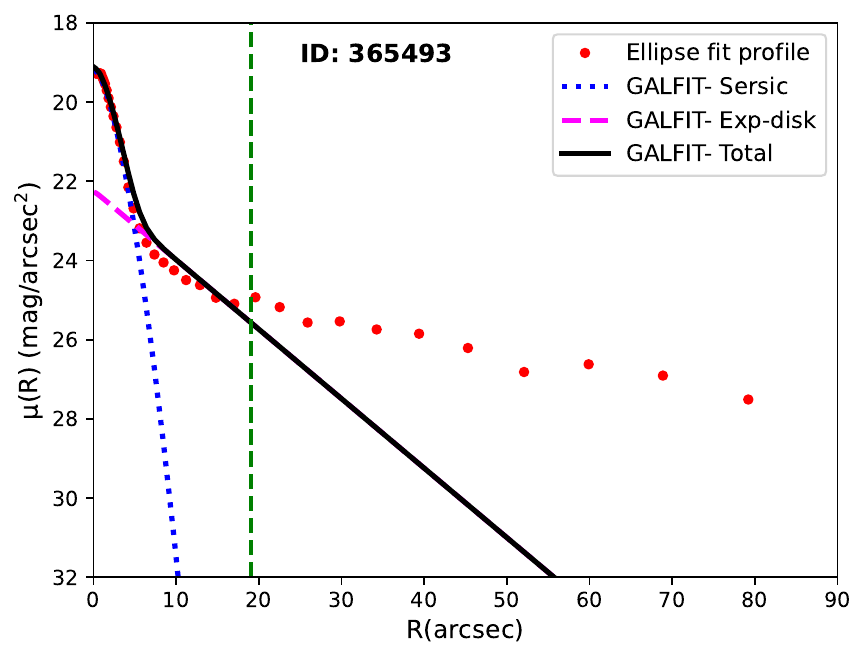}
\includegraphics[width = 0.33\textwidth, height=0.24\textwidth]{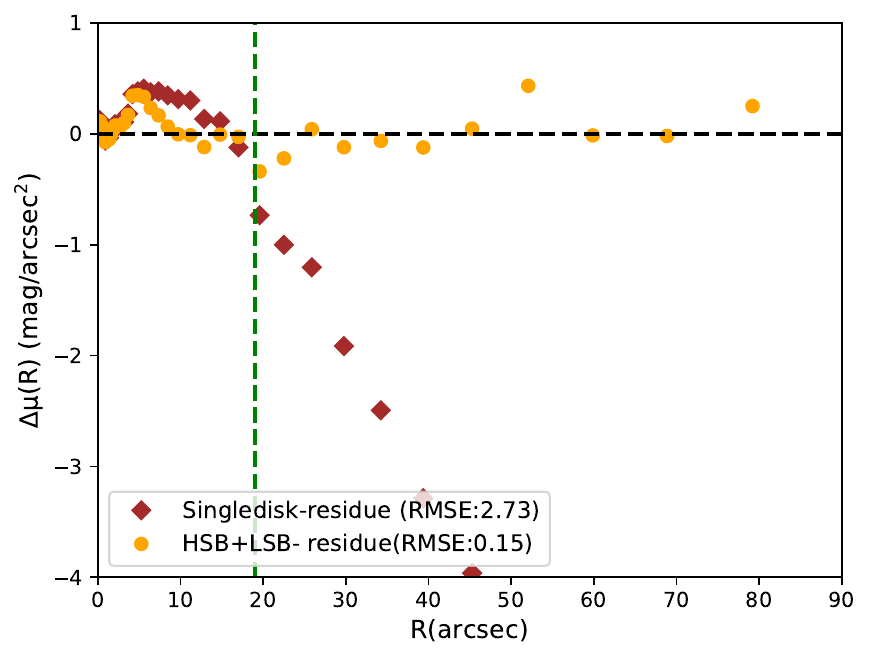}
\medskip
\includegraphics[width = 0.33\textwidth, height=0.24\textwidth]{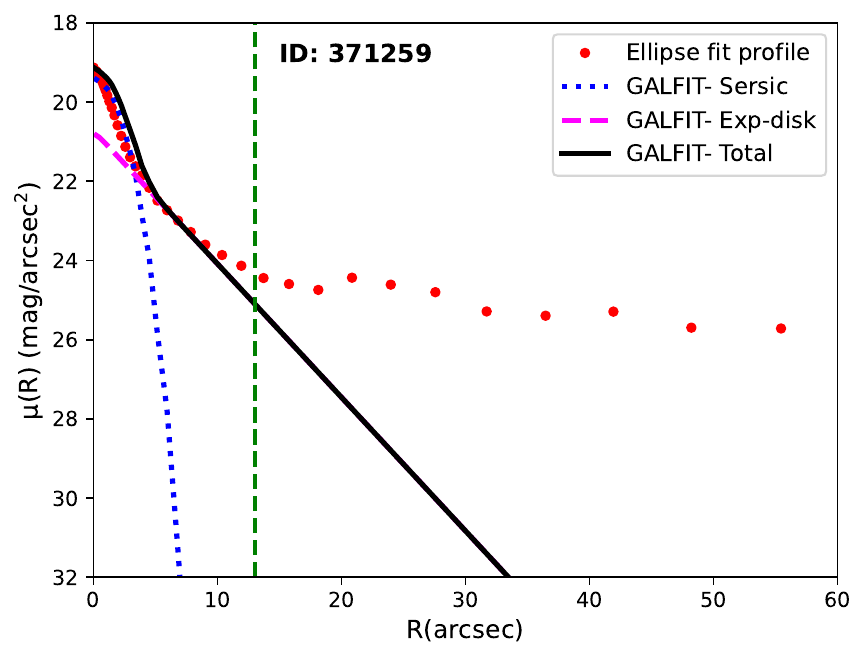}
\includegraphics[width = 0.33\textwidth, height=0.24\textwidth]{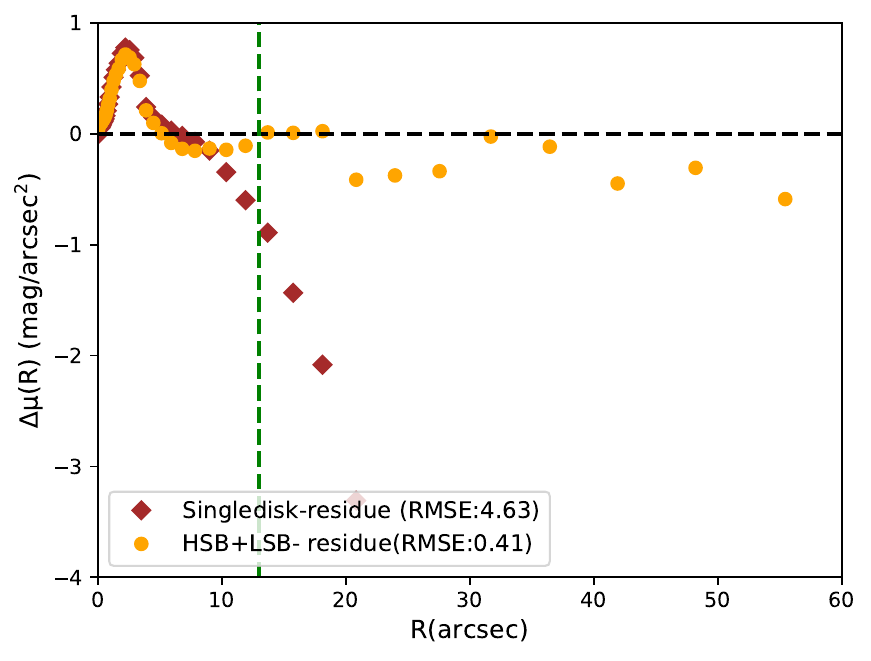}
\medskip
\includegraphics[width = 0.33\textwidth, height=0.24\textwidth]{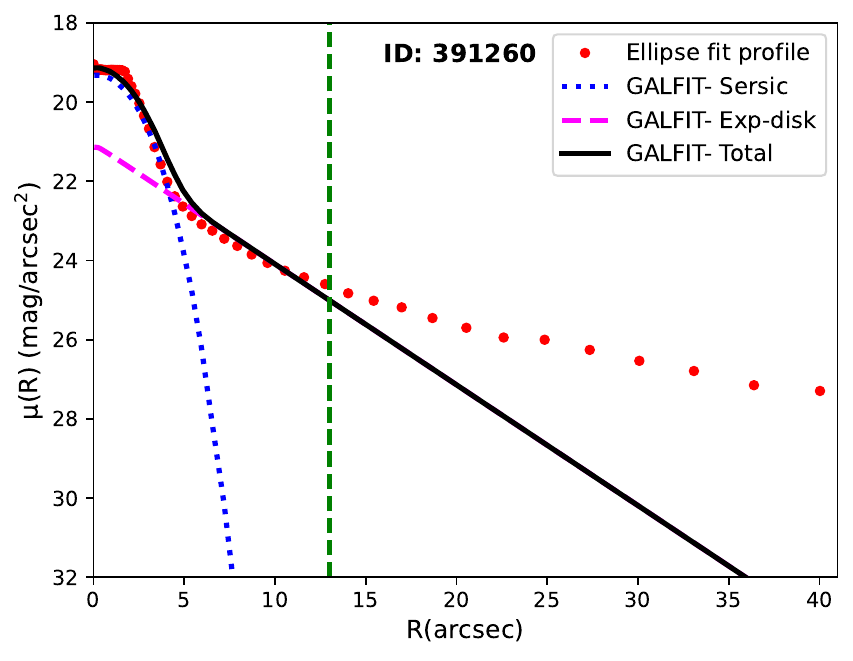}
\includegraphics[width = 0.33\textwidth, height=0.24\textwidth]{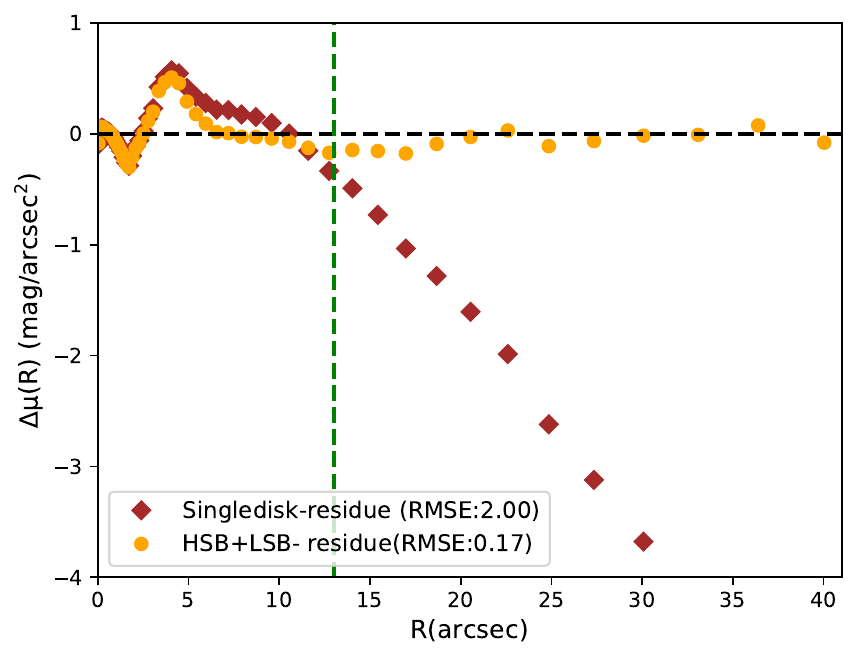}\par
Continuation of the above figure
\end{figure*}

\begin{figure*}
\centering
\includegraphics[width = 0.33\textwidth, height=0.24\textwidth]{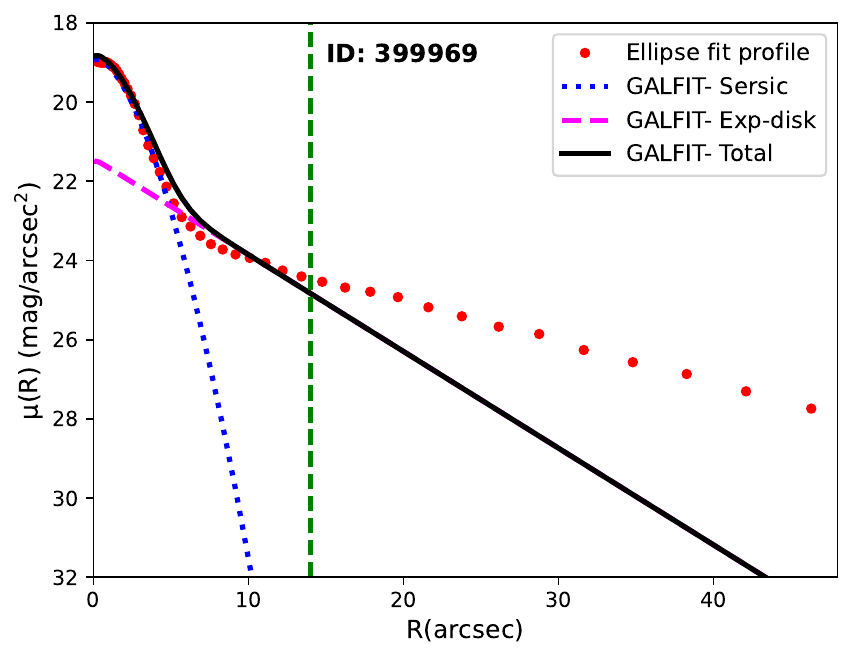}
\includegraphics[width = 0.33\textwidth, height=0.24\textwidth]{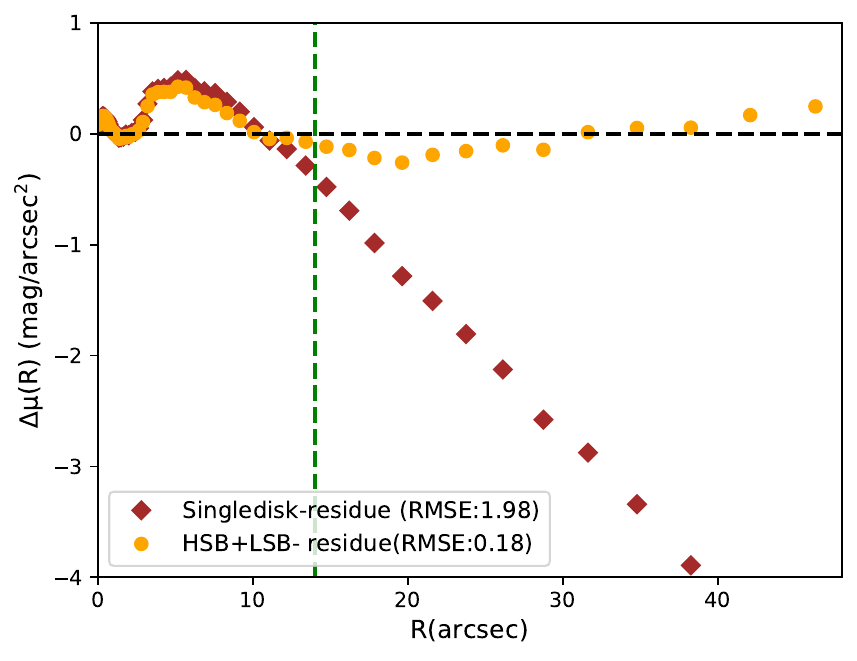}
\medskip
\includegraphics[width = 0.33\textwidth, height=0.24\textwidth]{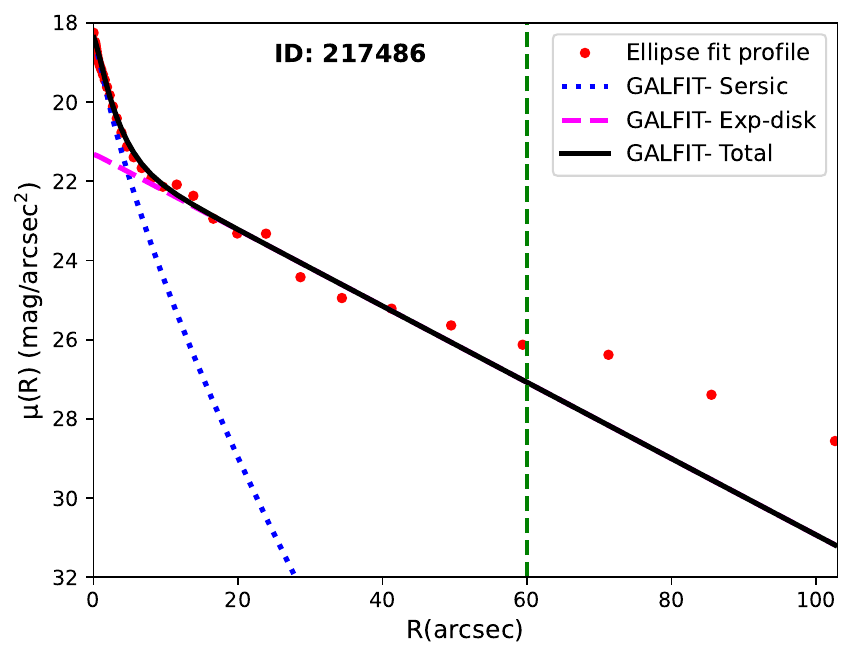}
\includegraphics[width = 0.33\textwidth, height=0.24\textwidth]{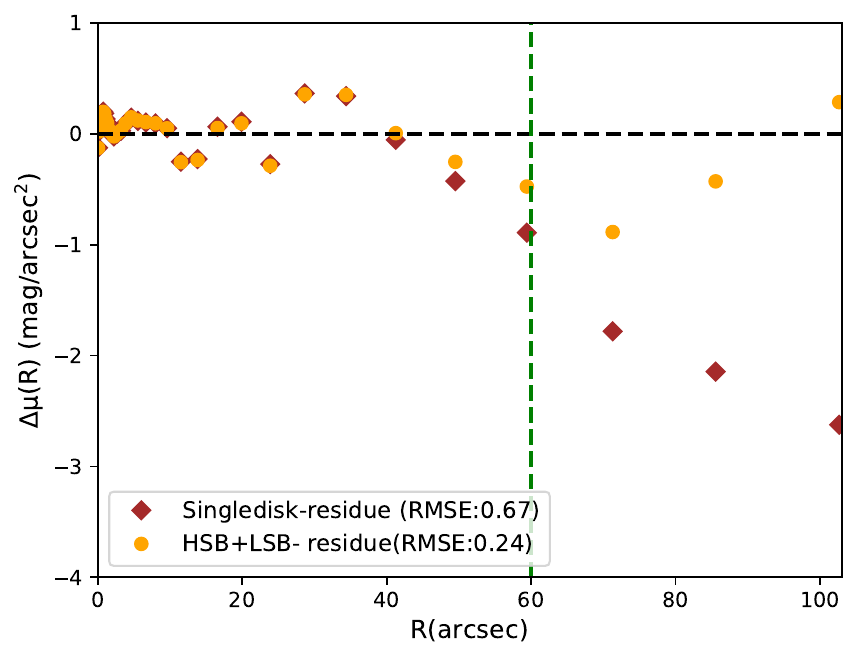}\par
Continuation of the above figure
\end{figure*}

\section{Massive, single exponential gLSBs from this work}
We show the surface brightness profiles along with the Sersic and the exponential disk components of the massive, single-exponential gLSBs detected in this work.
\begin{figure*}
\centering
\includegraphics[width = 0.31\textwidth, height=0.21\textwidth]{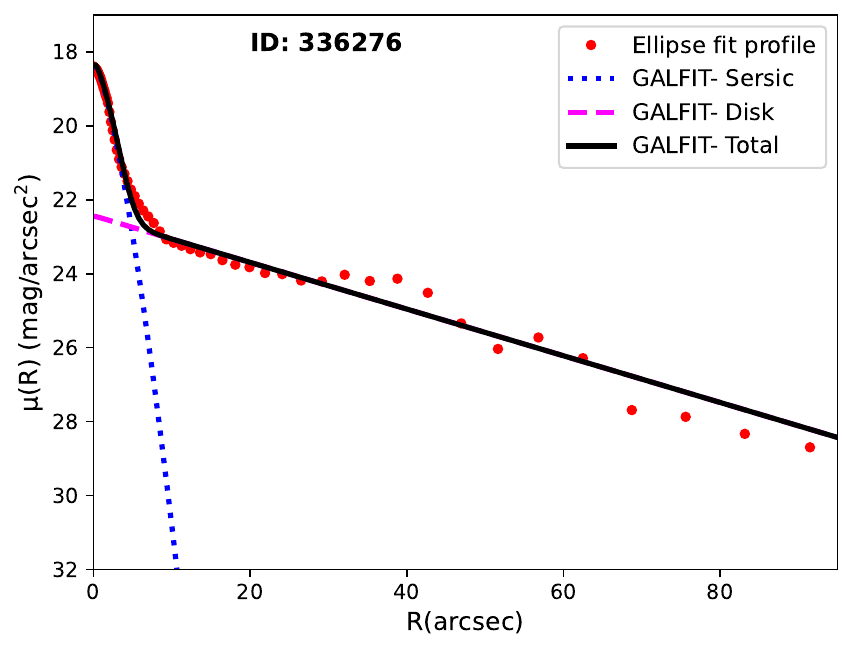}
\includegraphics[width = 0.31\textwidth, height=0.21\textwidth]{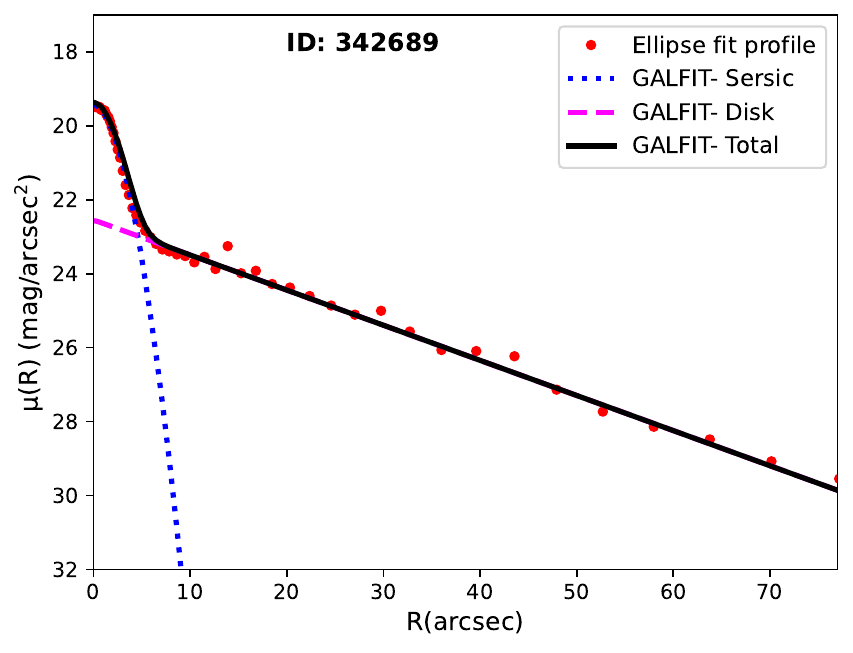}
\includegraphics[width = 0.31\textwidth, height=0.21\textwidth]{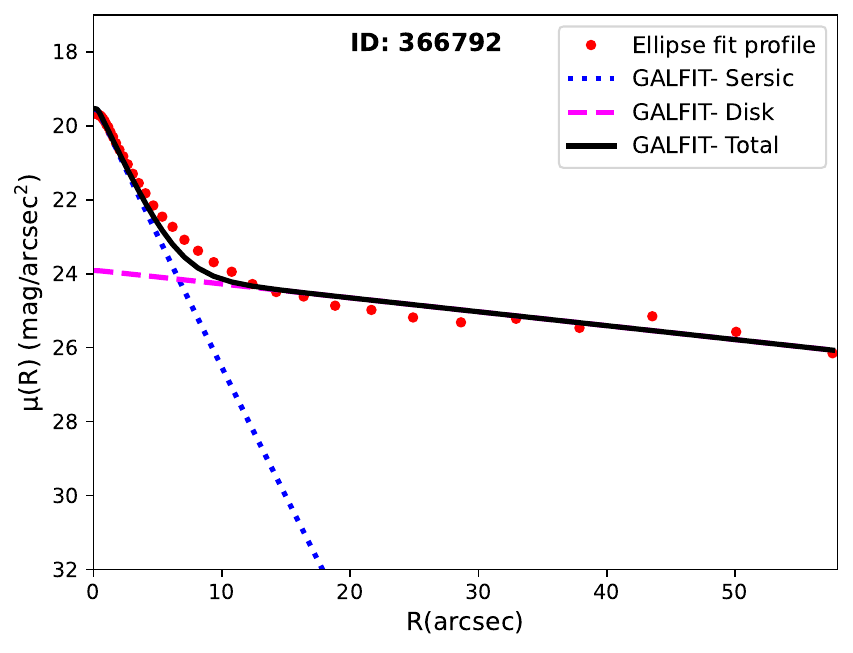}
\includegraphics[width = 0.31\textwidth, height=0.21\textwidth]{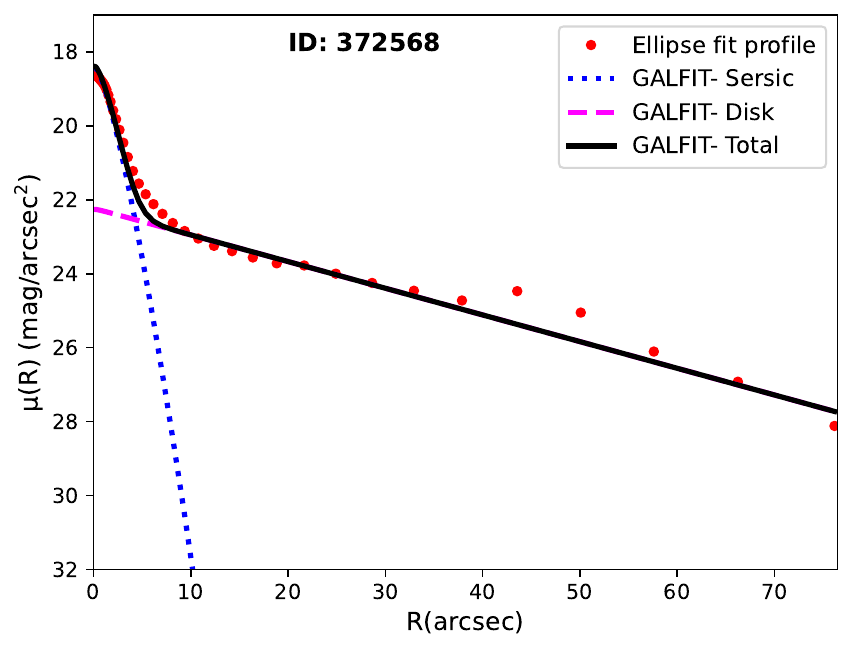}
\includegraphics[width = 0.31\textwidth, height=0.21\textwidth]{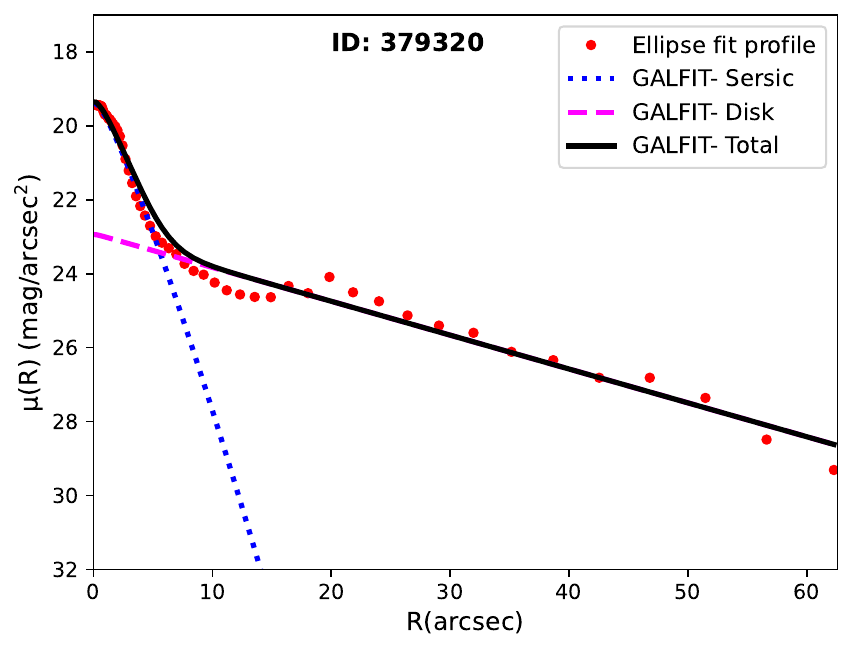}
\includegraphics[width = 0.31\textwidth, height=0.21\textwidth]{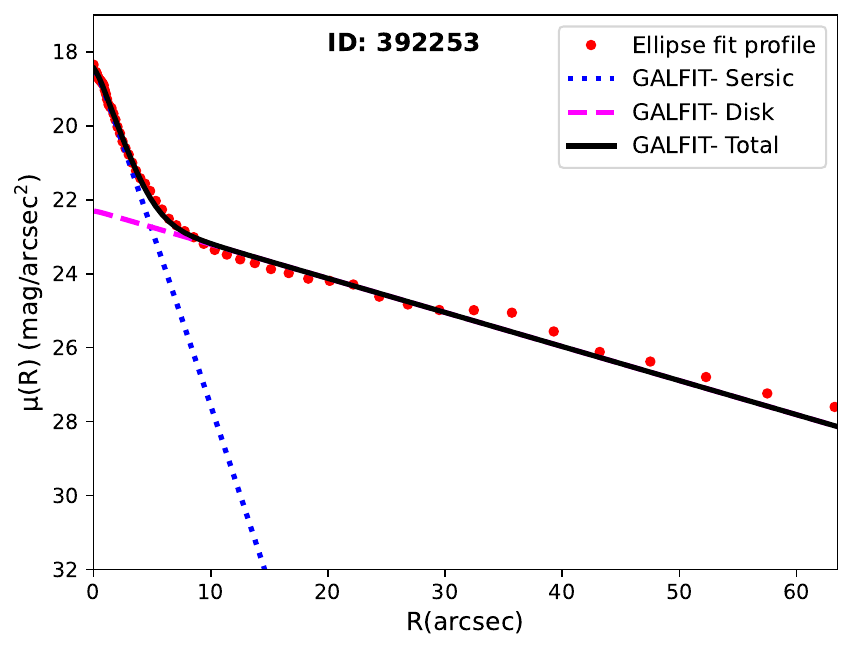}
\includegraphics[width = 0.31\textwidth, height=0.21\textwidth]{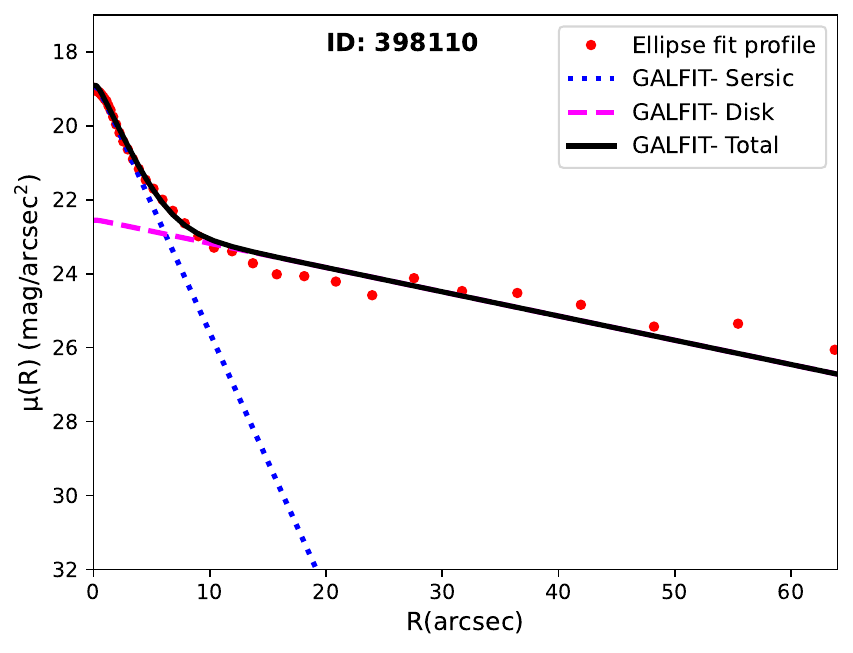}
\includegraphics[width = 0.31\textwidth, height=0.21\textwidth]{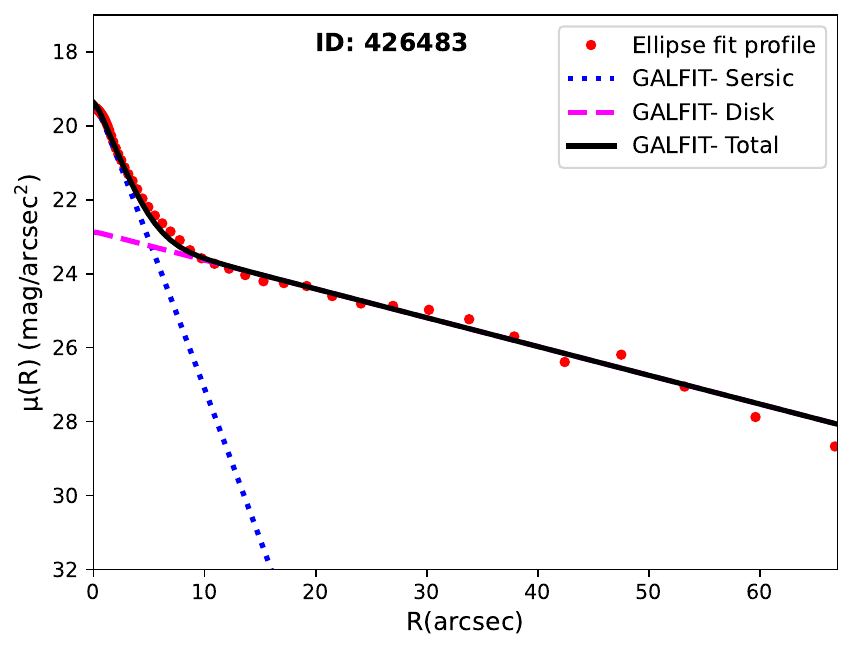}
\includegraphics[width = 0.31\textwidth, height=0.21\textwidth]{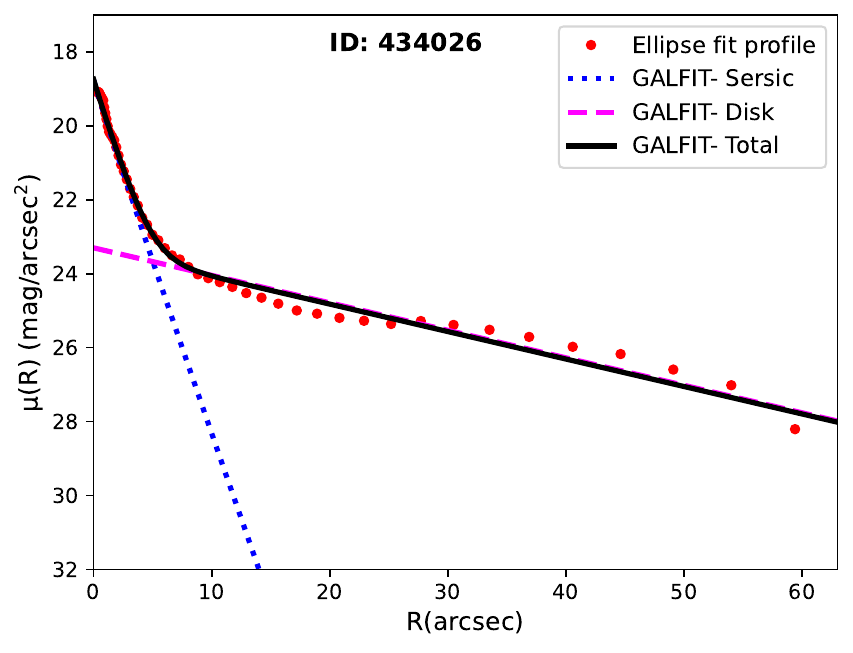}
\includegraphics[width = 0.31\textwidth, height=0.21\textwidth]{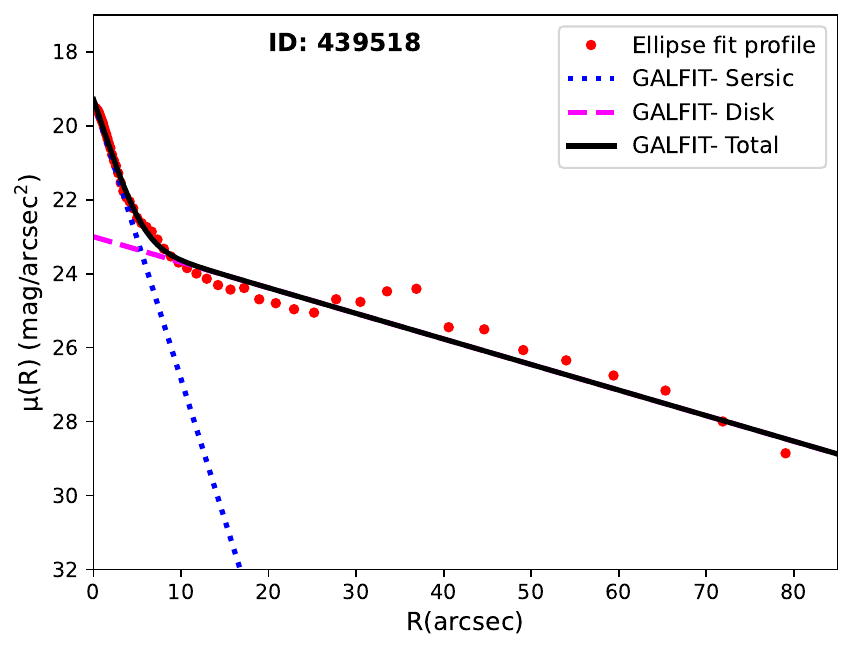}
\includegraphics[width = 0.31\textwidth, height=0.21\textwidth]{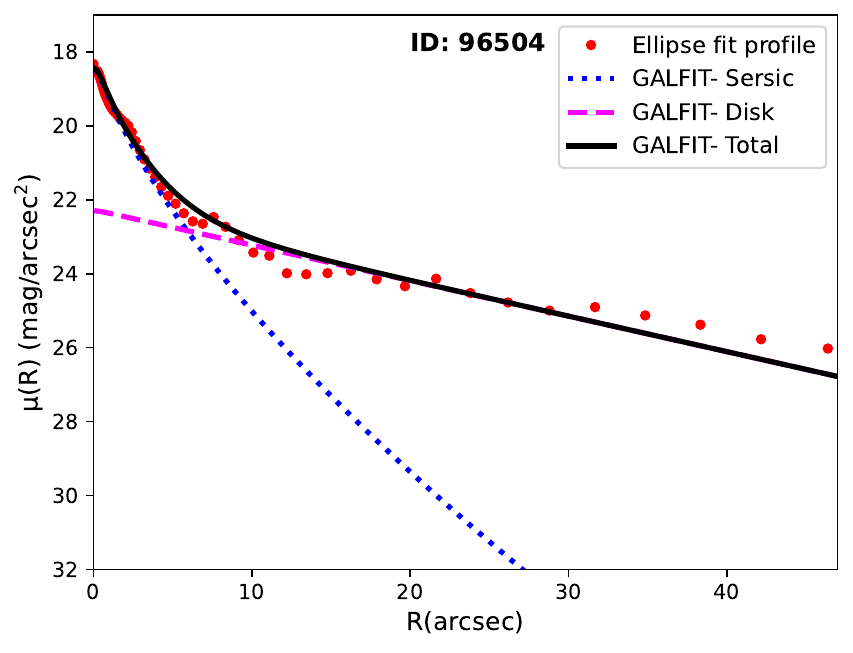}
\includegraphics[width = 0.31\textwidth, height=0.21\textwidth]{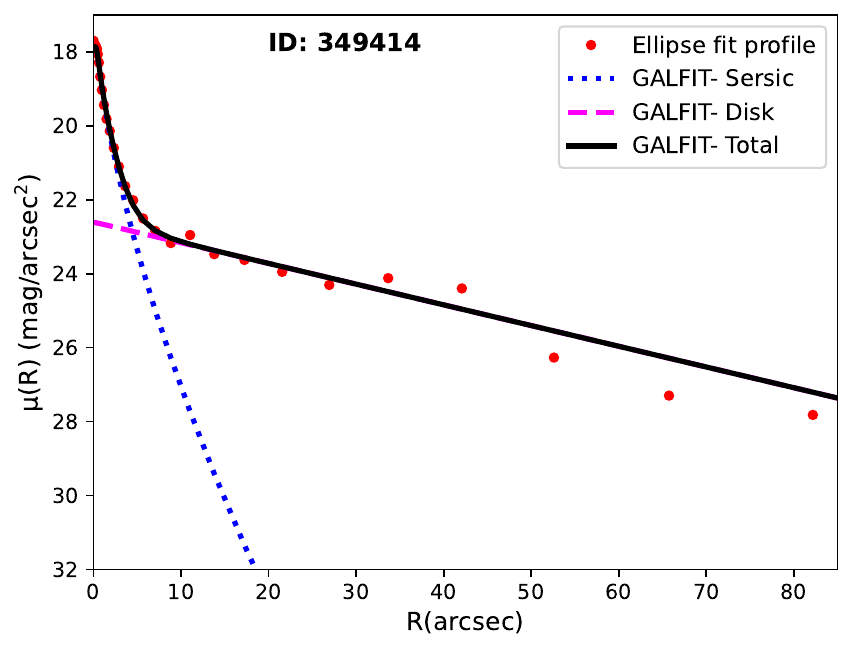}
\includegraphics[width = 0.31\textwidth, height=0.21\textwidth]{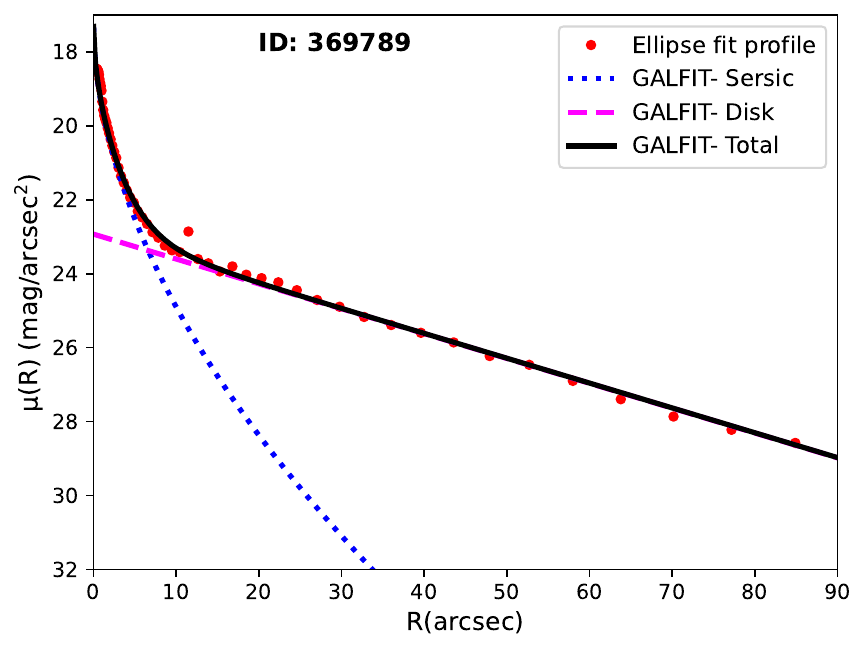}
\medskip
\includegraphics[width = 0.31\textwidth, height=0.21\textwidth]{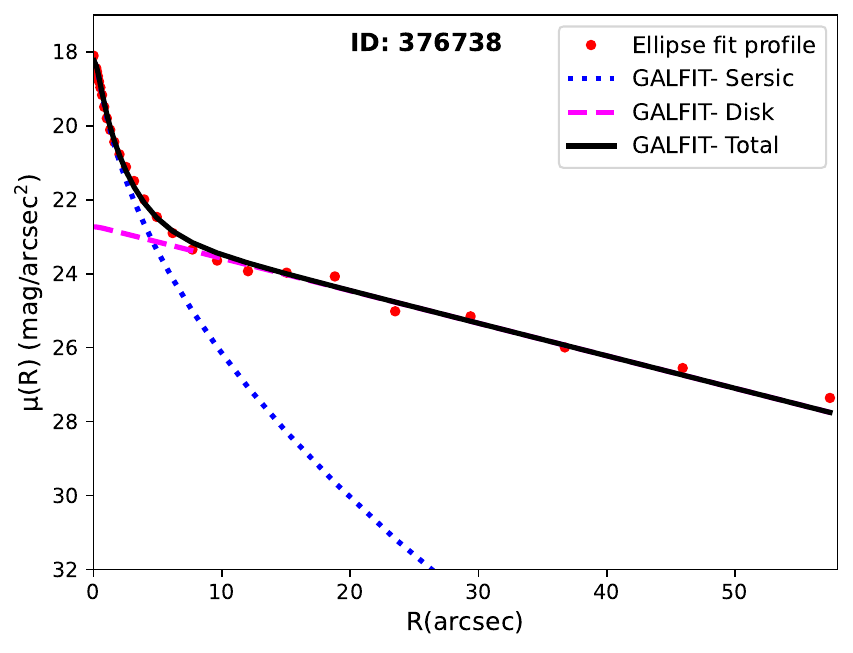}
\includegraphics[width = 0.31\textwidth, height=0.21\textwidth]{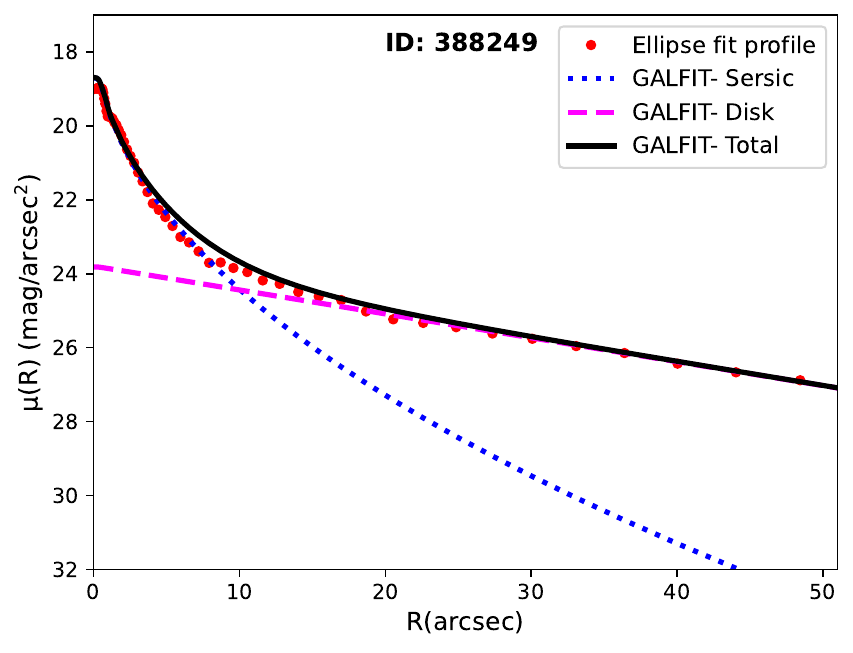}
\medskip
\includegraphics[width = 0.31\textwidth, height=0.21\textwidth]{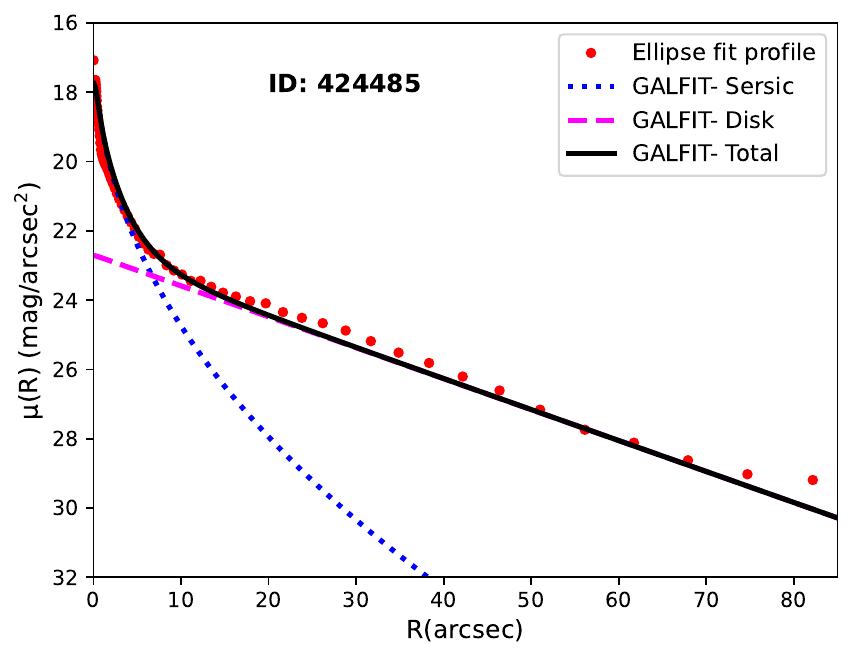}
\includegraphics[width = 0.31\textwidth, height=0.21\textwidth]{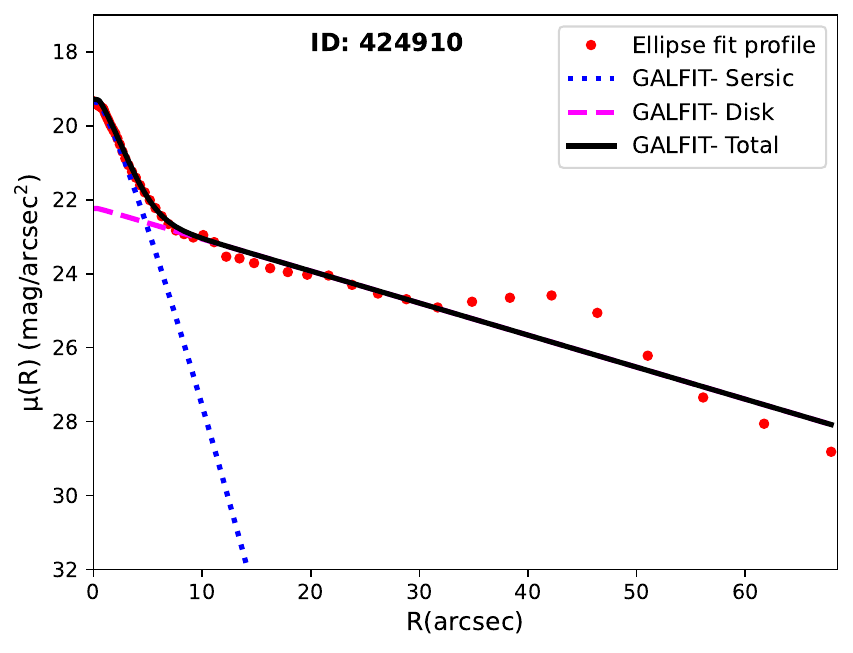}
\includegraphics[width = 0.31\textwidth, height=0.21\textwidth]{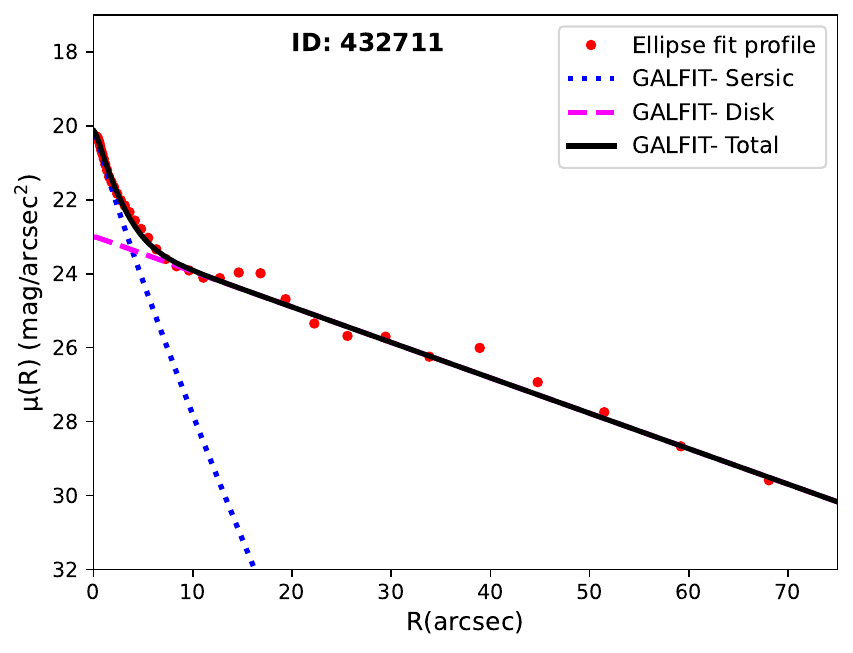}\par
\textbf{Figure B:}Single disk gLSBs from all the disk samples.
\end{figure*}

\begin{figure*}
\centering
\includegraphics[width = 0.31\textwidth, height=0.21\textwidth]{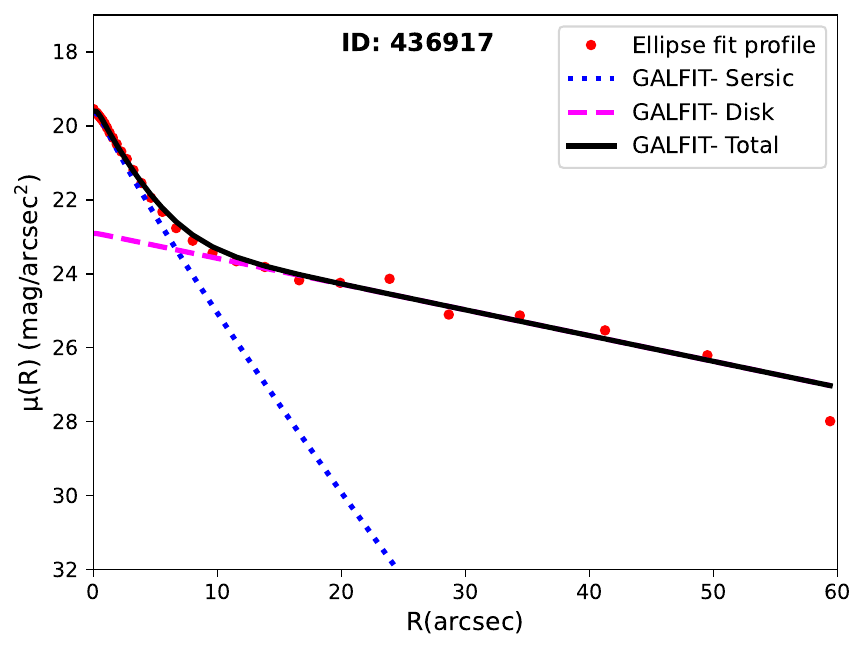}
\includegraphics[width = 0.31\textwidth, height=0.21\textwidth]{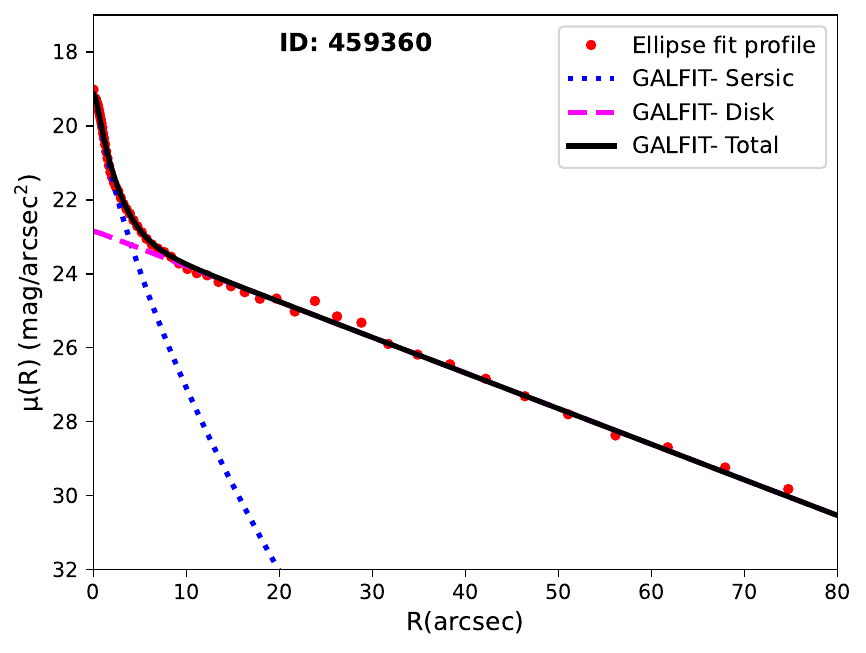}
\includegraphics[width = 0.31\textwidth, height=0.21\textwidth]{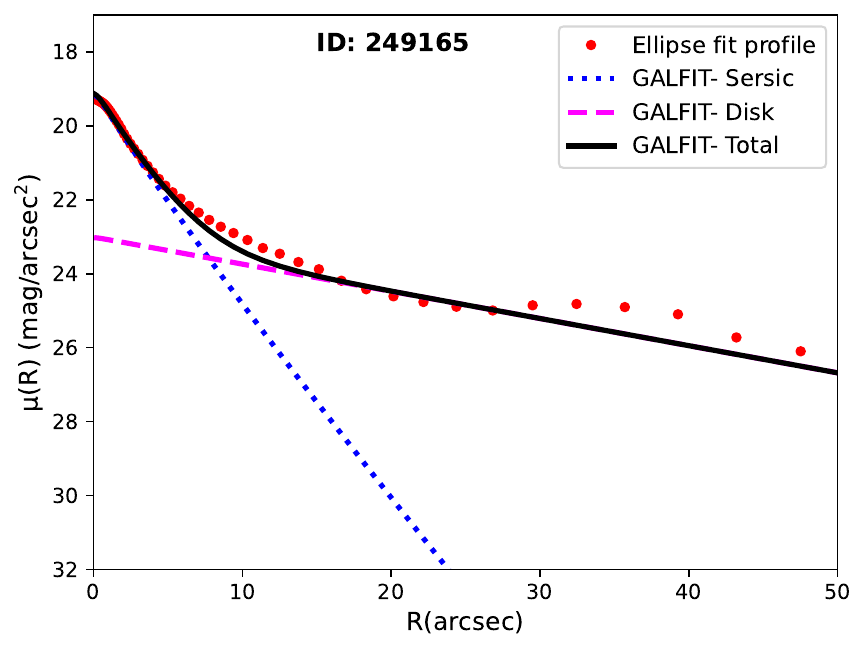}
\includegraphics[width = 0.31\textwidth, height=0.21\textwidth]{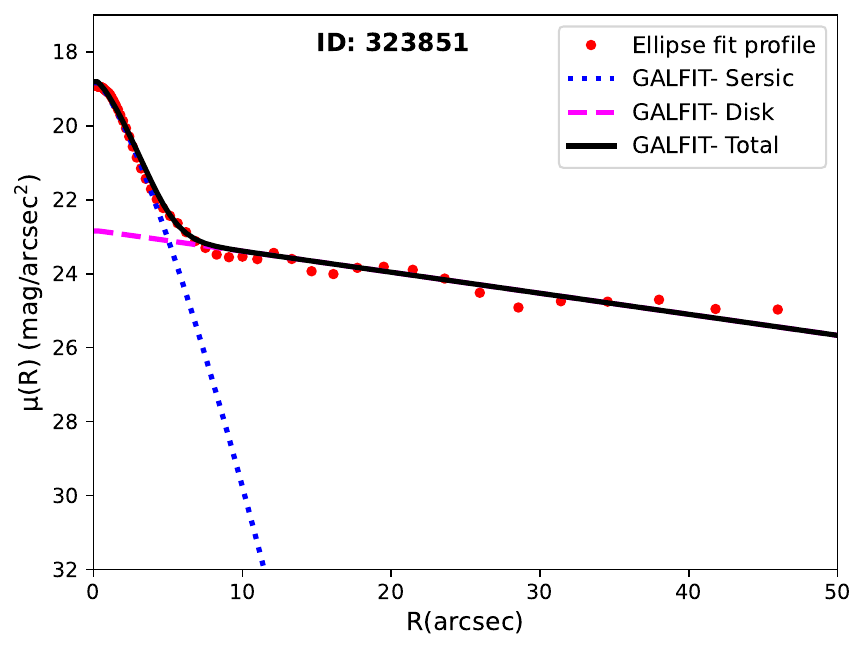}
\includegraphics[width = 0.31\textwidth, height=0.21\textwidth]{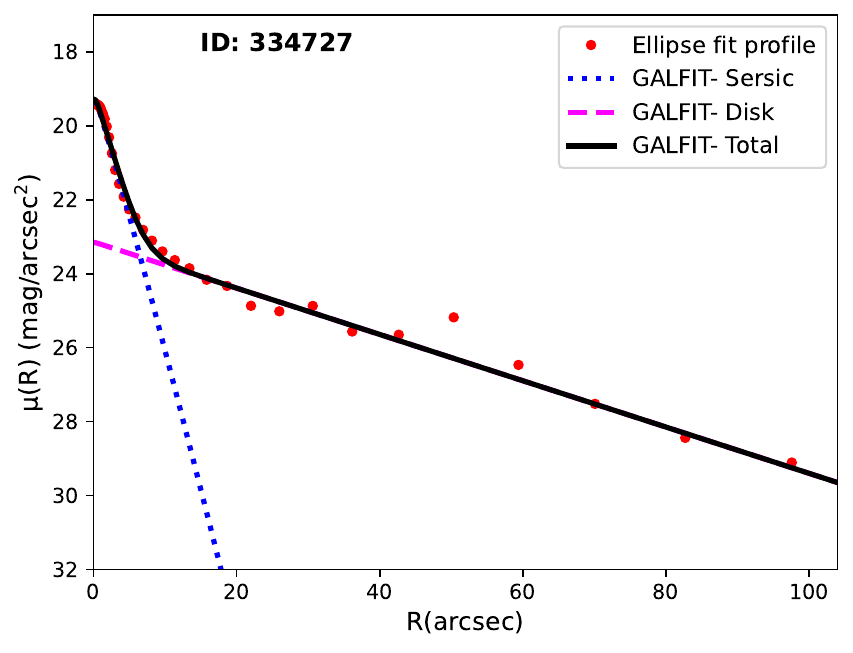}
\includegraphics[width = 0.31\textwidth, height=0.21\textwidth]{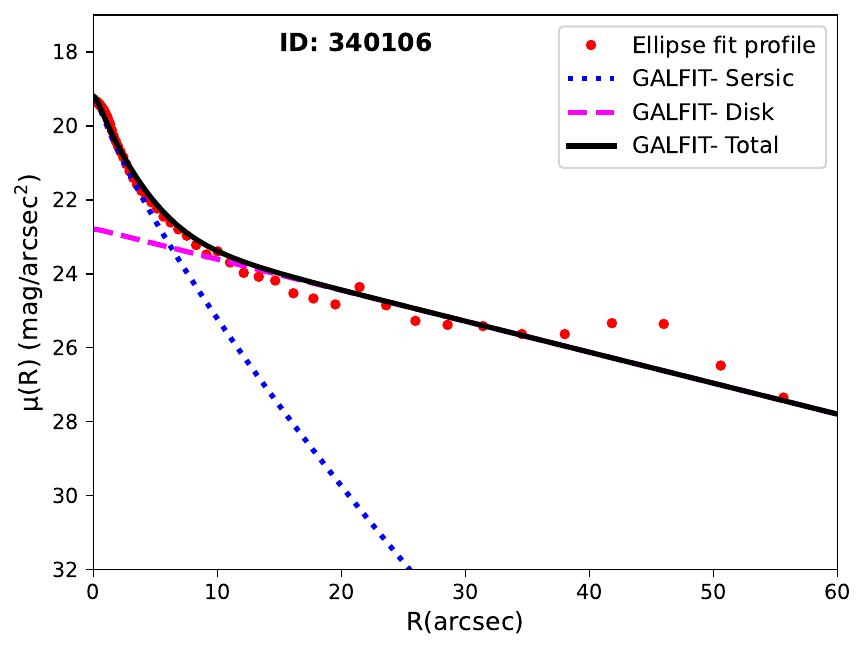}
\includegraphics[width = 0.31\textwidth, height=0.21\textwidth]{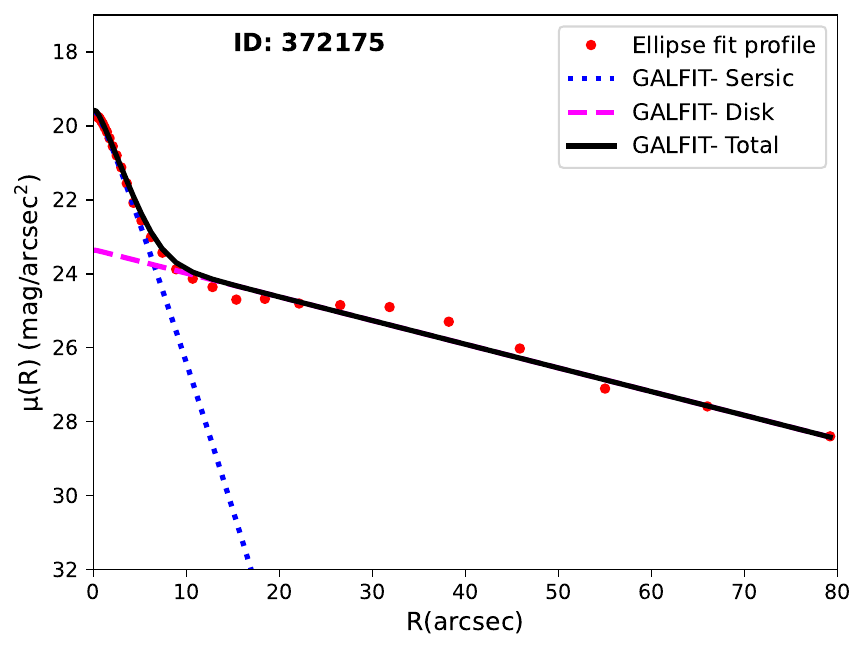}
\includegraphics[width = 0.31\textwidth, height=0.21\textwidth]{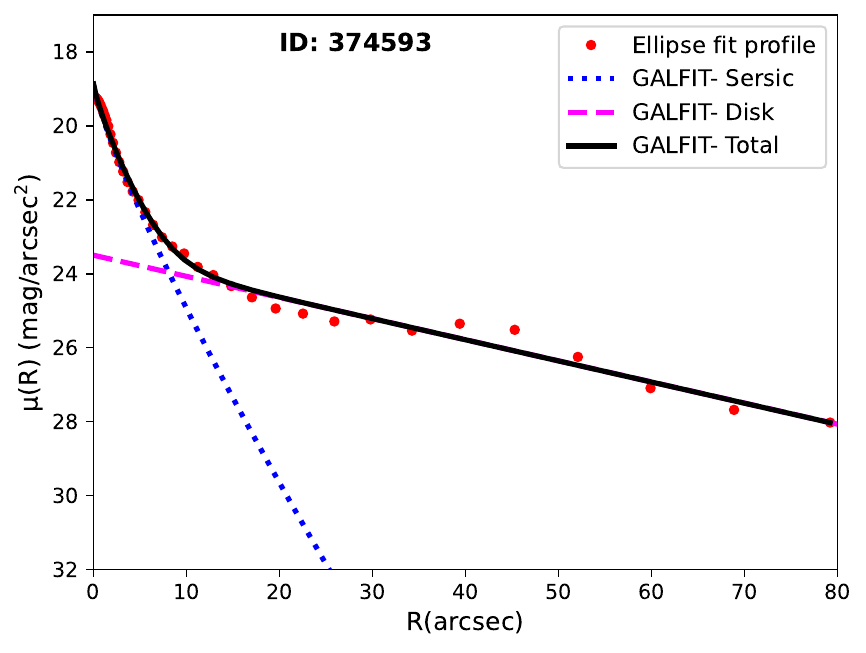}
\medskip
\includegraphics[width = 0.31\textwidth, height=0.21\textwidth]{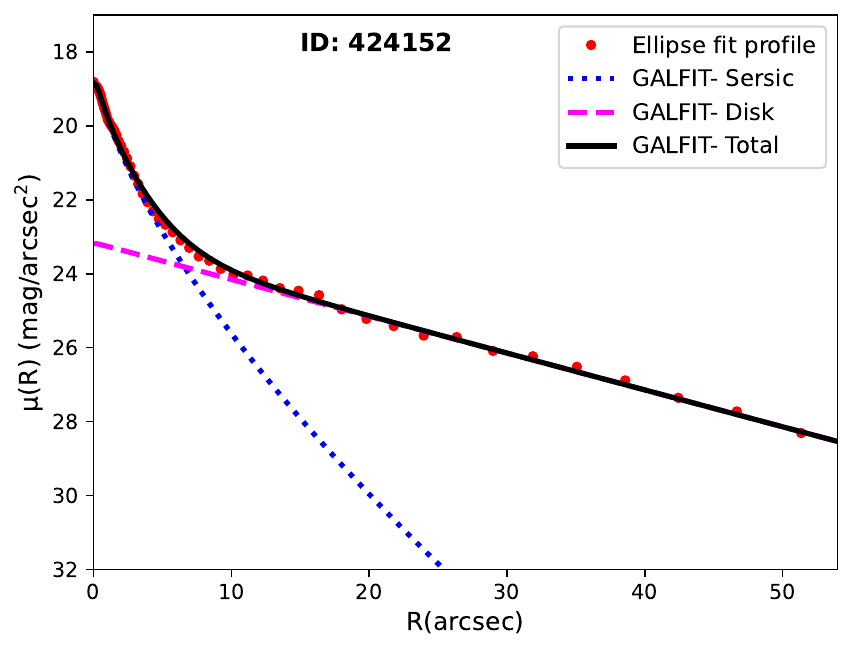}
\includegraphics[width = 0.31\textwidth, height=0.21\textwidth]{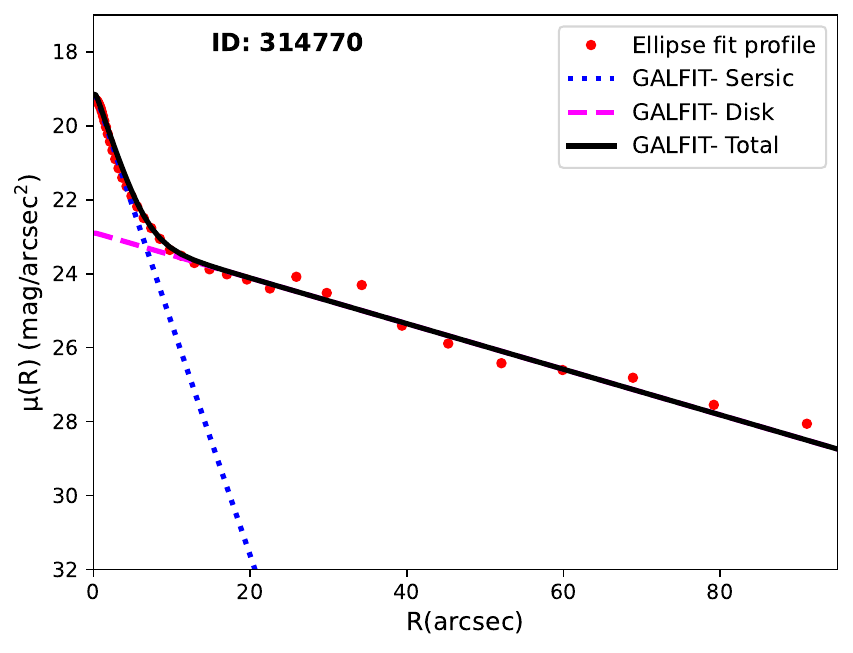}
\includegraphics[width = 0.31\textwidth, height=0.21\textwidth]{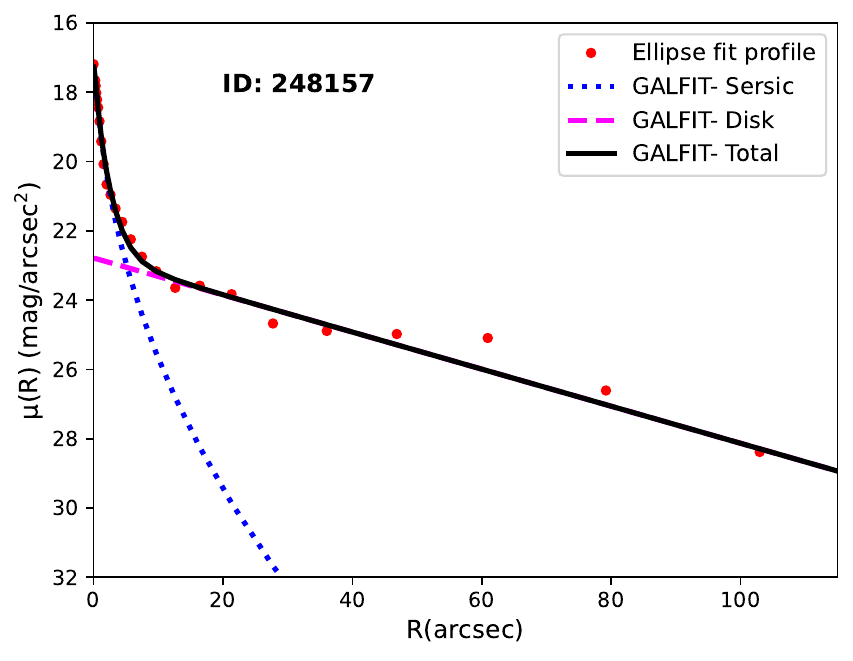}
\includegraphics[width = 0.31\textwidth, height=0.21\textwidth]{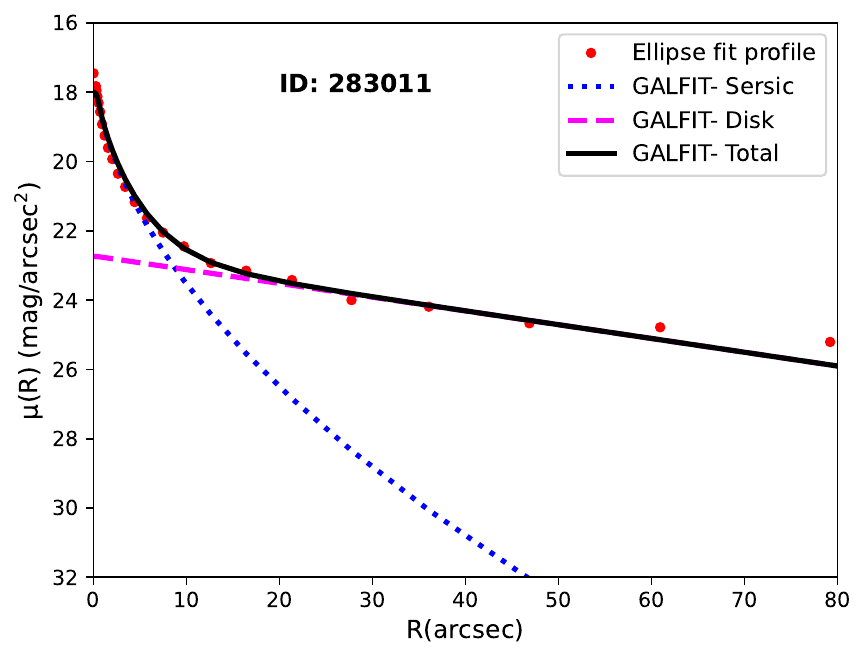}
\includegraphics[width = 0.31\textwidth, height=0.21\textwidth]{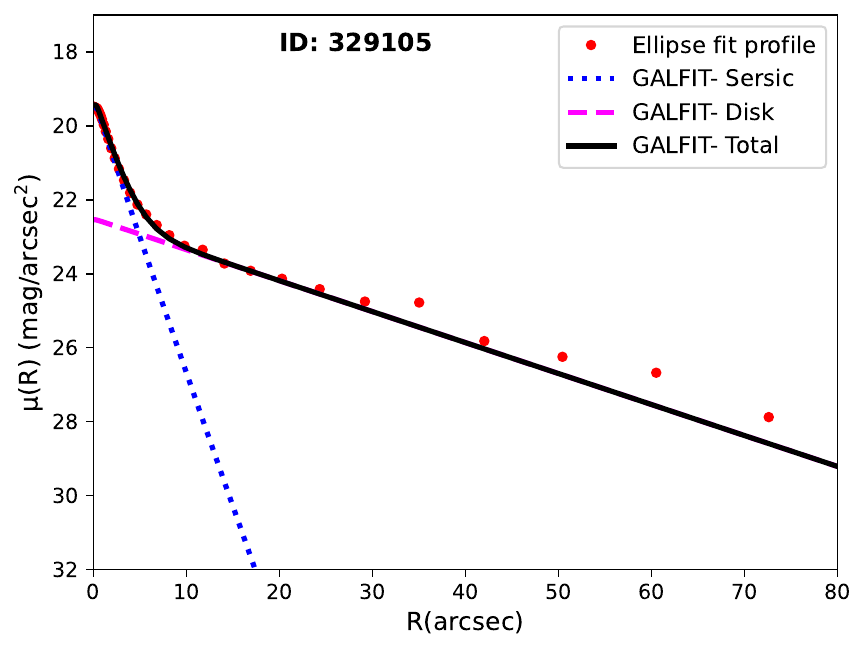}
\includegraphics[width = 0.31\textwidth, height=0.21\textwidth]{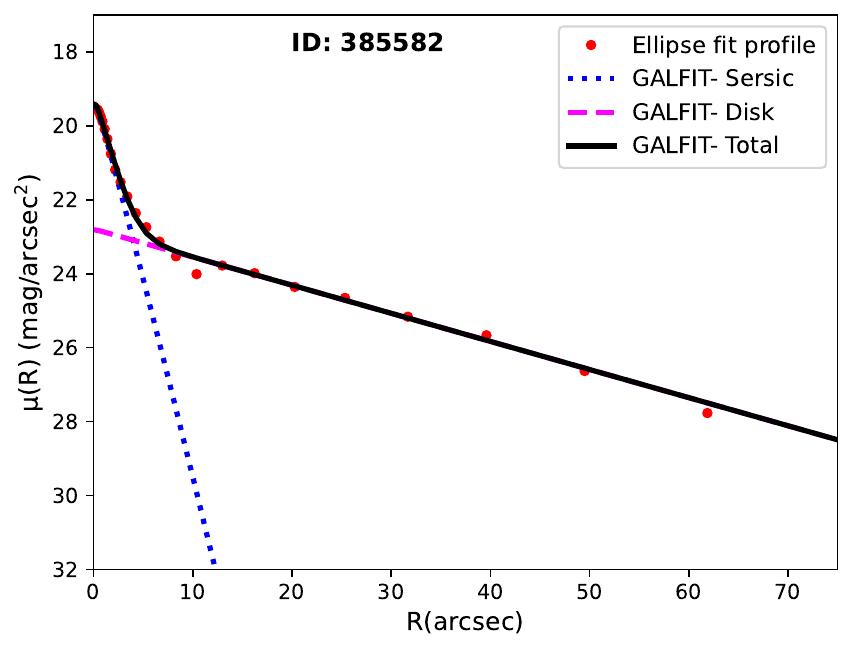}\par
Continuation of the above figure
\end{figure*}

\bibliography{ms}{}
\bibliographystyle{aasjournal}

\end{document}